\documentclass{aastex62}

\graphicspath{{./}{figures/}}
\received{December 2022}
\revised{March 2023}
\accepted{March 2023}
\submitjournal{ApJ}

\shorttitle{}
\shortauthors{Lacy}

\begin{document}

\title{Self-consistent Models of Y Dwarf Atmospheres with Water Clouds and Disequilibrium Chemistry}

\correspondingauthor{Brianna Lacy}
\email{brianna.lacy@utexas.edu}

\author[0000-0002-0786-7307]{Brianna Lacy}
\altaffiliation{51 Pegasi b Fellow}
\affil{Department of Astronomy, University of Texas at Austin, Austin, TX, 78712}
\author[0000-0002-3099-5024]{Adam Burrows}
\affil{Department of Astrophysical Sciences, Princeton University, Princeton, NJ, 08544}

\begin{abstract}

Y dwarfs are the coolest spectral class of brown dwarf. They have effective temperatures less than 500 K, with the coolest detection as low as $\sim$250 K. Their spectra are shaped predominantly by gaseous water, methane, and ammonia. At the warmer end of the Y dwarf temperature range, spectral signatures of disequilibrium carbon monoxide have been observed. Cooler Y dwarfs could host water clouds in their atmospheres. Since they make up the low-mass tail of the star formation process, and are a valuable analogue to the atmospheres of giant gaseous exoplanets in a temperature range that is difficult to observe, understanding Y dwarf atmospheric compositions and processes will both deepen our understanding of planet and star formation and provide a stepping stone towards characterizing cool exoplanets. JWST spectral observations are anticipated to provide an unprecedented level of detail for these objects, and yet published self-consistent model grids do not accurately replicate even the existing HST and ground-based observations. In this work, we present a new suite of 1-d radiative-convective equilibrium models to aid in the characterization of Y dwarf atmospheres and spectra. We compute clear, cloudy, equilibrium-chemistry and disequilibrium-chemistry models, providing a comprehensive suite of models in support of the impending JWST era of panchromatic Y dwarf characterization. Comparing these models against current observations, we find that disequilibrium CH$_4$-CO and NH$_3$-N$_2$ chemistry and the presence of water clouds can bring models and observations into better, though still not complete, agreement. 

\end{abstract}


\section{Introduction} \label{sec:intro}

Brown dwarfs are the low-mass tail of the star formation process. They form by direct collapse, but have too little mass to attain the high pressures and temperatures needed to fuse hydrogen to helium in their cores. Without the energy provided by long-term nuclear fusion, they continually cool over time, at a rate depending on metallicity and mass \citep{Burrows1997}. Detecting and characterizing brown dwarfs is an essential piece of the star formation puzzle. Furthermore, understanding the processes which shape their atmospheres will enable useful comparisons to exoplanets, since they occupy similar regimes in temperature and composition (\citealt{Faherty2016}; \citealt{Burrows1997};\citealt{Kirkpatrick2005}).

Y dwarfs are the oldest and smallest brown dwarfs which have had time to reach temperatures less than $\sim$500 K (\citealt{Burrows2000}; \citealt{Cushing2011}; \citealt{Kirkpatrick2011}). At these low temperatures, CO is converted into CH$_4$, ammonia absorption increases, and water clouds can begin to form (\citealt{Lodders2002}; \citealt{Burrows2003b}; \citealt{Burrows1999}). The Y spectral class is bounded at higher temperatures by a reversal in the J - K and J - H color trend seen for later T dwarfs (\citealt{Burrows2003b}; \citealt{Cushing2011}; \citealt{Kirkpatrick2012}). Such cold objects, when nearby, present an opportunity to study temperate atmospheres at a finer level of detail than we will likely be able to do for exoplanets in the near future. The coolest Y dwarf detected, WISE 0855,  has a temperature of only $\sim$250 K (\citealt{Luhman2011}; \citealt{Luhman2014a}). It is thus similar in effective temperature to Earth and the future exoearths we hope to characterize, albeit in an H$_2$-He-dominated rather than N$_2$ dominated atmosphere. 

JWST's large collecting area, broad infrared wavelength coverage, and location at L2 all make it perfect for observing these cool faint objects. Several early-mission programs in GTO and Cycle 1 are already set to collect spectra and photometry of a large sample of Y dwarfs (GTO 1189\footnote{https://www.stsci.edu/jwst/science-execution/program-information.html?id=1189}, GTO 1230\footnote{https://www.stsci.edu/jwst/science-execution/program-information.html?id=1230}, GO 2124\footnote{https://www.stsci.edu/jwst/science-execution/program-information.html?id=2124}, GO 2302\footnote{https://www.stsci.edu/jwst/science-execution/program-information.html?id=2302}, GO 2327\footnote{https://www.stsci.edu/jwst/science-execution/program-information.html?id=2327}). In anticipation of a new era of Y dwarf characterization with JWST, we present a comprehensive set of models for Y dwarf atmospheres incorporating an updated suite of molecular and atomic opacities, water clouds, and disequilibrium chemistry.

Detections of Y dwarfs accelerated with the advent of all-sky infrared surveys, particularly the Wide-Field Infrared Survey Explorer (WISE, \citealt{Wright2010}; \citealt{Reid2008}; \citealt{Cushing2011}). Y dwarfs are usually identified by their extremely red colors and high proper motions, implying low temperatures and small distances (\citealt{Meisner2020}). \cite{Kirkpatrick2019} provide Spitzer-based parallaxes for a sample of $\sim$50 known Y dwarfs. A set of Hubble Space Telescope near-infrared spectra are currently available for fifteen objects, published by \citealt{Schneider2015}, including re-reductions of some previously published data. \cite{Miles2020} collected low-resolution (R $\sim$ 300) M band spectra for a range of seven late T through Y dwarf substellar objects and Jupiter, using Gemini/GNIRS. Three objects in their sample were Y dwarfs. Other works provided spectral observations of one or two objects at a time (\citealt{Cushing2011}; \citealt{Pinfield2014}; \citealt{Cushing2021}; \citealt{Leggett2016}; \citealt{Morley2018}). Multi-band photometry using WISE, GEMINI and Spitzer has been published in a number of studies (\citealt{Cutri2013}; \citealt{Leggett2013}; \citealt{Leggett2015}; \citealt{Leggett2017}; \citealt{Kirkpatrick2019}; \citealt{Martin2018}). We anticipate this sample of Y dwarf observational data will expand quickly with the advent of JWST.

There are two large sets of cloud-free spectra and evolutionary model grids available in the literature, which extend down to Y dwarf temperatures and have been computed using recent advances in opacity data: Sonora-Bobcat\footnote{https://zenodo.org/record/5063476\#.YRiJOtNKhHQ} (\citealt{Marley2021}) and ATMO2020\footnote{https://www.erc-atmo.eu/?page\_id=322} (\citealt{Phillips2020}). These model grids build on the older evolutionary model grids which covered L and T type brown dwarfs (\citealt{Lunine1986}; \citealt{Burrows1997}; \citealt{Baraffe2002}; \citealt{Saumon2008}; \citealt{Allard2012}). Another set of spectral models is available\footnote{https://www.carolinemorley.com/models} in \citealt{Morley2014}, which includes patchy water clouds.  

Clear equilibrium chemistry models from ATMO2020 and Sonora-Bobcat fail to replicate Y dwarf photometry and spectra.
 This implies that additional physical processes are at play than those captured in current models. A well-established additional physical processes at play is disequilibrium chemistry for CO-CH$_4$ and N$_2$-NH$_3$ (\citealt{Fegley1996}; \citealt{Hubeny2007}; \citealt{Visscher2011}). The presence of CO absorption in observations (\citealt{Miles2020}) indicates that vertical mixing can pull CO and N$_2$ up from deeper in the atmosphere, making them overabundant in the photosphere relative to equilibrium values, and CH$_4$ and NH$_3$ underabundant. ATMO2020 includes two disequilibrium chemistry options, one set with moderate vertical mixing rates and one set with more vigorous vertical mixing rates. \cite{Phillips2020} showed that disequilibrium chemistry improved agreement with observations in many cases, but does not reconcile all portions of the spectrum. Sonora-Bobcat covers only clear equilibrium-chemistry models. Sonora-Cholla, the non-equilibrium counter part of Sonora-Bobcat, extends down to effective temperatures of 400 K, but no lower \citep{Karalidi2021}.

\cite{Morley2018} present a case study of the coolest detected Y dwarf: WISE 0855, computing a fine grid of models centered on the properties inferred for WISE 0855 from previous studies, varying metallicity, C/O ratio, surface gravity, and temperature. They collected a new GNIRS/Gemini L band spectrum and paired it with the M band spectrum later published in \citealt{Miles2020} to assess its properties. They find evidence of water ice clouds and evidence that methane is depleted relative to solar. To explain the found mismatch between models which fit M and L band spectra well and models which replicate the NIR photometry, they suggest a deep continuum opacity source may be obscuring the NIR flux. 

\cite{Tang2021} consider the impact of adding effects from water's latent heat, since it plays a significant role in solar-system gas giants \citep{Lunine1987}. They concluded that including effects from latent heat could make a marked difference in cases of super-solar metallicity, when water is likely more abundant, but has only a minor effect at solar or sub-solar metallicity. Since free floating Y dwarfs are thought to form from direct collapse, they are unlikely to attain the high metallicities needed for latent heat to play a role. But, for temperate giant exoplanets or substellar companions enriched in a disk, latent heat may need to be taken into account.

\cite{Mang2022} explore microphysical models of water clouds in Y Dwarf Atmospheres. They use the 1D Community Aerosol and Radiation Model for Atmospheres (CARMA, \citealt{Toon1988}; \citealt{Turco1979}; \citealt{Jacobson1994}; \citealt{Ackerman1995}) to model cloud spatial positions and particle sizes, accounting for a variety of the processes involved in cloud formation. They found that water cloud properties differ depending on whether they predominantly form via homogeneous nucleation or via heterogeneous nucleation onto up-welled KCl particles. Homogeneously nucleated water clouds have a greater vertical extent, whereas heterogeneously nucleated water clouds on KCl seeds have a more compressed scale height due to KCl's higher mass. In both cases, particle sizes tend to be around 10-100 $\mu$m near cloud bases at 1 - 0.01 bar, and decrease to sub-$\mu$m sizes at higher altitudes, (0.01 - 10$^{-6}$ bars). Their study provides valuable insight into how realistic water cloud properties might shape Y dwarf spectra. However, CARMA takes a fixed pressure-temperature (P-T) profile as input and post-processes the spectral effects of clouds rather than running in concert as one converges an atmosphere model to radiative-convective equilibrium.

In some radiative-convective equilibrium models, convection is treated as an adiabatic process where pressure P and temperature T are related by P$^{(1 - \gamma)}$T$^{\gamma}$ = constant (see e.g., \citealt{Marley2015}). For an ideal gas, $\gamma$ is the ratio of specific heats at constant pressure and volume. \cite{Leggett2021} posit that the adiabatic ideal gas $\gamma$ should apply at depth, but that higher in the atmosphere, effects such as rapid rotation and the associated atmospheric dynamics unaccounted for in 1D models will alter the effective $\gamma$ (\citealt{Zhang2014}; \citealt{Augustson2019}). They improve the agreement between models and observations significantly with an ad-hoc approach wherein they take ATMO2020 vigorous mixing disequilibrium models as a starting point and then fine-tune P-T profiles by modifying the value of $\gamma$ higher in the atmosphere as a free parameter. They still do not perfectly match the data, and a precise quantitative physical motivation for the change to $\gamma$ has yet to be provided. 

\cite{Zalesky2019} depart from the equilibrium model approach and instead do retrievals (Bayesian parameter estimation) for the \citealt{Schneider2015} objects. In most cases, their P-T profiles agree with equilibrium models (given the uncertainties), but they would also agree with the shifted P-T profiles of \citealt{Leggett2021}. It should be noted that they fit only the NIR data, not the M band spectra and longer-wavelength photometric points where available. One of their most unambiguous findings was strong support for rainout chemistry, rather than local equilibrium chemistry. It is expected that retrievals will provide more insights when HST spectra can be paired with infrared data from the JWST.

The model grid we present in this work fills a gap in the existing literature: self-consistent cool models which incorporate both nonequilibrium chemistry and water clouds, with updated molecular opacities. This provides an opportunity to examine whether the P-T profiles favored in \citealt{Leggett2021} and \citealt{Zalesky2019} can be explained by the combined impact of clouds and the dramatic shift in molecular opacity that accompanies nonequilibrium chemistry. The remainder of this paper is organized as follows: in \S \ref{sec:methods}, we describe the  methods for computing our models, in \S \ref{sec:results_grid}, we illustrate model dependence on fundamental parameters, in \S \ref{sec:results_observations}, we compare models to observations from the literature, and finally in \S \ref{sec:conclusions}, we summarize and discuss our conclusions. We also include an appendix describing the updates to our molecular opacity database in more detail and comparing our clear equilibrium  models with the Sonora-Bobcat model set.

\section{Methods}\label{sec:methods}

\subsection{coolTLUSTY}
Models presented in this paper are computed using \textit{coolTLUSTY} (\citealt{Burrows2008}; \citealt{acceleratedLambda}; \citealt{sudarsky2005}). \textit{coolTLUSTY} solves for 1D radiative convective equilibrium in a plane-parallel atmosphere. In the case of modeling isolated objects, the free parameters are: effective temperature, surface gravity, and metallicity, as well as any additional settings for clouds or disequilibrium chemistry. \textit{coolTLUSTY}'s basic iteration loop uses a hybrid of Accelerated Lambda Iteration and Complete Linearization to simultaneously solve hydrostatic equilibrium, radiative-convective equilibrium, and radiative transfer equations. It is essentially a Newton-Raphson solver, which begins with an initial guess and makes corrections based on partial derivatives until the desired convergence criteria are reached. Convection is treated with leaky-blob mixing-length theory. Single-scattering is accounted for, but assumed to be isotropic. Opacities are treated as if the resolution provides line-by-line sampling. In our calculations we use 100 atmospheric layers and 30000 frequency points spaced evenly in log from 0.4 to 300 $\mu$m. Models typically extend from at least 10$^{-4}$ bars to hundreds of bars or more, depending upon the exact combination of metallicity, surface gravity, and effective temperature, although there is some variation where altering the optical depth boundaries of the atmosphere aided convergence. Models were converged to 1 part in 10000 where possible, but, in some regions of parameter space, we settled for 5 parts in 1000.

We show a comparison between \textit{coolTLUSTY} and Sonora-Bobcat, a published model grid, in appendix \ref{sec:literature}.

\subsection{Gaseous Opacities}

\begin{figure}
    \centering
    \includegraphics[width=\textwidth]{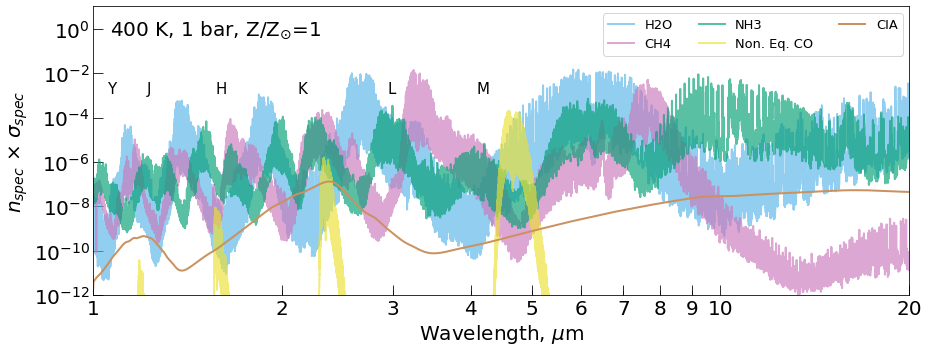}
    \caption{Relative contributions of major opacity sources in Y dwarf atmospheres. 400 K and 1 bar correspond to the photosphere for our clear, equilibrium model with $T_{\mathrm{eff}}$ = 350, log$_{10}$(g) = 4.25, and Z/Z$_{\odot}$ = 1. CO is negligible at 400 K, but we have enhanced its abundance in the figure, as one might expect in cases of strong vertical mixing, in order to illustrate which wavelengths are affected.}
    \label{fig:opacities}
\end{figure}

We pair \textit{coolTLUSTY} with an updated version of the opacity database described in \citealt{Sharp2007}. This update is outlined in more detail in appendix \ref{sec:opacity_appendix}. This includes line absorption for 15 molecules (H$_2$O, H$_2$, CH$_4$, CO, H$_2$S, NH$_3$, PH$_3$, TiO, VO, SiO, FeH, CaH, CrH, TiH, MgH) and 6 atoms (Li, Na, K, Fe, Rb, Cs), H$^{-}$ free-free and bound-free absorption, and collision-induced absorption from both H$_2$-H$_2$  and H$_2$-He. These various absorption opacity sources are compiled into pre-mixed thermochemical equilibrium tables following the chemistry calculations of \citealt{Sharp2007} and \citealt{Burrows1999}. In addition to the absorption opacity sources, we account for Rayleigh scattering off of H$_2$ and He. 

\subsection{Water Clouds}
Where water clouds are included, we assign their optical properties using Mie theory for homogeneous spheres of water ice, with complex indices of refraction from \citep{Warren1984}. Particle size distributions are Deirmendjian:  
\begin{equation}
    n(a) \propto \big(\frac{a}{a_0}\big)^6 exp \big[-6\big(\frac{a}{a_0}\big)\big] \quad\quad ,
\end{equation} where $n(a)$ is the number fraction of particles with size $a$ and $a_0$ is the modal particle size. A Deirmendjian particle size distribution reproduces the distribution in Earth's cumulus water clouds fairly well when $a_0$ = 4 \citep{Deirmendjian1964}. Our modal particle size is typically set at 10 $\mu$m throughout this work, roughly following findings for particles sizes near water cloud bases in the microphysical models of \citealt{Mang2022}.

We assume clouds form in a spatially homogeneous layer covering the entire object. Cloud vertical extents are specified in the same manner as previous works which published cloudy models using \textit{coolTLUSTY} (\citealt{Madhusudhan2011}, \citealt{Burrows2006}, \citealt{Sudarsky2000}). In this parameterization, clouds form near the intersection of the Clausius-Clapeyron line and the P-T profile, then taper off towards higher and lower pressures. Mathematically, this ``cloud shape parameter" follows the relation:
\begin{equation}
f(P) = \left\{
        \begin{array}{ll}
            \big(\frac{P}{P_{top}}\big)^{TCUP}\quad, & \quad P < P_{top} \\
            1\quad, & \quad P_{top} \leq P \leq P_{base} \\
            \big(\frac{P}{P_{base}}\big)^{-TCLOW}\quad, & \quad P > P_{base}
        \end{array}
    \right.
\end{equation} where $P_{top}$ is the pressure of the upper cloud base, which we set to the intersection of the P-T profile and water's Clausius-Clapeyron line, $P_{base}$ is the pressure of the lower cloud base, which we set initially based on where the clear model with the same $T_{\mathrm{eff}}$, $g$ and $Z$ intersected water's Clausius-Clapeyron line, then update gradually until $P_{base}$ and $P_{top}$ are equal, $TCUP$ is a free parameter which dictates how quickly the cloud tapers off from the upper cloud base, and $TCLOW$ dictates how quickly the cloud tapers off from the lower cloud base. We always keep $TCLOW$ at a steep cut-off of 10. We consider cloudy atmospheres with four values of $TCUP$, keeping with the nomenclature of previous work: ``E" type clouds are the most vertically compact with $TCUP$ = 6, ``A" type clouds are the most vertically extended with $TCUP$ = 0. Then, there are intermediate cloud types: ``AEE" with $TCUP$ = 2 and ``AE" with $TCUP$ = 1.

The total cloud opacity in each layer is limited by the amount of water vapor present at each pressure level and the supersaturation assumed necessary for condensation along with this cloud shape factor. Cloud opacity in cm$^2$ per gram of total atmospheric material can thus be expressed as:
\begin{equation}
    \kappa_{cloud}(\nu,P) = \chi_{cloud}\frac{\mu_{cloud}}{\mu} \times \sigma(\nu,a_0) \times S_{cloud}f(P) 
\end{equation} where $P$ is the pressure of a given layer in the atmosphere model, $\nu$ is the frequency, $\chi_{cloud}$ is the molar mixing ratio of the cloud species in gaseous form, $\mu_{cloud}$ is the molar mass of an individual molecule of cloud species in gaseous form, $\mu$ is the mean molecular weight of the atmosphere (also in molar masses), $\sigma(\nu,a_0)$ is the cross section at frequency $\nu$ per gram of cloud species assuming a Deirmendjian distribution with modal size $a_0$ and homogeneous spheres, $S_{cloud}$ is the super-saturation parameter (set to 0.01 for water clouds in this work), and $f(P)$ is the cloud shape factor, described above.

\subsection{Disequilibrium Chemistry}
We incorporate disequilibrium chemistry due to vertical mixing in the manner described in \citealt{Hubeny2007}. This involves comparing the local chemical timescales for CO-CH$_4$ reactions and N$_2$-NH$_3$ reactions against the vertical mixing timescales throughout the atmosphere. If the chemical timescales, $t_{chem}$, are longer than mixing timescales, $t_{mix}$, at any layer, we expect the atmosphere to depart from thermochemical equilibrium in the overlying layers \citep{Prinn1977} and take on the abundances equal to the layer where $t_{chem}$ = $t_{mix}$.  

In the convective layers of the atmosphere the mixing timescale is set by the convective velocity, $v_c$, and the convective mixing length, $H_c$:
\begin{equation}\label{eq:tmix_conv}
    t_{mix} = 3H_c/v_c \quad .
\end{equation}
In the radiative layers of the atmosphere the mixing timescale is not well defined. We parameterize it in the standard manner \citep{Saumon2006}:
\begin{equation}\label{eq:tmix_rad}
    t_{mix} = H^2/K_{zz} \quad ,
\end{equation}
where $H$ is the pressure scale height, and K$_{zz}$ is the coefficient of eddy diffusion. K$_{zz}$ is the only additional free parameter in our disequilibrium models. For nearly all of the effective temperatures and gravities that we consider, the intersection of the mixing timescale and the chemical timescales occur in the convective regions of the atmospheres. This means that varying K$_{zz}$, has no effect on many of the disequilibrium models. However, for some combinations of $T_{\mathrm{eff}}$ and $g$, there are multiple convective regions separated by radiative zones, and occasionally the intersection of mixing timescale and chemical timescale will fall into one of these detached radiative zones. Mixing timescales in radiative zones tend to be orders of magnitude slower than in convective zones. As such, there is not always a perfectly smooth monotonic behavior in the disequilibrium model grid, and for some grid points it would be worthwhile to consider variations in K$_{zz}$. This exploration is left to future work.

\begin{figure}
    \centering
    \includegraphics[width=0.5\textwidth]{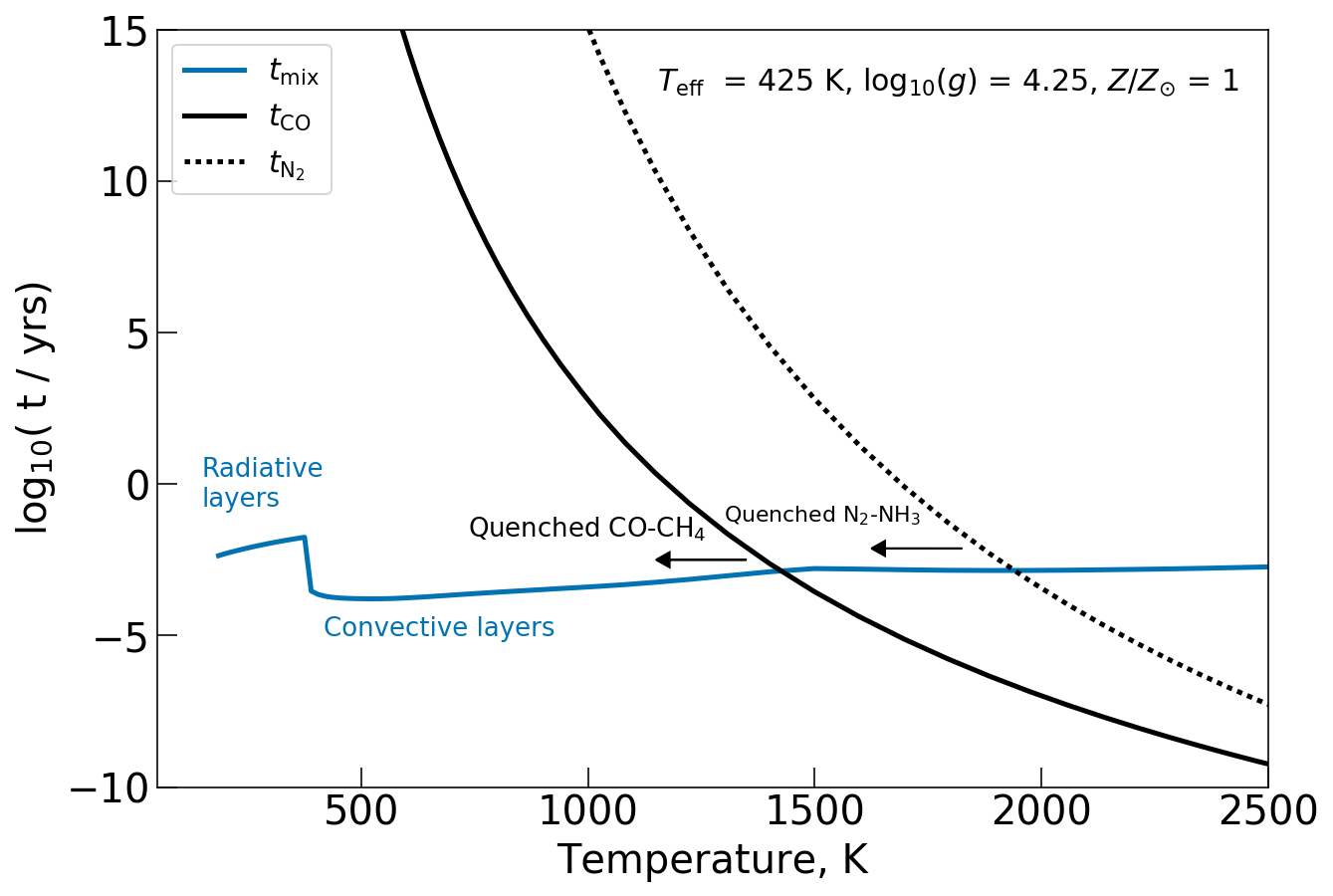}
    \caption{Mixing and chemical timescales for an atmosphere with $T_{\mathrm{eff}}$=425, log$_{10}$(g) = 4.25, and Z/Z$_{\odot}$ = 1, assuming disequilibrium chemistry. The blue line indicates mixing timescales as described in equations \ref{eq:tmix_conv} and \ref{eq:tmix_rad}. The discontinuity around 400 K corresponds to the change from radiative layers in the upper atmosphere to convective layers deeper in the atmosphere. The black lines show CO-CH$_4$ timescales (solid black line) and N$_2$-NH$_3$ timescales (dotted black line). For higher temperatures and pressures than the intersection of chemical timescale and mixing timescale, the atmosphere is expected to be in thermochemical equilibrium for those species. For lower temperatures and pressures than the intersection the atmosphere is expected to have the mixing ratios of those species at the intersection or quench point.}
    \label{fig:chem_demo}
\end{figure}

Figure \ref{fig:chem_demo} shows an example of how chemical timescales and mixing timescales compare. These curves must be computed for a particular P-T profile, in this case one with: $T_{\mathrm{eff}}$=425, log$_{10}$(g) = 4.25, and Z/Z$_{\odot}$ = 1, and disequilibrium chemistry. Each temperature thus corresponds to a specific pressure and altitude in the atmosphere. The solid blue curve shows mixing timescales. The steep change between longer timescales in radiative layers to shorter timescales in convective zones is apparent around 400 K. In this case the atmosphere doesn't have any detached radiative/convective zones. The carbon chemistry timescale is indicated by the solid black line and the nitrogen chemistry timescale is indicated by a dotted black line. For all our Y dwarf models, the carbon chemistry timescale always intersects with the mixing timescale at a lower temperature and pressure than the nitrogen chemistry. In some cases the intersection occurs very deep in the atmosphere, so it is important to ensure that all models are computed to an adequate depth so that there are some layers below this intersection.

For N$_2$-NH$_3$ the timescale comes from \citealt{Lodders2002}:
\begin{equation}
    t_{chem} \equiv t_{N_2} = \frac{1}{\kappa_{N_2}N(H_2)} \quad ,
\end{equation} where $\kappa_{N_2}$ is the rate constant for the N$_2$-NH$_3$ reaction and $N(H_2)$ is the number density of molecular hydrogen. In units of cm$^3$/seconds $\kappa_{N_2}$ is:
\begin{equation}
    \kappa_{N_2} = 8.54 \times 10^{-8}exp\big(-\frac{81515}{T}\big) \quad .
\end{equation}
For CO-CH$_4$, the adopted timescale comes from \citealt{Prinn1977}:
\begin{equation}
    t_{chem} \equiv t_{CO} = \frac{N(CO)}{\kappa_{CO}N(H_2)N(H_2CO)} \quad ,
\end{equation}
with $\kappa_{CO}$ defined as:
\begin{equation}
    \kappa_{CO} = 2.3\times10^{-10}exp\big(-\frac{36200}{T}\big) \quad ,
\end{equation} with units of cm$^3$/seconds.

A thorough exploration of the chemical networks governing CO, CH$_4$, and NH$_3$ in gaseous planets and brown dwarfs was conducted by \citealt{Zahnle2014}. They performed 1d kinetic chemical calculations for a suite of metallicities, C/O ratios, effective temperatures and surface gravities and then derived effective timescales which best match kinetic model results when paired with the quenching approximation described above. Their results showed that, for N$_2$-NH$_3$, the reaction rates of \citealt{Lodders2002} and the quenching approximation provide an excellent match to the results of detailed 1-d kinetic model results. \citealt{Zahnle2014} found that there was not one single limiting reaction for CO-CH$_4$ which can be used within the quenching approximation, but they derive an effective reaction rate which can be paired with the quenching approximation to replicate the results of the 1-d kinetic models. For CO-CH$_4$, the \citealt{Prinn1977} approach results in excellent agreement for CO and reasonable agreement for CH$_4$ abundances, although it is getting the right results using the wrong chemical reaction as the limiting timescale. 

\subsection{Synthetic Photometry}

\begin{figure}
    \centering
    \includegraphics[width=\textwidth]{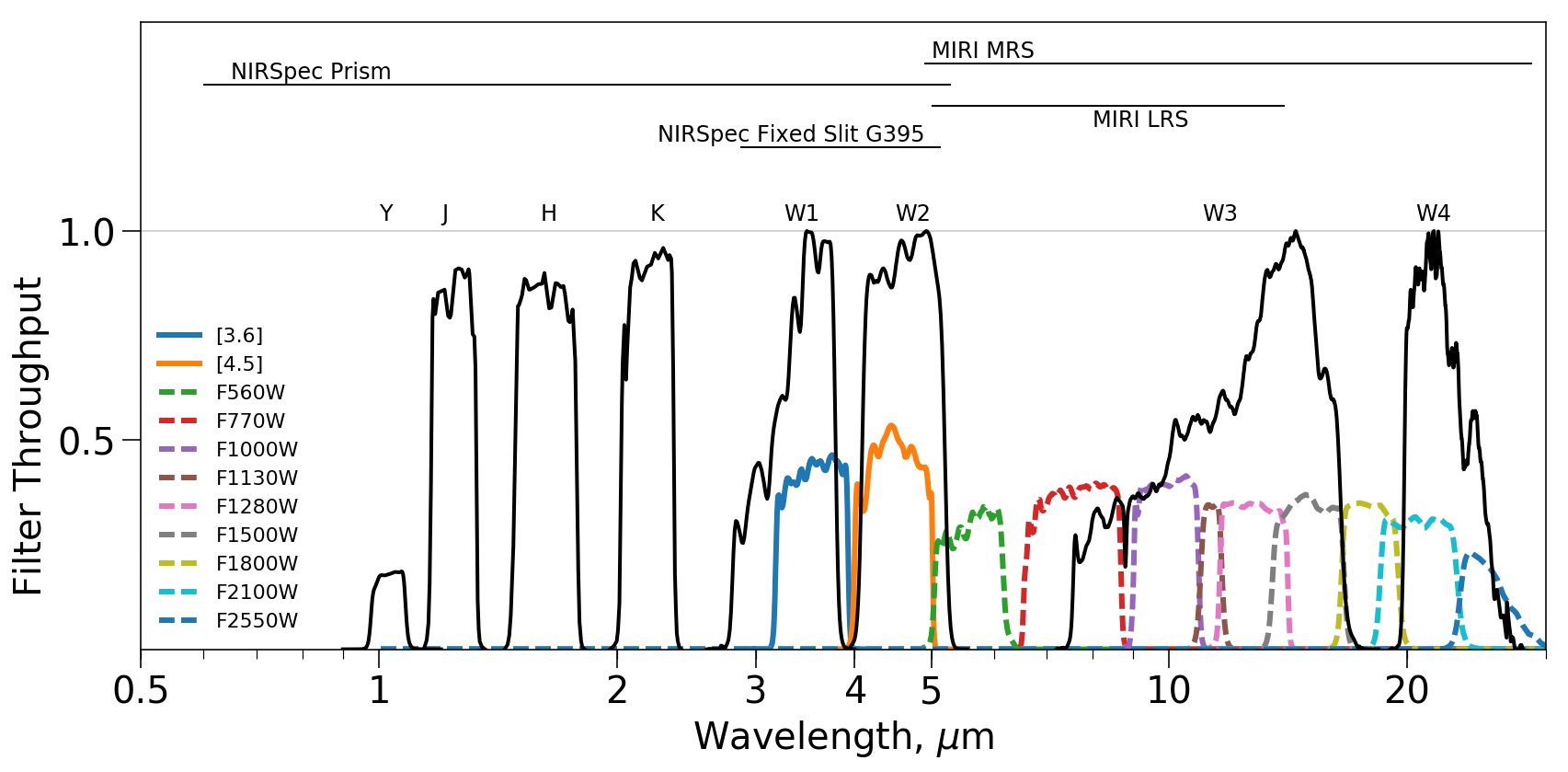}
    \caption{Spectral response functions for filters used to observe Y dwarfs. Solid black lines are ground-based and have their filter name labeled above them. Y, J, H, and K are on the Mauna-Kea Observatories system, but collected from various sources. W1, W2, W3, and W4 are from the WISE survey. Colored lines are space-based with names given in the legend. [3.6] and [4.5] are Spitzer filters, and the rest are JWST's MIRI imaging filters. The spectral coverage of NIRSpec and MIRI are also indicated at the top of the figure, with resolutions summarized in Table \ref{tab:jwst_modes}.}
    \label{fig:filters}
\end{figure}

Figure \ref{fig:filters} shows the spectral response functions for all the bandpasses we use for computing synthetic photometry to compare with observed photometry. Observations are not available in all filters for all objects. We compute the flux density for a photometric point as a weighted average across the band, weighted by the transmission function. We convert the observed photometry from apparent magnitude $m$,  to a flux density $F_{\lambda}$ as: $F_{\lambda} = F_{\lambda,zero} \times 10^{(-m/2.5)}$, where $F_{\lambda,zero}$ is the zero-point flux density for that band.

\begin{figure}
    \centering
    \includegraphics[width=0.5\textwidth]{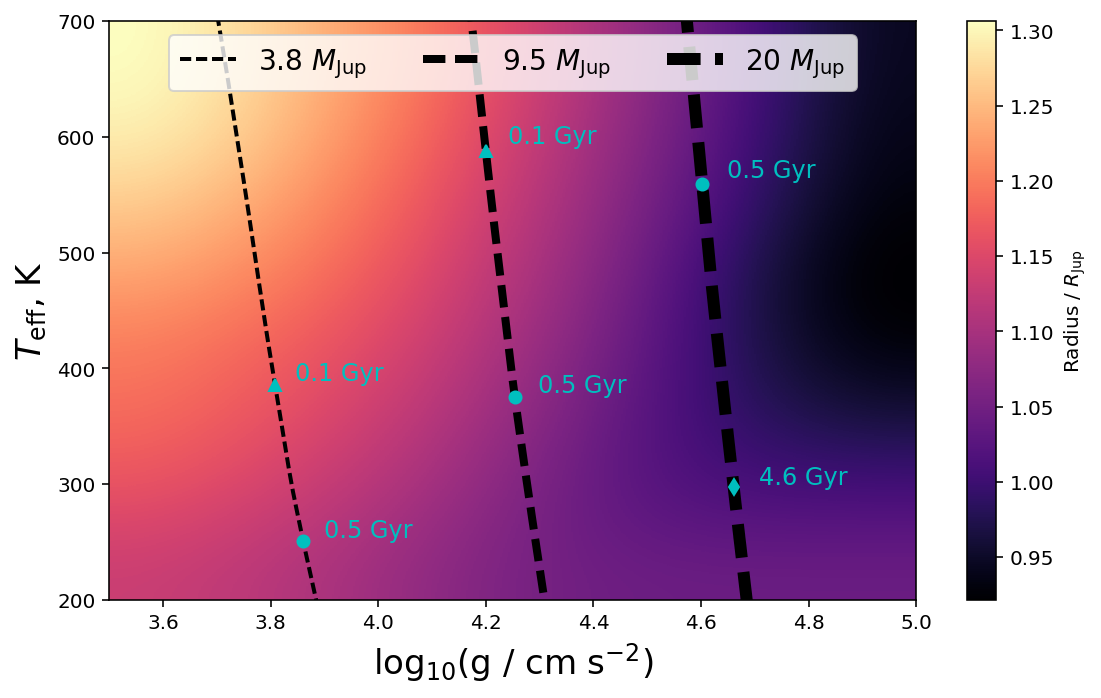}
    \caption{When comparing models to photometry we assign radii for effective temperature, $T_{\mathrm{eff}}$, and surface gravity, $g$, by interpolating within the evolutionary tracks of \citealt{Burrows1997}. The relation is shown in this figure. Objects span only a small range in radius, from around 0.9 to 1.3 Jupiter radii. We have also plotted some evolutionary tracks for 3.8, 9.5 and 20 Jupiter mass objects over the top.}
    \label{fig:radii}
\end{figure}

Across the top of Figure \ref{fig:filters}, we also mark the wavelength coverage of various JWST spectrographs, and dashed colored lines correspond to MIRI photometric filters. Table \ref{tab:jwst_modes} summarizes some of the instruments and observing modes which various groups are planning to use for Y dwarfs with JWST GTO and Cycle 1 GO time.

\begin{table}[]
    \centering
    \begin{tabular}{lccc}
        Instrument + Mode  & Wavelength Coverage & Spectral Resolution & Reference\\
        \hline
        NIRSpec Prism/Clear & 0.6-5.3 $\mu$m & R$\sim$100&\\
        NIRSpec Fixed Slit G395M & 2.87-5.14 $\mu$m & R$\sim$2700&\\
        MIRI LRS   & 5-14 $\mu$m & R$\sim$100&\\
        MIRI MRS   & 4.9-28.8 $\mu$m & R ranging from 1550-3250&\\
        MIRI imaging   & 5-28 $\mu$m & 9 bands make R$\sim$5\footnote{http://ircamera.as.arizona.edu/MIRI/pces.htm}&\\
    \end{tabular}
    \caption{Selection of JWST observing modes well-suited to characterization of Y dwarfs}
    \label{tab:jwst_modes}
\end{table}

\section{Presentation of Model Grids} \label{sec:results_grid}
We compute four classes of models: (1) clear equilibrium chemistry models, (2) clear disequilibrium chemistry models, (3) cloudy equilibrium chemistry models, and (4) cloudy disequilibrium chemistry models. For the clear models, we consider 16 temperatures in steps of 25 K from 200 to 600 K, 7 surface gravities at log$_{10}$($g$/cm s$^{-2}$) = 3.5, 3.75, 4.0, 4.25, 4.5, 4.75, and 5.0, and three metallicities: 0.316, 1.0 and 3.16 times solar. Cloudy models are computed for a subset of this range covering effective temperatures from 225-400 K, log$_{10}$($g$/cm s$^{-2}$) = 3.75-5.0, and 0.316, 1.0 and 3.16 times solar. We consider two different vertical extents of water cloud. There are additional models with a smaller modal particle size computed at solar metallicity with equilibrium chemistry. All models are publicly available online\footnote{https://doi.org/10.5281/zenodo.7779180}. 

Previous works have indicated that disequilibrium chemistry is strongly favored, but we still include equilibrium chemistry models for completeness and to enhance understanding (\citealt{Phillips2020}; \citealt{Leggett2021}). We will begin by examining the clear equilibrium models and build up to the cloudy disequilibrium models.

\begin{table}[]
    \centering
    \begin{tabular}{lcccccc}
        Name  & Chemistry & Modal Cloud  & Vertical Extent & Temperature  & Surface Gravity & Metallicity \\
         & & Particle Size &  ($TCUP$ in eq. (2)) & Range & Range & Range \\
        \hline
        Clear EQ & equilibrium & none & none & 200-600 K & 3.5-5.0  & -[0.5], [0], [0.5]\\
        Clear NEQ & nonequilibrium & none & none & 200-600 K & 3.5-5.0 & -[0.5], [0], [0.5]\\  
        E10 Cloudy EQ & equilibrium &10 $\mu$m & 6 & 200-400 K & 3.75-5.0 & -[0.5], [0], [0.5]\\     
        AEE10 Cloudy EQ & equilibrium &10 $\mu$m & 2 & 200-400 K & 3.75-5.0 & -[0.5], [0], [0.5]\\    
        E1 Cloudy EQ & equilibrium &1 $\mu$m & 6 & 200-400 K  & 3.75-5.0  &  [0] \\     
        AEE1 Cloudy EQ & equilibrium &1 $\mu$m & 2 & 200-400 K & 3.75-5.0 & [0] \\   
        E10 Cloudy NEQ & nonequilibrium &10 $\mu$m & 6 & 200-400 K & 3.75-5.0 & -[0.5], [0], [0.5]\\     
        AEE10 Cloudy NEQ & nonequilibrium &10 $\mu$m & 2 & 200-400 K & 3.75-5.0 & -[0.5], [0], [0.5]\\   
        WISE 0855 mini-grid & equilibrium & 1, 3.16, 10 $\mu$m & 0, 1, 2, 6 & 250 K & 4.25 & -[0.5]\\ 
    \end{tabular}
    \caption{A summary of the suite of models presented in this work. Models can be accessed online at: https://doi.org/10.5281/zenodo.7779180. For some combinations of chemistry, metallicity, effective temperature, and surface gravity, we expand the cloudy models up to 450 K or down to surface gravities of log(g) = 3.5. For the super-solar metallicity cloudy models, we only reach up to 350 K.}
    \label{tab:grid_summary}
\end{table}	

\subsection{Clear Equilibrium Chemistry Models}\label{sec:clear_eq_models}

\begin{figure}
    \centering
    \includegraphics[width=0.9\textwidth]{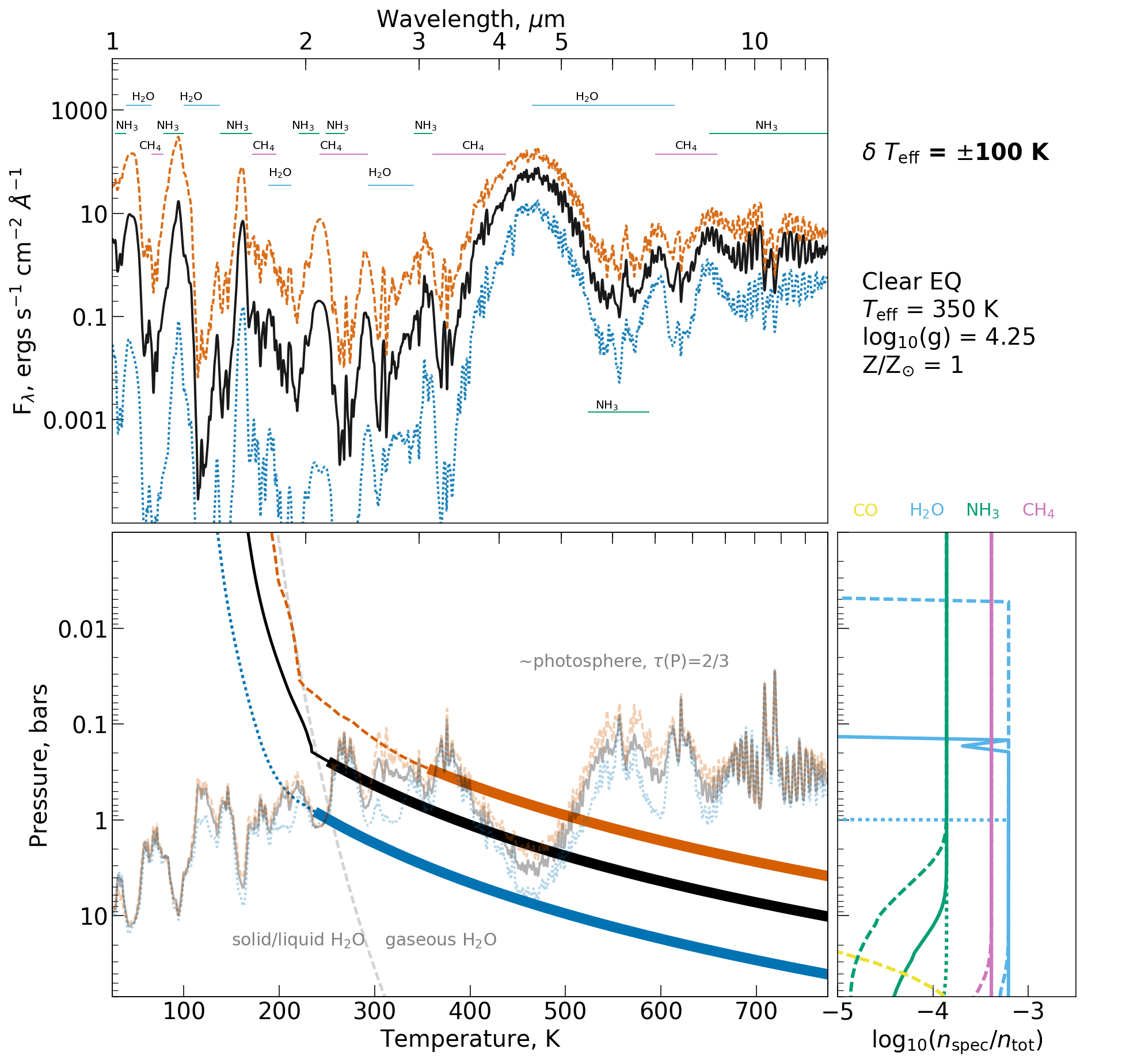}
    \caption{Demonstration of model sensitivity to effective temperature, $T_{\mathrm{eff}}$. We compare a fiducial model with effective temperature of 350 K, log$_{10}$(g) = 4.25, and Z/Z$_{\odot}$ = 1, to a model with $T_{\mathrm{eff}}$ = 450 K and a model with 250 K. The top panel shows spectra in the form of surface flux density. The bottom left panel shows the P-T profiles and the pressure levels where $\tau$ = 2/3 as a function of wavelength, which is a good approximation for the photosphere. Convective regions of the atmospheres are marked with thicker lines on the P-T profile. We have also overlaid the Clausius-Clapeyron line for water (light gray). The bottom right panel shows the mixing ratios of major absorbing species. Different colors correspond to different species as labeled, solid lines correspond to the fiducial model, dashed lines to a perturbation upward in effective temperature, and dotted lines to a perturbation downward in effective temperature.}
    \label{fig:vary_temp}
\end{figure}

\begin{figure}
    \centering
    \includegraphics[width=0.9\textwidth]{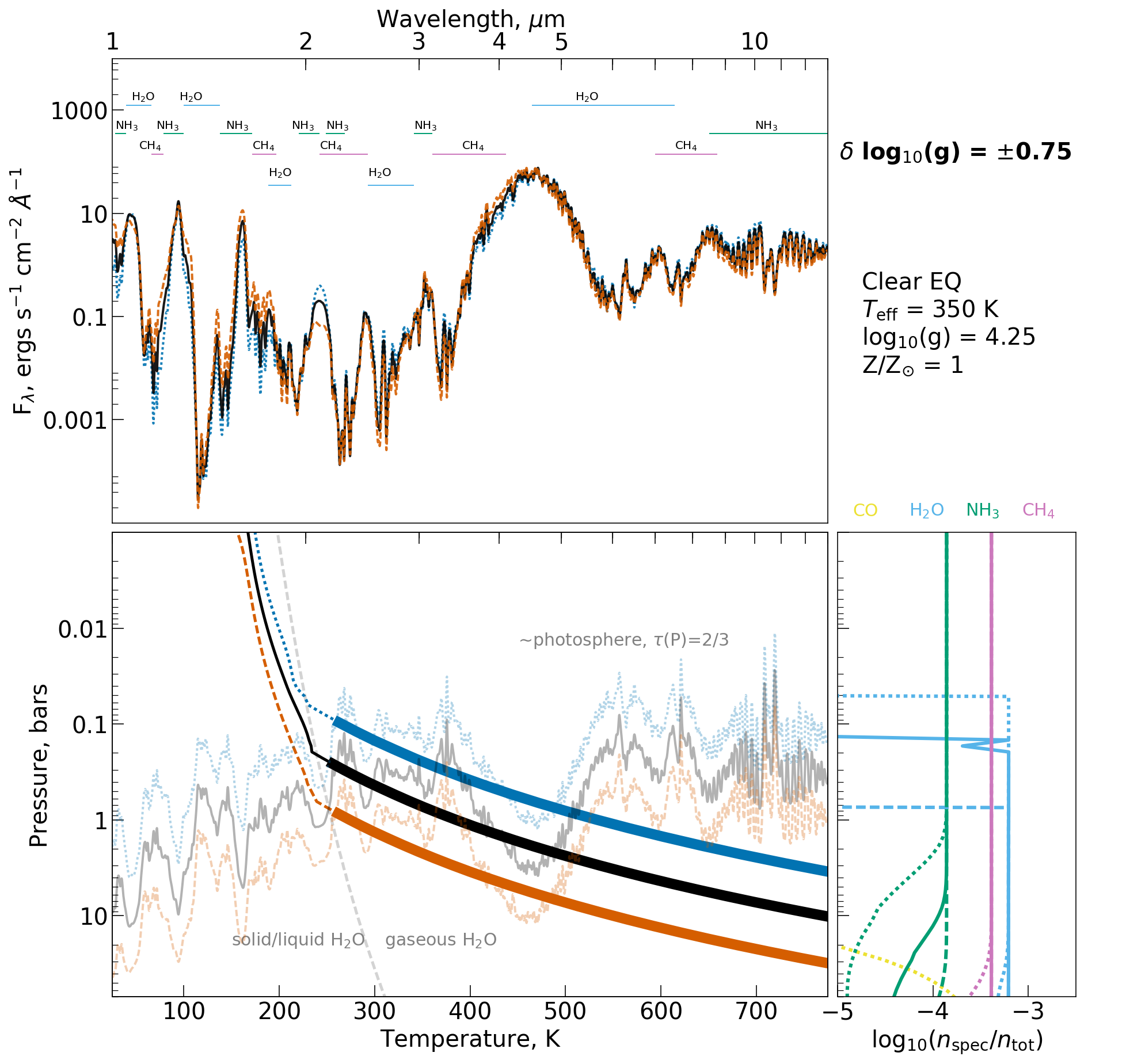}
    \caption{Demonstration of model sensitivity to surface gravity, $g$. This figure has the same form as Figure \ref{fig:vary_temp}. Once again, the fiducial model has effective temperature of 350 K, log$_{10}$(g) = 4.25, and Z/Z$_{\odot}$ = 1 (black solid lines). This time we compare the fiducial model against models with a surface gravity of log$_{10}$(g) = 3.5 (blue dotted lines), and a surface gravity of log$_{10}$(g) = 5 (orange dashed lines).}
    \label{fig:vary_g}
\end{figure}

\begin{figure}
    \centering
    \includegraphics[width=0.9\textwidth]{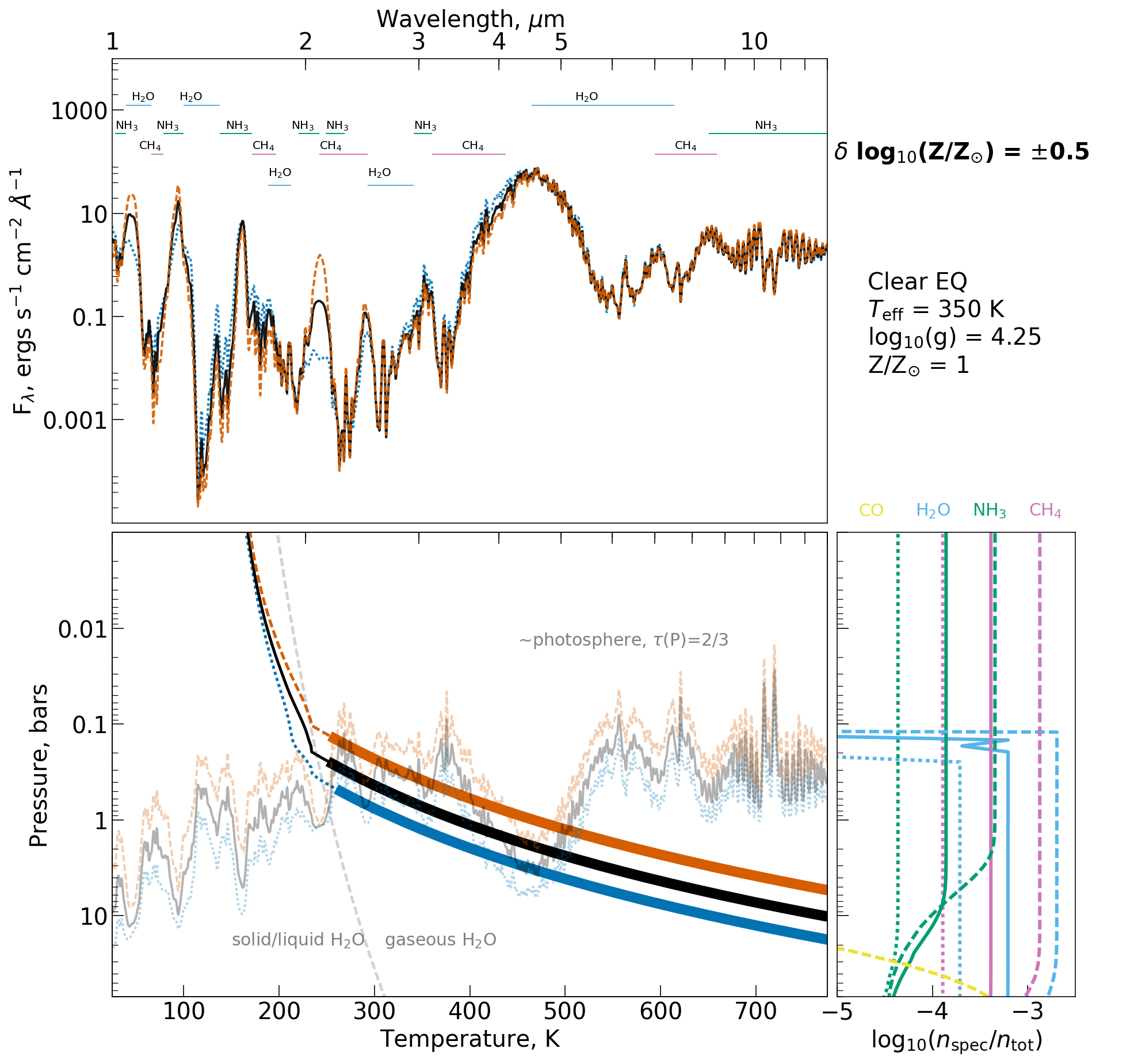}
    \caption{Demonstration of model sensitivity to metallicity, $Z$. This figure has the same format as Figure \ref{fig:vary_temp} and Figure \ref{fig:vary_g}. Black solid lines show the fiducial model with effective temperature of 350 K, log$_{10}$(g) = 4.25, and Z/Z$_{\odot}$ = 1. Orange dashed lines show a perturbation upwards in metallicity to Z/Z$_{\odot}$ = 3.16, and blue dotted lines show a perturbation downward in metallicity to Z/Z$_{\odot}$ = 0.316.}
    \label{fig:vary_Z}
\end{figure}

Figures \ref{fig:vary_temp} - \ref{fig:vary_Z} demonstrate model sensitivity to the fundamental input parameters: effective temperature, surface gravity, and metallicity. We use a fiducial model with $T_{\mathrm{eff}}$=350 K, log$_{10}$($g$/cm s$^{-2}$)=4.25, and Z/Z$_{\odot}$=1, then perturb up and down one parameter at a time. In all figures, the top panel shows the spectrum as surface flux density, the bottom left panel shows the P-T profile of the atmosphere overlaid with an approximate measure of the photospheric pressure as a function of wavelength, and the bottom right panel shows the mixing ratios of major absorbing species. Along with P-T profiles, we show the Clausius-Clapeyron line for water as a gray dashed line, and we note the convective layers of the atmosphere with thicker lines. 

By looking at a given pressure in the bottom panel of Figure \ref{fig:vary_temp} and tracing across the figure horizontally we can see which portions of the spectrum tend to form at that pressure. This pressure then corresponds to a specific temperature shown by the P-T profile in the same panel. In general, this shows that the peak emission in Y, J, and H tends to emerge from deepest in the atmosphere around 5-10 bars, then emission in the M band around 2-4 bars, then emission in the K band around 1 bar, and finally emission out in the MIR below 1 bar. In reality photons are emerging across a range of pressures for each wavelength. We show the depth at which $\tau$=2/3 which corresponds to the level where photons are statistically more likely to emerge from the atmosphere in the Eddington Approximation.

Opacities for major absorbing species were shown earlier in Figure \ref{fig:opacities}, and we also attribute spectral features to different species above the spectra in Figures \ref{fig:vary_temp} - \ref{fig:vary_Z}. The blue edge of the Y band is dominated by ammonia and the red edge by water. The J band has ammonia over its center, then methane (pink) on the blue edge and water on the red edge. The H band has ammonia on the blue edge and methane red edge. The K band has ammonia on the blue edge, then methane and ammonia on the blue edge. It is also near a peak in CIA, which does not dominate at pressures of 1 bar, as we show in Figure \ref{fig:opacities}, but can start to dominate at higher pressures as it scales in strength with density$^2$ rather than density, as for other opacity sources. The M band has methane on the blue edge and water on the red edge. In this section we are looking at equilibrium chemistry models, so the CO is much less abundant than in that shown in Figure \ref{fig:opacities}. 

We'll begin by looking at the impacts of varying effective temperature in Figure \ref{fig:vary_temp}. In the chemical profiles shown in the bottom right panel dotted lines correspond to a 250 K atmosphere, solid lines correspond to the fiducial 350 K atmosphere, and dashed lines correspond to a 450 K atmosphere. The 250 K and 350 K atmospheres don't contain any CO in these layers, while the 450 K object starts to have a bit of CO, but still only below the photosphere (yellow curves). The amount of ammonia for all three temperatures is similar above $\sim$1 bar, but starts to vary below one bar, in the layers where the Y, J, and H band peaks predominantly form. This contributes to changes in the relative heights of the Y, J, H, and K peaks and their widths with effective temperature. Ammonia also has a strong absorption feature out at 10.5 $\mu$m, but this portion of the spectrum tends to form at lower pressures in the atmosphere where ammonia abundances for all three temperatures are more similar. Methane abundances remain relatively constant for all objects throughout the layers of the atmospheres shown here. One can just start to see a reduction in the 450 K atmosphere around when CO rises as the two carbon bearing species trade off. However, this is again occurring below the photosphere.

Water abundances are similar for all three effective temperatures below a pressure of 1 bar, but, at lower pressures, we see that water rains out deeper in the atmosphere for the 250 K object and higher in the atmosphere for the 450 K object. This rainout occurs where the P-T profile intersects the Clausius-Clapeyron line. The $\tau$=2/3 curve for the 250 K atmosphere (blue line) looks substantially different than the 350 K and 450 K atmospheres because the water has rained out below/within the photosphere in this case, whereas for 350 K and 450 K water rains out mostly above the photosphere. Water rainout in the 250 K atmosphere allows greater emission through portions of the spectrum where gaseous water normally dominates, making methane and ammonia absorption features appear deeper in comparison. Looking at the $\tau$=2/3 curves in the bottom left panel of Figure \ref{fig:vary_temp}, one can see that varying effective temperature keeps the photosphere at the same level for the most part, aside from the effects of water rainout.

Figure \ref{fig:vary_g} shows the impact of changing surface gravity on the atmospheric structure and the emergent spectrum. Note that in these figures we are comparing the spectrum as the surface flux, so we do not have any relation between surface gravity and radius. Since effective temperature is the same for all three models, their overall emission integrated across wavelengths is the same regardless of surface gravity, and we just see subtle changes in shape to the spectrum due to the fact that the photosphere occurs at higher pressures for higher surface gravity objects and lower pressures for lower surface gravity objects. The lower left panel of Figure \ref{fig:vary_g} illustrates how changing surface gravity shifts the P-T profile up or down in pressure while maintaining a very similar shape in these layers of the atmosphere, with the convective zone ending at nearly the same temperature. If this plot extended up to lower pressures, all three curves would asymptote to nearly the same isothermal curve in the upper atmosphere. 

Changing the surface gravity influences the relative peaks in the NIR and their widths, and the amount of methane absorption seen on the blue edge of the M band and redward of the H band. For the high surface gravity atmosphere with log$_{10}$(g) = 5.0, there is weaker methane absorption in the M band and redward of the H band, a lower Y band peak, a broader J band peak, a higher H band peak and a lower K band peak. This reduced K band is due to stronger CIA significantly suppressing emission. Conversely the lower surface gravity of log$_{10}$(g) = 3.5 has a narrower J band peak, a lower H band peak, and a higher K band peak. It has more methane absorption in the M band and redward of the H band. The pressure and temperature where water rains out varies with surface gravity, but, since the photospheric pressure also varies, resulting changes are less dramatic than the changes between the 250 K model and 350 K model. In all three surface gravities, the water rainout occurs above the photosphere for most wavelengths. 

Figure \ref{fig:vary_Z} shows the effect of changing the metallicity of the atmosphere by a half a dex up and down. To first order, mixing ratios of major absorbing species H$_2$O, CH$_4$, and NH$_3$ scale linearly with metallicity, while CO scales with the square of metallicity. This is seen in the bottom right panel where abundance profiles of major absorbers have a similar shape relative to each other but are all shifted to higher or lower abundances with metallicity (with the exception of ammonia at pressures greater than $\sim$1 bar). Because of this, increasing the metallicity increases the average opacity and mean molecular weight of the atmosphere, so the photosphere moves up to lower pressures and the atmosphere has a slightly smaller scale height for higher metallicities, and vice versa for low metallicities.  

The lower metallicity atmosphere has Y and J band peaks that are more similar in height to the H band, while the K band peak is flattened. There is weaker methane absorption redward of the H band and in the M band. Conversely the higher metallicity has more emission in the Y and J band relative to the H band and more emission from the K band. There is stronger methane absorption redward of the J band and on the blue edge of the M band. 

\subsection{Disequilibrium Chemistry Models}\label{sec:clear_neq_models}

\begin{figure}
    \centering
     \includegraphics[width=0.9\textwidth]{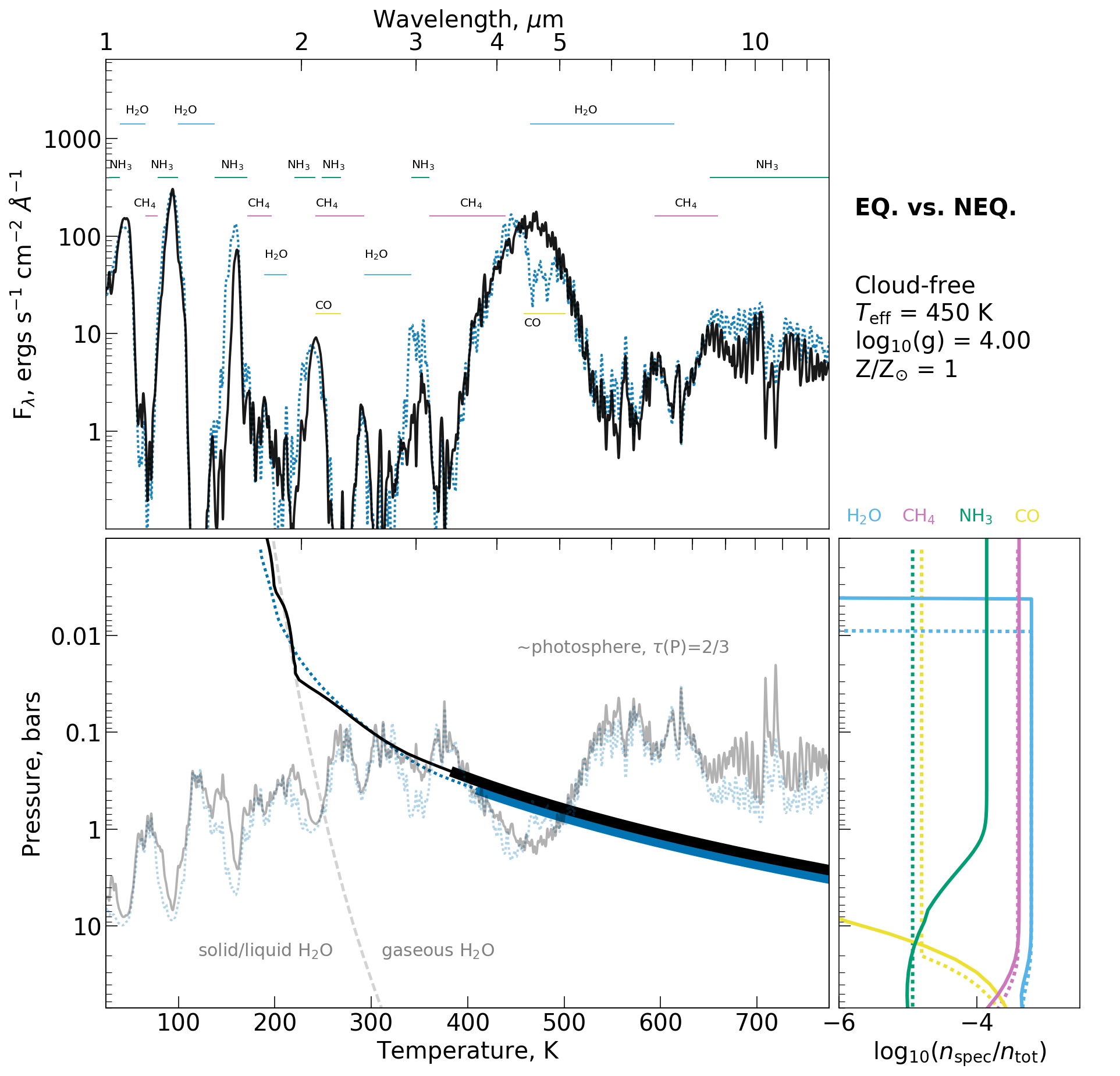}
    \caption{Comparison of equilibrium chemistry (black solid lines) and disequilibrium chemistry (blue dotted lines) for atmospheres with $T_{\mathrm{eff}}$=450 K, log$_{10}$($g$) = 4, and Z/Z$_{\odot}$ = 1. Figure has the same format as Figures \ref{fig:vary_temp}-\ref{fig:vary_Z}.}
    \label{fig:NEQ_vs_EQ_450K}
\end{figure}

\begin{figure}
    \centering
    \includegraphics[width=\textwidth]{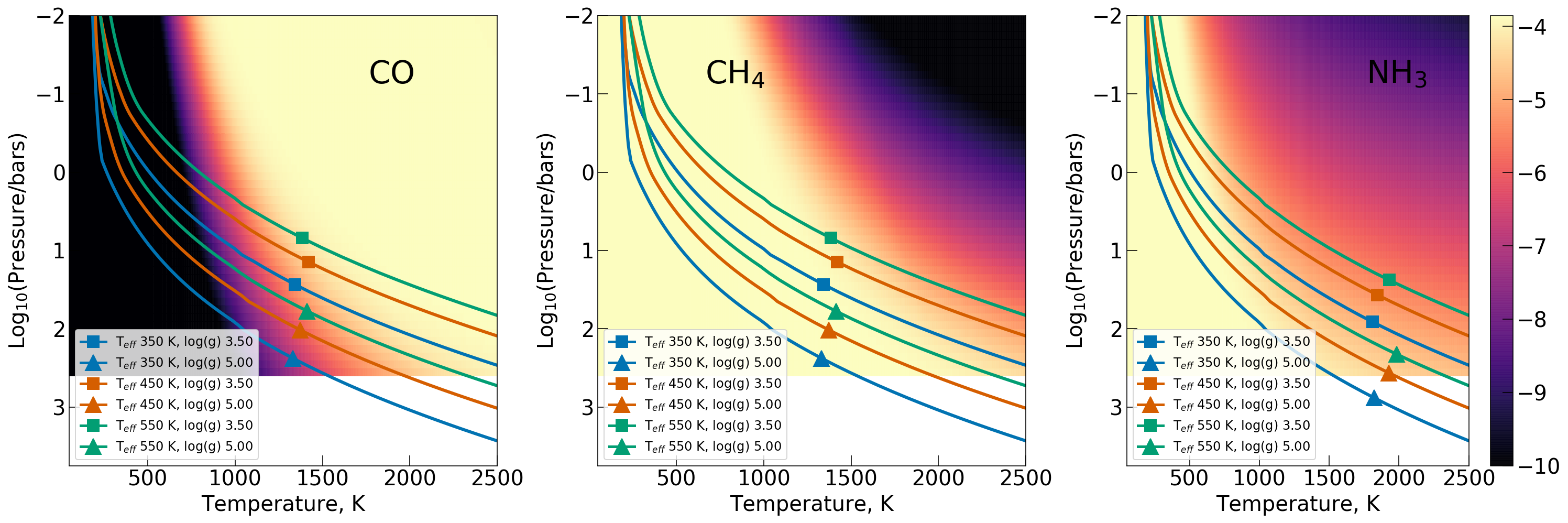}
    \caption{Illustration of how P-T profiles and quench points compare to equilibrium abundances of CO (left), CH$_4$ (center) and NH$_3$ (right). Each panel has temperature on the x-axis and pressure on the y-axis. Colored background shading indicates log$_{10}$(n$_{spec}$/n$_{tot}$), following the color bar at the far right of the figure, where the the color map's minimum value has been set to -10. The P-T profiles and quench points for a selection of clear nonequilibrium models are plotted over the top. }
    \label{fig:chem_abundances_quench_points}
\end{figure}

In Figure \ref{fig:NEQ_vs_EQ_450K} we show a comparison of a model computed with equilibrium chemistry and a model computed with disequilibrium chemistry. Both have $T_{\mathrm{eff}}$=450 K, log$_{10}$($g$) = 4, and Z/Z$_{\odot}$ = 1. For the disequilibrium model, the NH$_3$ abundance is decreased, leading to a broader peak on the blue side of the K band, and more flux around 3 and 10 $\mu$m. The degree to which NH$_3$ is depleted relative to equilibrium values is roughly an order of magnitude at pressures lower than 1 bar. As one moves deeper, the NH$_3$ abundance in the disequilibrium model gradually approaches the equilibrium values. This means that there is still similar NH$_3$ absorption in the Y and the J bands, which are probing deeper in the atmosphere. The CO abundance is enhanced, leading to an absorption feature around 4.5-5 $\mu$m. At this temperature and surface gravity, the change in CH$_4$ abundance is less pronounced than the change in NH$_3$ and CO abundance (see bottom right panel and Figure \ref{fig:chem_abundances_quench_points}), and correspondingly, we see only slight differences between the equilibrium and non-equilibrium spectra in CH$_4$ features on this log-scale plot. 

CO and N$_2$ are much less opaque than CH$_4$ and NH$_3$, so the nonequilibrium model's P-T profile is slightly cooler in the deeper atmosphere and up through most of the photosphere. At pressures lower than 0.1 bars, the upper atmospheres of nonequilibrium models are actually a bit warmer than the equilibrium models.

Generally, models with higher temperatures and lower surface gravities tend to show stronger signatures of disequilibrium chemistry: a deeper CO absorption feature in the M band and weaker NH$_3$ and CH$_4$ absorption. For lower effective temperatures and higher surface gravities, the CO enhancement is not always enough to make the absorption feature in the M band appear, but reductions in CH$_3$ and NH$_3$ still manifest. Figure \ref{fig:chem_abundances_quench_points} illustrates why this occurs. At cooler temperatures and higher surface gravities quench points occur where the difference in equilibrium abundances between the deep atmosphere and the upper atmosphere are less extreme. However, it should be noted that there are non-monotonic relations  with $T_{\mathrm{eff}}$ and surface gravity for mixing ratios of CO, CH$_4$, and NH$_3$. This is due to the occasional emergence of zones with slower convective velocities than their surroundings. This makes for longer mixing timescales, moving quench points to cooler temperatures. 

\subsection{Models with Water Clouds}\label{sec:cloudy_eq_models}
\begin{figure}
    \centering
     \includegraphics[width=0.9\textwidth]{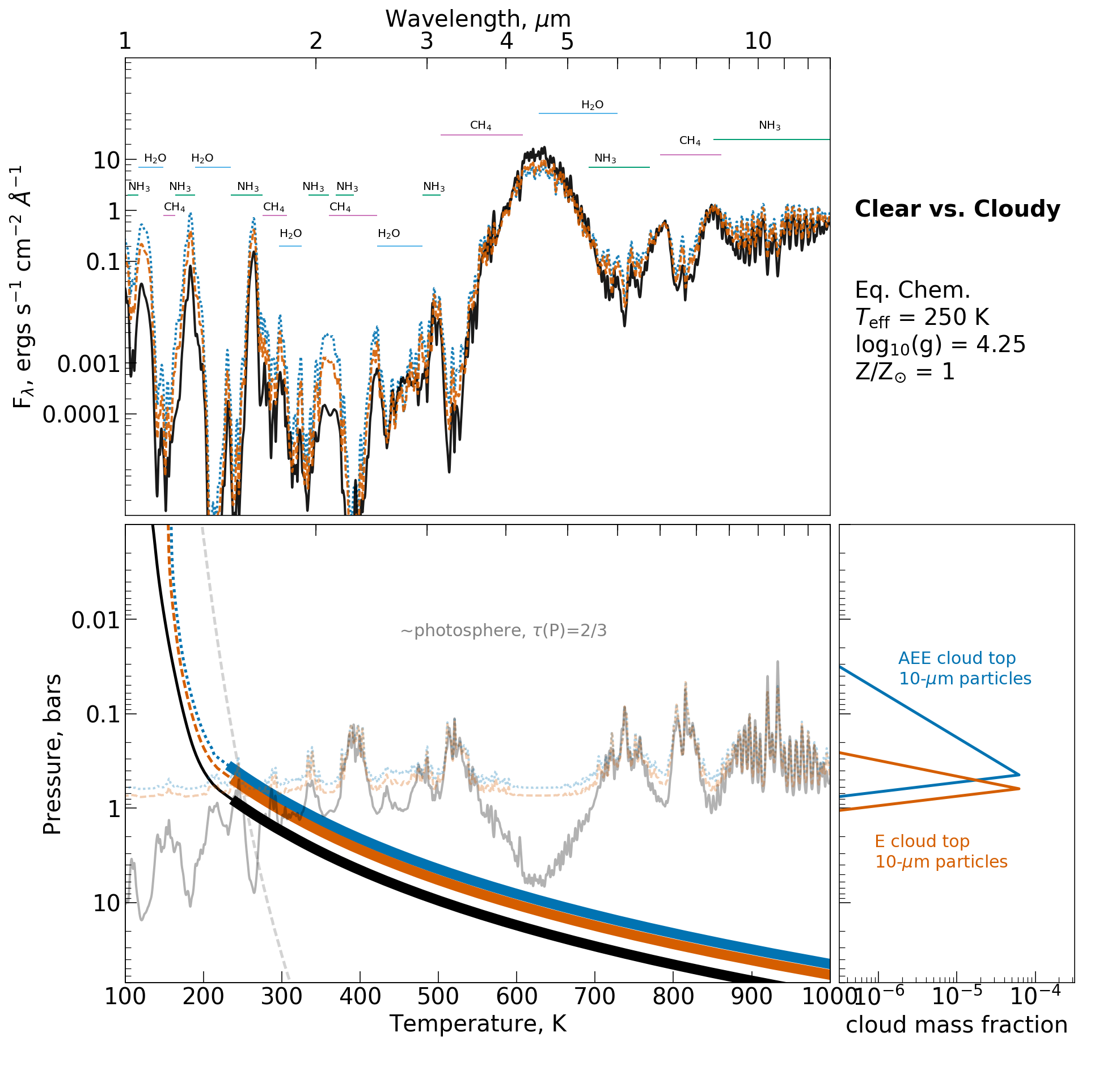}
    \caption{A comparison of clear and cloudy atmosphere models. Similar format to Figures \ref{fig:vary_temp}-\ref{fig:vary_Z}, but now the lower right panel shows the cloud positions rather than the chemical abundance profiles. The dotted blue lines correspond to a more vertically extended AEE type cloud top, the orange dashed lines correspond to a more compact E type cloud top, and the black solid lines correspond to the clear atmosphere. Notably the cloudy models $\tau$=2/3 curve indicating the approximate photosphere are centered around the cloud base.}
    \label{fig:clear_vs_cloudy}
\end{figure}

Figure \ref{fig:clear_vs_cloudy} shows a comparison of models computed with and without water clouds. All models have an effective temperature of 250 K, log$_{10}$(g) = 4.25, and solar metallicity. The clear model is shown by a black solid line and the cloudy models by a blue dotted line and an orange dashed line. Instead of the chemical abundance profiles shown in previous figures, the bottom right panel now shows a representation of where the water cloud forms. The cloudy models both have modal particle sizes of 10 $\mu$m, but have different vertical extents in the atmosphere.  

At this effective temperature and surface gravity, the cloud base forms around $\sim$0.4 bars, shifting the photosphere up to lower pressures than the clear model and warming the atmosphere in general. The more vertically extended cloud provides more heating. Comparing the spectra in the top panel of Figure \ref{fig:clear_vs_cloudy}, one can see that the cloudy models' M band emission peaks are suppressed, and absorption features across the full spectrum are not as deep as the clear model's spectrum. Y, J, H, and K band emission are all higher for the cloudy model relative to the clear model. In general, more vertically extended clouds have higher overall cloud optical depths, warming the atmosphere so that the cloud base forms at a lower pressure. 

\begin{figure}
    \centering
     \includegraphics[width=0.9\textwidth]{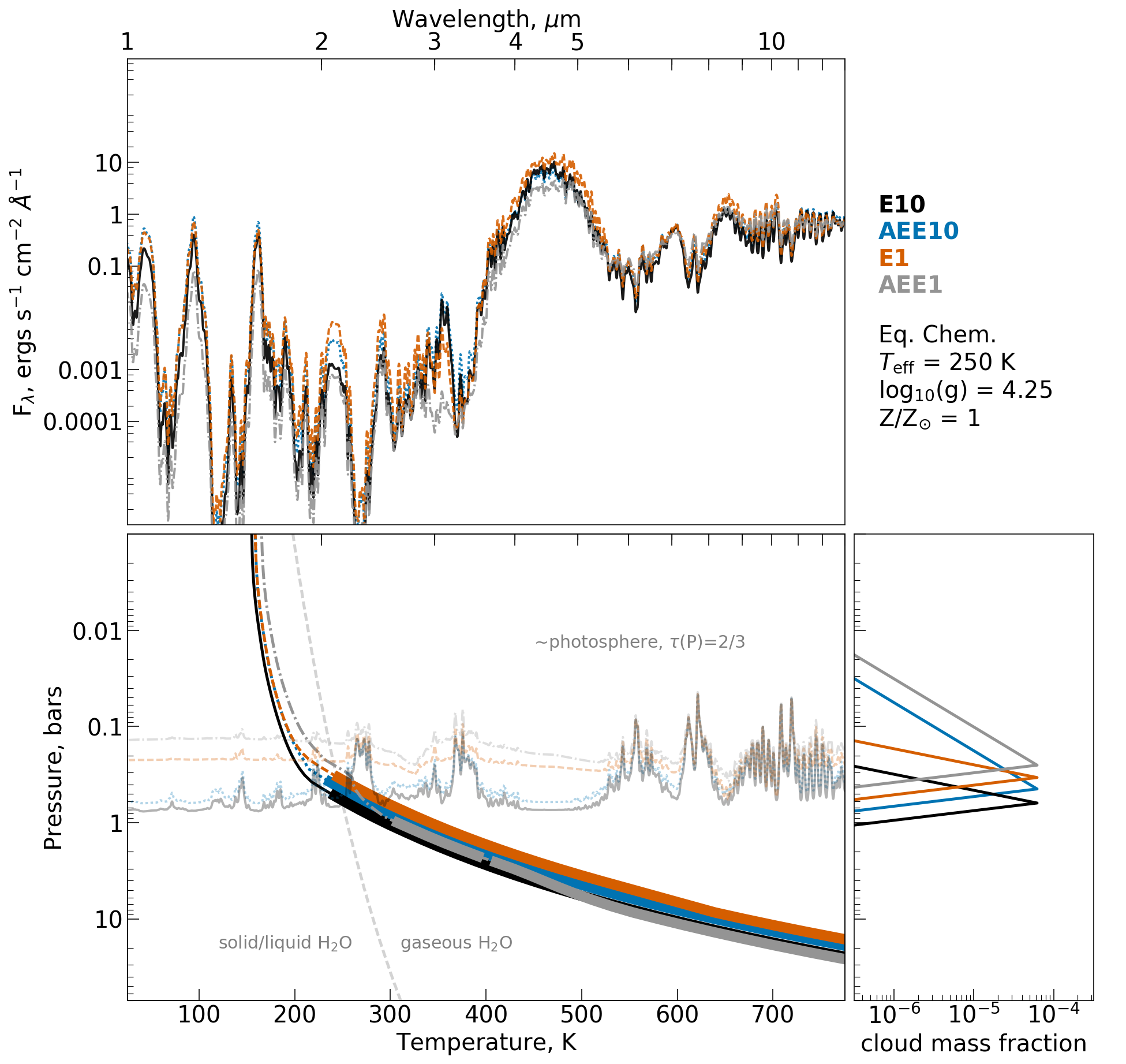}
    \caption{Similar format to Figure \ref{fig:clear_vs_cloudy}, but now comparing different cloud particles sizes and spatial extensions.}
    \label{fig:cloudy_compare_4}
\end{figure}

In Figure \ref{fig:cloudy_compare_4} we now compare the two cloudy models with 10-$\mu$m modal particle size from Figure \ref{fig:clear_vs_cloudy} (blue and black lines) against models with 1-$\mu$m modal particle size (gray and orange lines). At 10-$\mu$m modal particle size, the clouds provide a very gray opacity. At 1-$\mu$m modal particle size there is a variation in water's optical properties with wavelength. This most apparent in the $\tau$=2/3 curves around 2.5-3.5 $\mu$m (faint gray dash-dotted and orange dashed lines in the bottom left panel of Figure \ref{fig:cloudy_compare_4}), and the corresponding absorption in the AEE1 spectrum at 3$\mu$m in particular. The most optically thick cloud, an AEE cloud profile with 1-$\mu$m modal particle size, induces a detached convective zone, significantly altering the slope of the P-T profile through the photosphere. Smaller particle sizes for a fixed total mass of water result in higher cloud opacity and a slightly warmer atmosphere, which in turn causes a cloud base at lower pressures. 

As effective temperature rises, the cloud bases also form at lower pressures. Once effective temperatures rise to $\sim$425 K, clouds are forming in the upper atmosphere, well above the photosphere. They are too tenuous to warm the deeper atmosphere much. Once temperatures are below $\sim$200 K, clouds are forming well below the photosphere and again have little effect on the emergent spectrum. For higher surface gravities, the cloud bases form deeper in the atmosphere at higher pressures. 

\subsection{Cloudy Disequilibrium Chemistry Models}\label{sec:cloudy_neq_models}

We now come to the culminating suite of models: disequilibrium chemistry with water clouds. Observations clearly favor disequilibrium chemistry, and effective temperatures indicate that P-T profiles should create atmospheric conditions where water condensation is possible. Therefore, we expect this class of models to be the best for interpreting data. Whether this is borne out will be explored in the subsequent section comparing models and observations.

\begin{figure}
    \centering
    \includegraphics[width=0.9\textwidth]{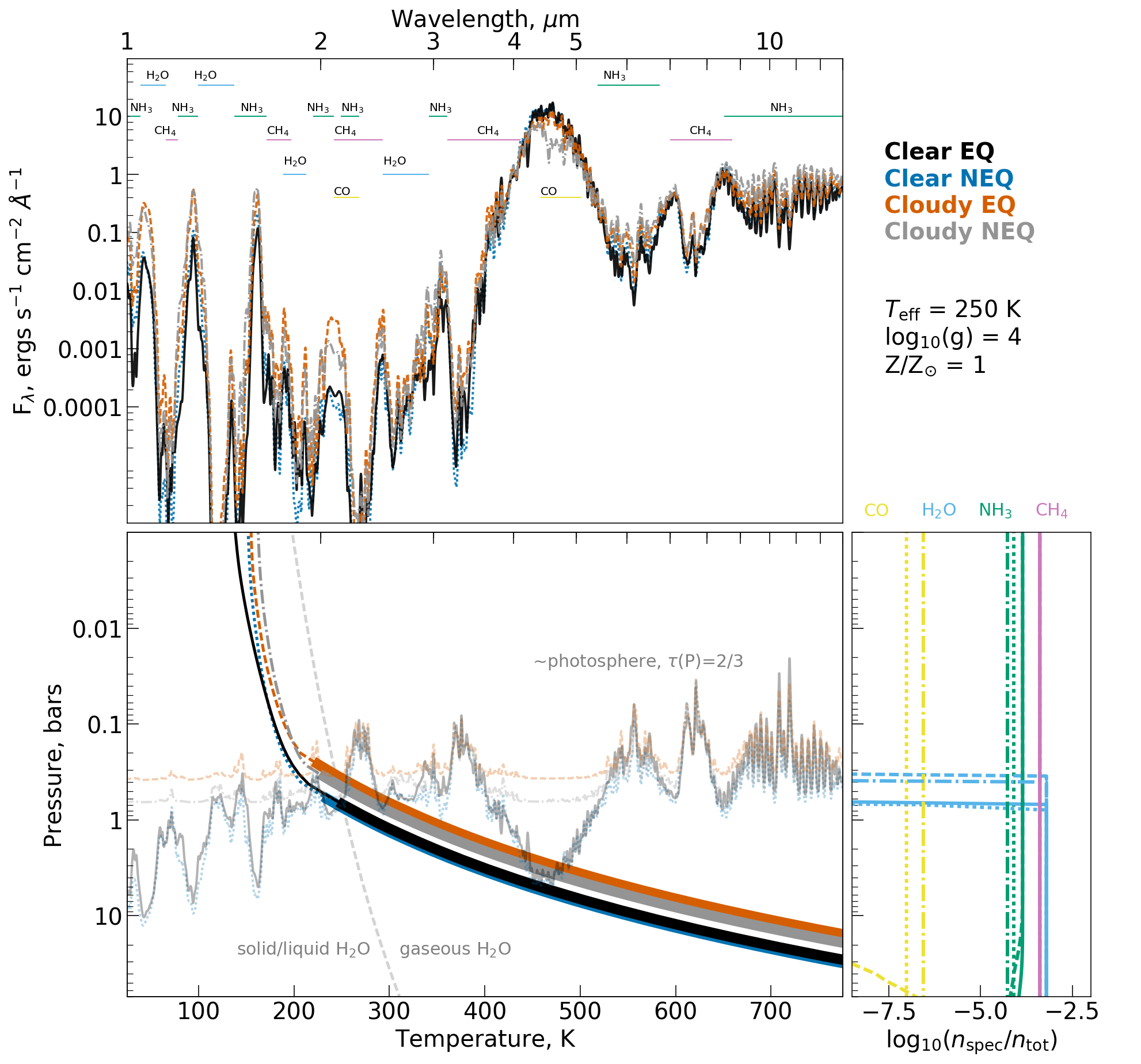}
    \caption{Comparison of all four classes of model for an atmosphere with $T_{\mathrm{eff}}$=250 K, log$_{10}$(g)=4, and Z/Z$_{\odot}$. }
    \label{fig:250K_four_models}
\end{figure}

In Figure \ref{fig:250K_four_models} we compare all four model classes: clear equilibrium chemistry (black lines), clear non-equilibrium chemistry (blue lines), cloudy equilibrium chemistry (orange lines), and cloudy non-equilibrium chemistry (gray lines) at an effective temperature of 250 K. The combination of water clouds and nonequilibrium chemistry results in an atmospheric structure and emergent spectrum with significant differences from the corresponding clear nonequilibrium and cloudy equilibrium models. This is not surprising. As we saw in the preceding sections, water clouds tend to warm the atmosphere when they are present with sufficient optical depth, and disequilibrium chemistry imprints a larger effect on the emergent spectrum when quench points occur in layers with higher temperatures. 

Meanwhile, disequilibrium chemistry tends to cool the atmosphere, but our results indicate that this is a smaller effect than the warming due to water clouds. At this effective temperature and surface gravity the reduction of CH$_4$ and the enhancement of CO cools the clear NEQ atmosphere only very very slightly relative to the clear EQ atmosphere. No CO absorption is seen at 4.5 $\mu$m, in fact the two clear spectra are nearly indistinguishable on this log scale plot. In contrast, with the addition of water clouds, disequilibrium chemistry mixes sufficient CO up into the upper atmosphere to create a strong feature at 4.5 $\mu$m, and depletes CH$_4$ and NH$_3$ sufficiently to both cool the cloudy NEQ P-T profile relative to the clear NEQ P-T profile and reduce CH$_4$ and NH$_3$ absorption. 

\begin{figure}
    \centering
    \includegraphics[width=0.475\textwidth]{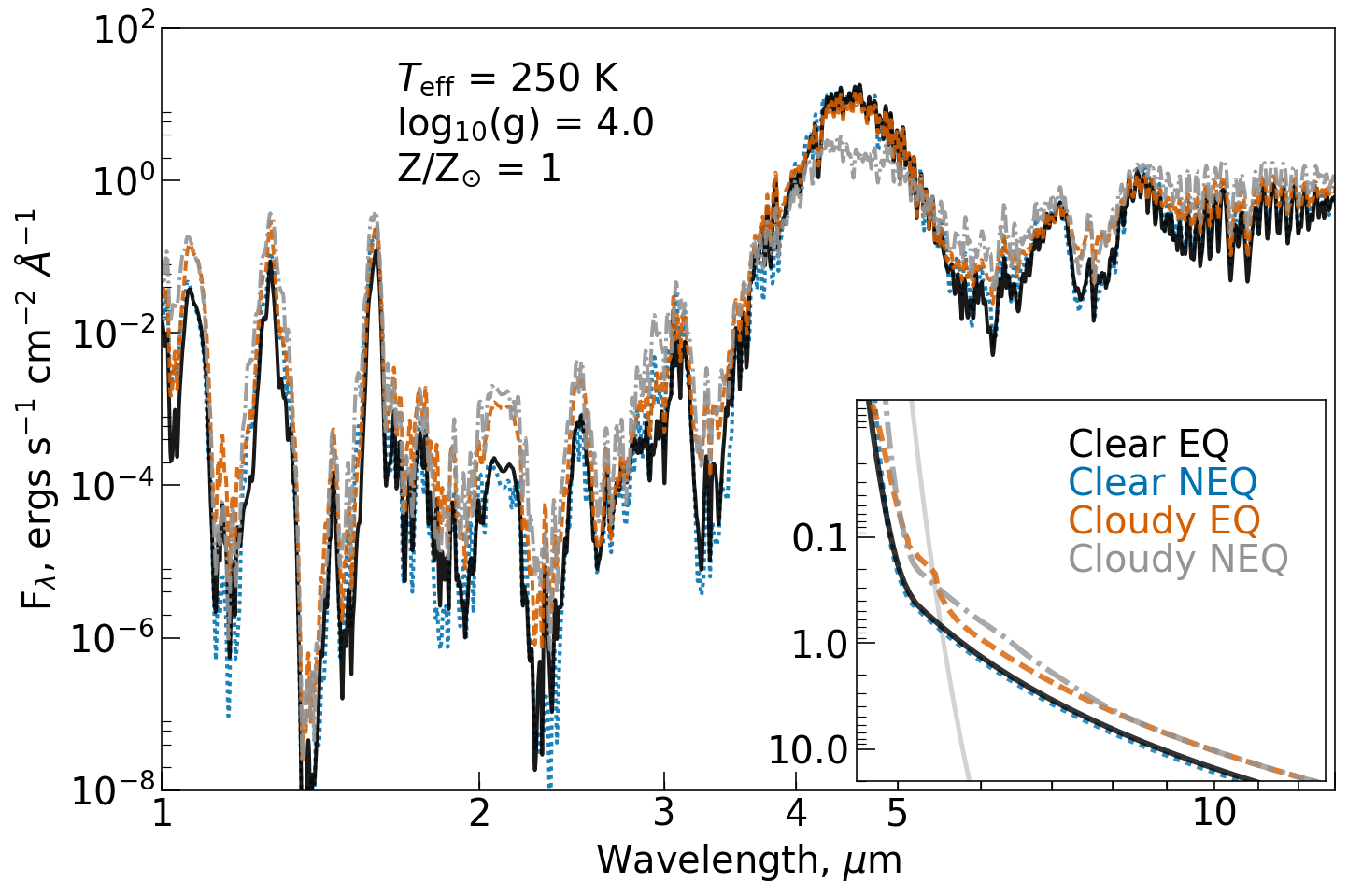}
    \includegraphics[width=0.475\textwidth]{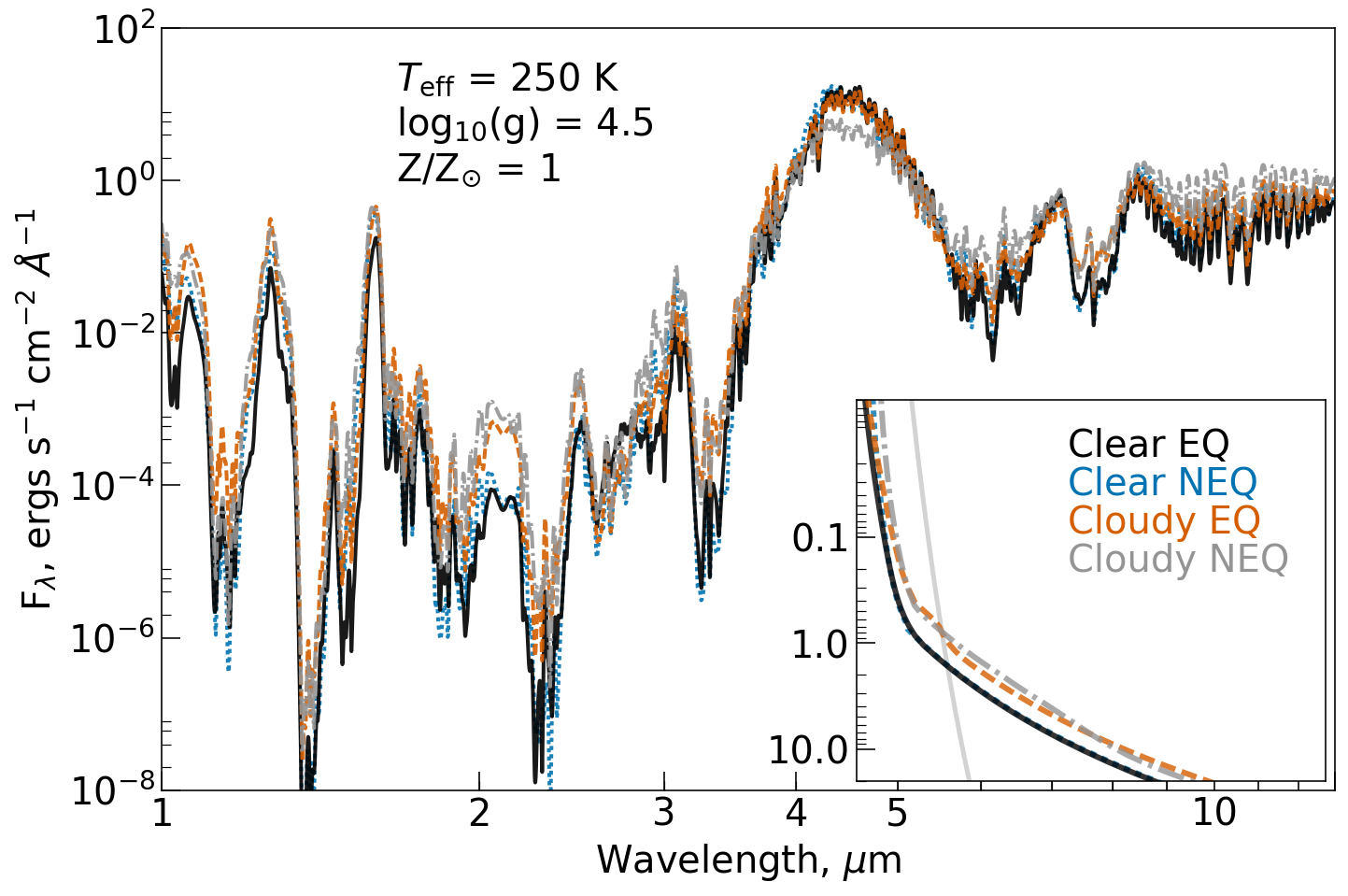}
    \includegraphics[width=0.475\textwidth]{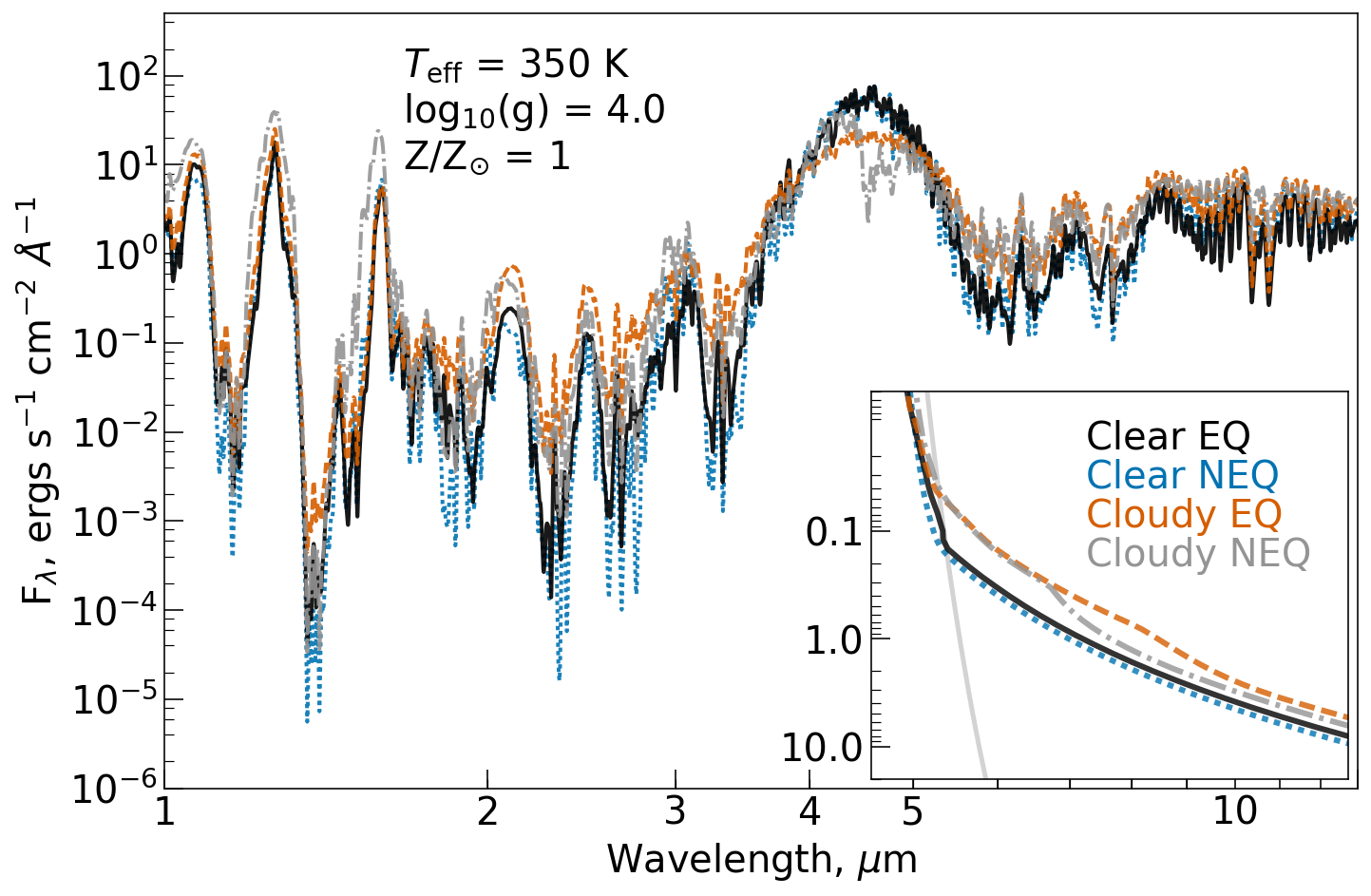}
    \includegraphics[width=0.475\textwidth]{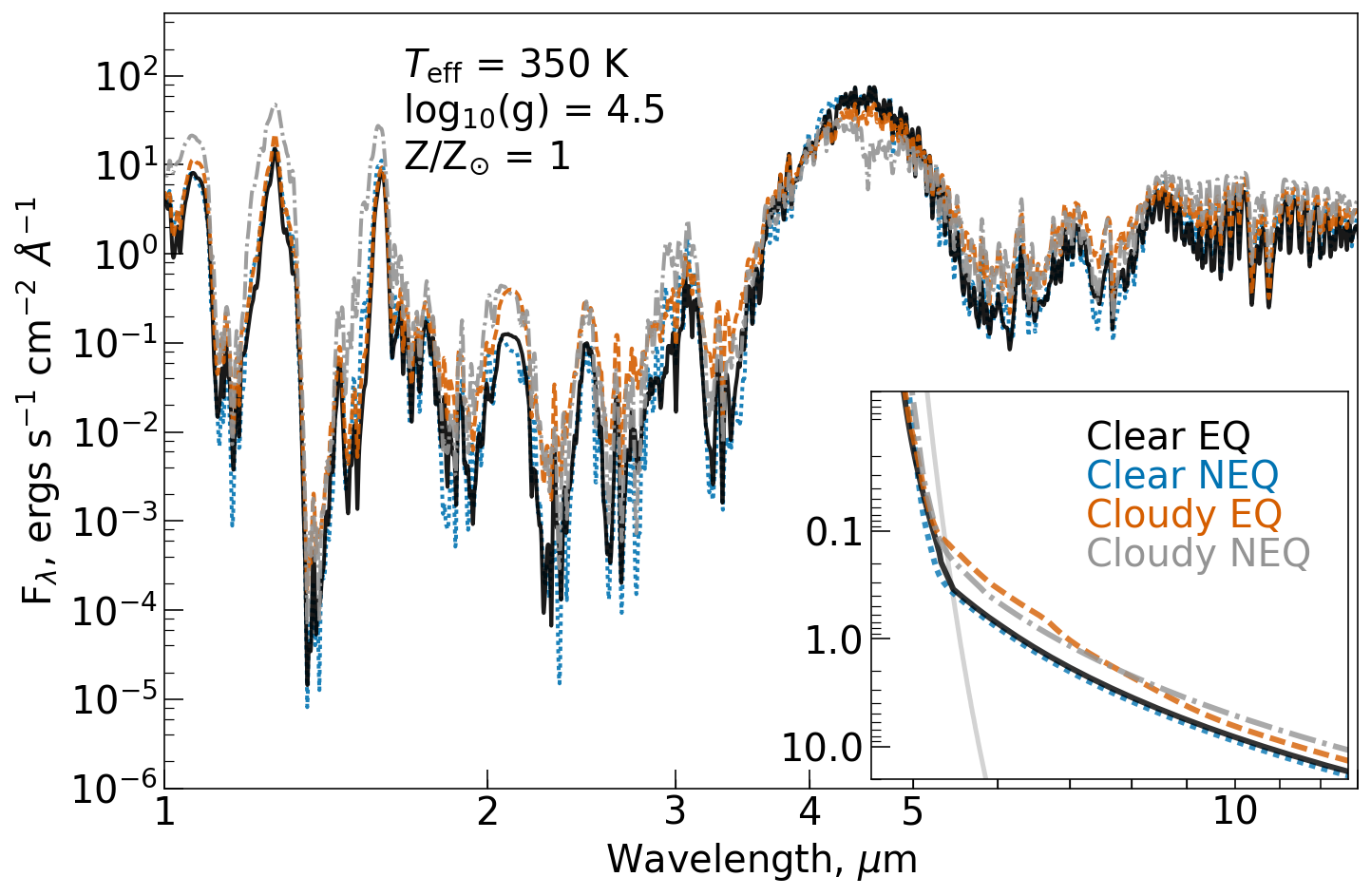}
    \includegraphics[width=0.475\textwidth]{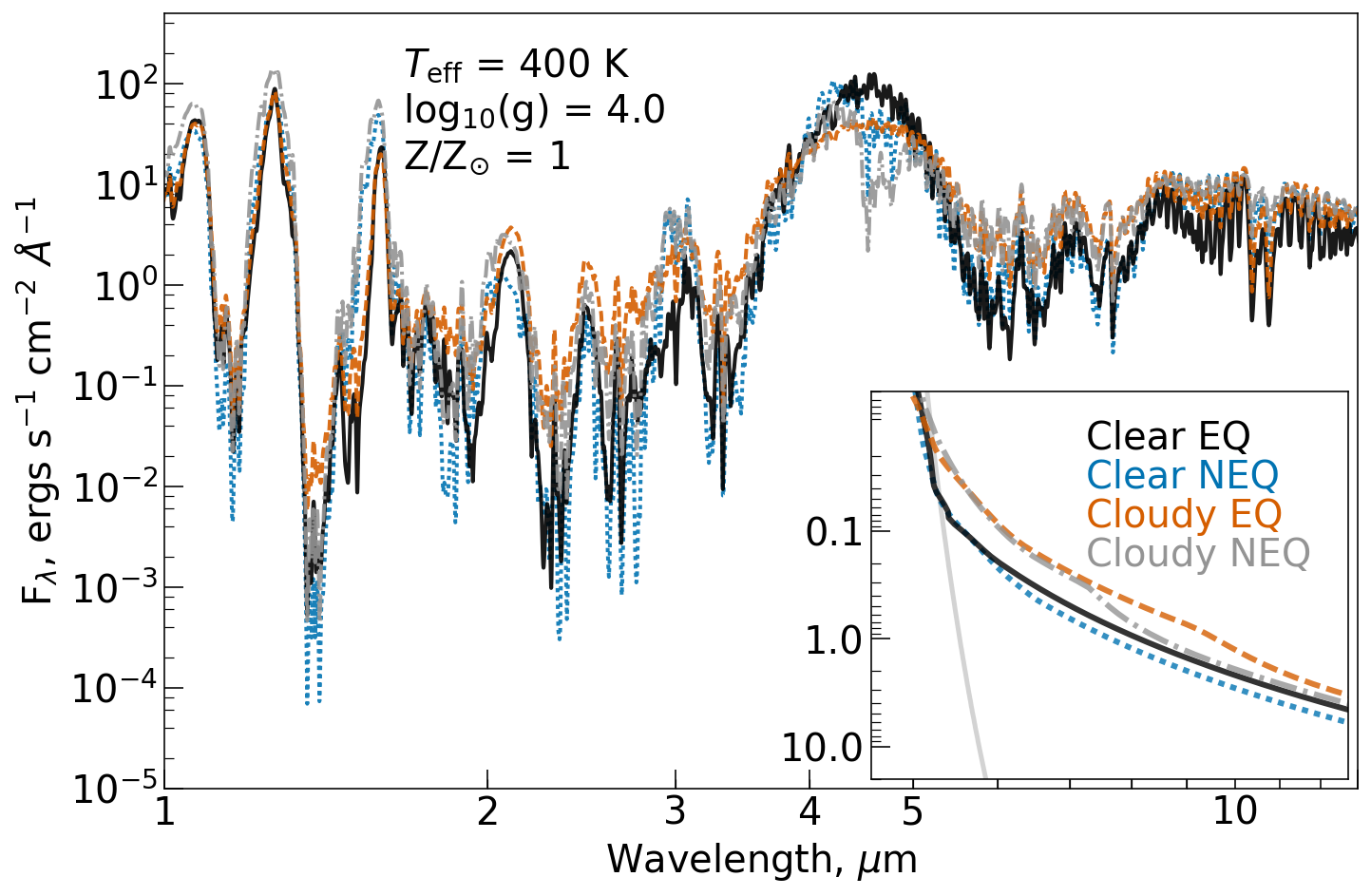}
    \includegraphics[width=0.475\textwidth]{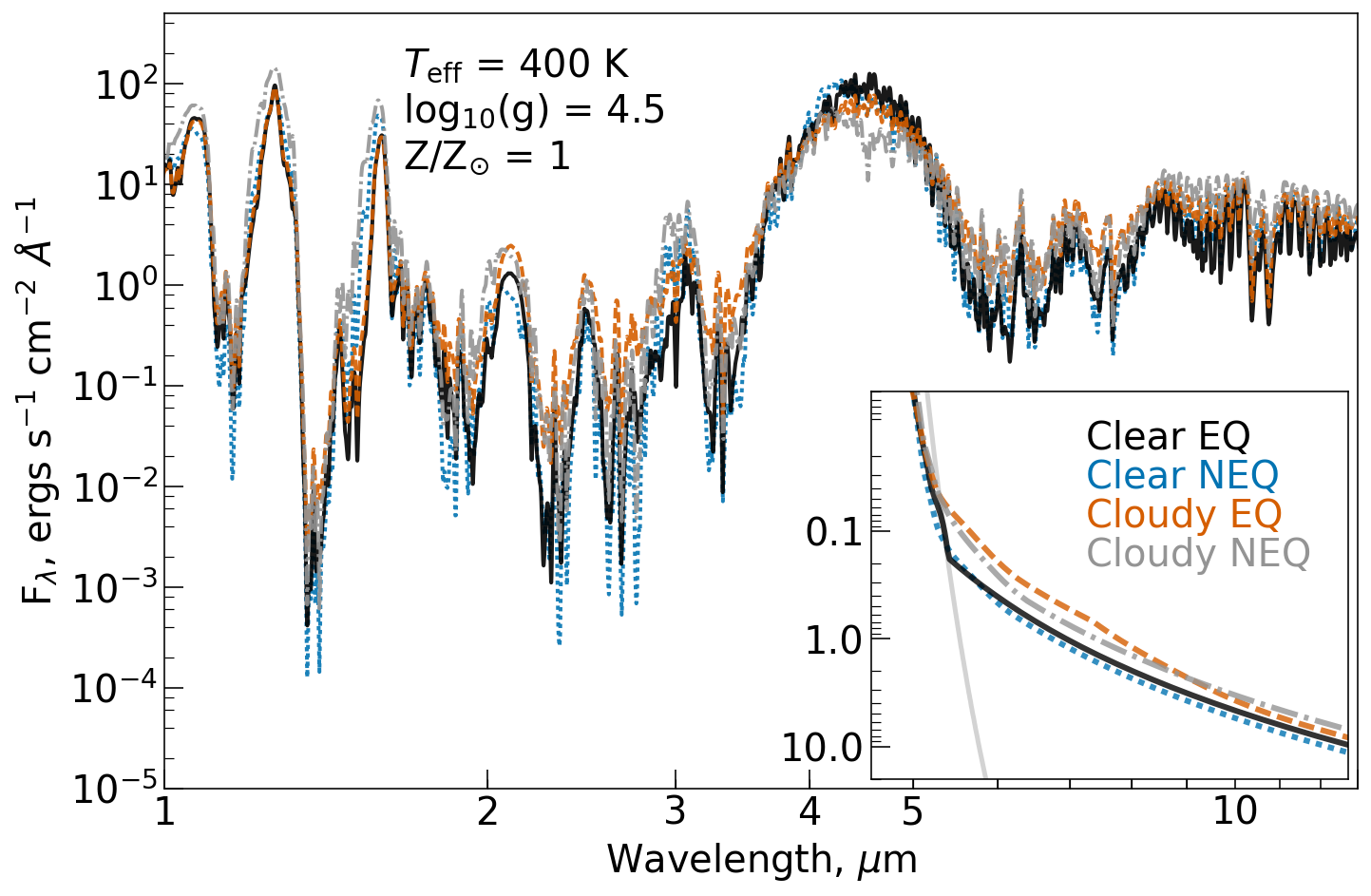}
    \caption{An illustration of the combined influence of water clouds and non-equilibrium chemistry on Y dwarf atmospheric structures and spectra. Each panel shows model spectra for clear equilibrium chemistry, clear nonequilibrium chemistry, and cloudy (type AEE10) nonequilibrium chemistry. The associated P-T profiles are shown as an inset. The top row has $T_{\mathrm{eff}}$ = 250 K, the middle row has $T_{\mathrm{eff}}$ = 350 K, and the bottom row has $T_{\mathrm{eff}}$ = 400 K. The left column has surface gravity log$_{10}$(g)=4 and the right column has surface gravity log$_{10}$(g)=4.5. Note that, in order to make the figure more legible, P-T profiles do not mark convective zones.}
    \label{fig:mega_4_models}
\end{figure}

Figure \ref{fig:mega_4_models} compares cloudy nonequilibrium models to cloudy equilibrium models, clear equilibrium, and clear nonequilibrium models for a range of effective temperatures (250 K, 350 K, and 400 K) and surface gravities (log$_{10}$(g)=4, and log$_{10}$(g)=4.5). These examples demonstrate overall trends in the cloudy nonequilibrium model grid: the warming effect of water clouds wins out over the cooling effect of disequilibrium chemistry, but the disequilibrium chemistry still cools the atmosphere relative to just a cloudy equilibrium chemistry atmosphere. This warmer than clear deep atmosphere enhances the signatures of vertical mixing. It is not shown here, but we find that, there can be non-monotonic spectral signatures of disequilibrium as a function of effective temperature and surface gravity, similar to the clear disequilibrium atmosphere case \cite{Mukherjee2022}.

\begin{figure}
    \centering
    \includegraphics[width=\textwidth]{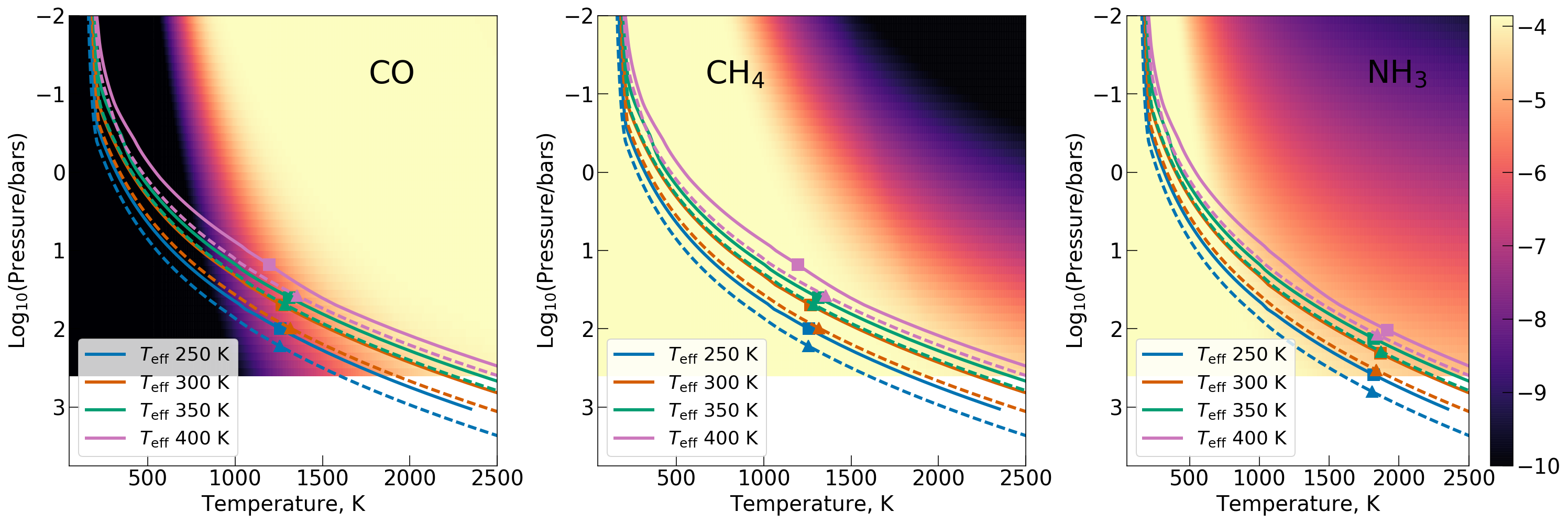}
    \caption{A demonstration of how adding clouds shifts quench points for disequilibrium chemistry. Abundances of disequilibrium species are indicated by the background colors. Dotted lines show clear models and solid lines show cloudy models. All models have log$_{10}$(g)=4.25.}
    \label{fig:cloud_effects_on_quench_points}
\end{figure}

\section{Comparison with Observations}\label{sec:results_observations}
In this section, we compare our new model grids to observations of Y dwarfs available in the pre-JWST literature. As mentioned in the introduction, it had been previously demonstrated that Y dwarf spectra show signatures of nonequilibrium chemistry due to vertical mixing (\citealt{Zalesky2019};\citealt{Miles2020}; \citealt{Phillips2020}; \citealt{Leggett2021}), and that typical 1D radiative-convective models produced too little flux in the K, [3.6]+W1, and W3 bands (\citealt{Phillips2020}; \citealt{Leggett2021}). The K, [3.6]+W1, and W3 band photospheres tend to reside higher in the atmospheres around pressures of ~1 bar, so one option to brighten these portions of Y dwarf spectra could be to invoke an additional process which warms the upper atmosphere relative to the deeper atmosphere. This is essentially what was done in \citealt{Leggett2021}, and achieves a reasonable match to observations. 

Rather than repeat their approach, which is agnostic to the source of the changes to the P-T structure, we construct models which include both nonequilibrium chemistry and water clouds, to see if their combined effects lead to a self-consistent atmosphere with warm enough upper layers relative to deep layers to replicate observations across the full spectral energy distribution.  As shown in the preceding sections, water clouds tend to warm the atmosphere while disequilibrium chemistry tends to cool the deeper layers, and nonequilibrium chemistry can further brighten the [3.6]+W1 photometry by reducing CH$_4$ absorption and brighten the W3 band by reducing NH$_3$ absorption. Both clouds and nonequilibrium chemistry can also alter the shapes of P-T profiles by affecting the locations and existence of detached convective zones/sandwiched radiative zones. 

Overall, we find that our nonequilibrium cloudy models are not able to fully replicate observations of Y dwarf spectra, nor the larger ensemble of photometric measurements. With our implementation of quenching chemistry and water cloud parameterization, the warming effect of water clouds wins out over the cooling effect of disequilibrium chemistry in the deep atmosphere. This leads to over-bright Y, J, and H band peaks in cloudy disequilibrium models, and, while the clouds do brighten the K, [3.6], and W3 bands, it is generally not a large enough effect to match with current observations. We walk through this in more detail in the remainder of this section, comparing to photometry in \S \ref{sec:photometry_compare} and to spectra in \S \ref{sec:spectra_compare}.

All observational data are taken from \citealt{Leggett2021}, which contributes new photometry, flux calibrates spectra from diverse literature sources, and compiles photometry from a number of previous studies outlined in their Table 10 (\citealt{2011AJ....141..203A}; \citealt{2010ApJ...718L..38A}; \citealt{2020ApJ...895..145B}; \citealt{2015ApJ...814..118B}; \citealt{2020AJ....159..257B}; \citealt{1999ApJ...522L..65B}; \citealt{2000ApJ...531L..57B}; \citealt{2002ApJ...564..421B}; \citealt{2003AJ....126.2487B}; \citealt{2004AJ....127.2856B}; \citealt{2006ApJ...637.1067B}; \citealt{2008ApJ...689L..53B}; \citealt{2008MNRAS.391..320B}; \citealt{2009MNRAS.395.1237B}; \citealt{2010MNRAS.404.1952B}; \citealt{2010MNRAS.406.1885B}; \citealt{2011MNRAS.414L..90B}; \citealt{2013MNRAS.433..457B}; \citealt{2006AJ....131.2722C}; \citealt{Cushing2011}; \citealt{2014AJ....147..113C}; \citealt{2016ApJ...823..152C}; \citealt{2008A&A...482..961D}; \citealt{2010A&A...518A..39D}; \citealt{2012ApJS..201...19D}; \citealt{2013Sci...341.1492D}; \citealt{2015ApJ...803..102D}; \citealt{2012ApJ...752...56F}; \citealt{2020ApJ...889..176F}; \citealt{2018A&A...616A...1G}; \citealt{2011AJ....142...57G}; \citealt{2010MNRAS.405.1140G}; \citealt{2019AJ....158..182G}; \citealt{2012AJ....144..148G}; \citealt{Kirkpatrick2011}; \citealt{Kirkpatrick2012}; \citealt{2013ApJ...776..128K}; \citealt{Kirkpatrick2019}; \citealt{2021ApJS..253....7K}; \citealt{2004AJ....127.3553K}; \citealt{2009ApJ...695.1517L}; \citealt{2010ApJ...710.1627L}; \citealt{2012ApJ...748...74L}; \citealt{2014ApJ...780...62L}; \citealt{Leggett2015}; \citealt{Leggett2017}; \citealt{2019ApJ...882..117L}; \citealt{2011ApJ...740..108L}; \citealt{2012ApJ...758...57L}; \citealt{2007MNRAS.379.1423L}; \citealt{2009MNRAS.397..258L}; \citealt{2012A&A...548A..53L}; \citealt{2007AJ....134.1162L}; \citealt{2010MNRAS.408L..56L}; \citealt{Luhman2011}; \citealt{2012ApJ...744..135L}; \citealt{Luhman2014a}; \citealt{2013ApJ...777...36M}; \citealt{2013ApJS..205....6M}; \citealt{2011ApJ...726...30M}; \citealt{2013A&A...560A..52M};\citealt{2010A&A...524A..38M}; \citealt{2020ApJ...888L..19M}; \citealt{Martin2018}; \citealt{2020ApJ...889...74M}; \citealt{Meisner2020}; \citealt{2011MNRAS.414..575M}; \citealt{2006ApJ...651..502P}; \citealt{2008MNRAS.390..304P}; \citealt{2012MNRAS.422.1922P};\citealt{2014MNRAS.437.1009P}; \citealt{Pinfield2014}; \citealt{2010A&A...510L...8S}; \citealt{2010A&A...515A..92S}; \citealt{2011A&A...532L...5S}; \citealt{2010A&A...511A..30S}; \citealt{1999ApJ...522L..61S}; \citealt{2004PASP..116....9S}; \citealt{2009AJ....137.4547S}; \citealt{2013PASP..125..809T}; \citealt{2003AJ....126..975T}; \citealt{2005AJ....130.2326T}; \citealt{2012ApJ...759...60T}; \citealt{2014ApJ...796...39T}; \citealt{2018ApJS..236...28T}; \citealt{2000ApJ...531L..61T}; \citealt{2004AJ....127.2948V}; \citealt{2007MNRAS.381.1400W}; \citealt{2013AJ....145...84W};).

\subsection{Comparison to Photometric Data}\label{sec:photometry_compare}

\begin{figure}
    \centering
    \includegraphics[width=0.9\textwidth]{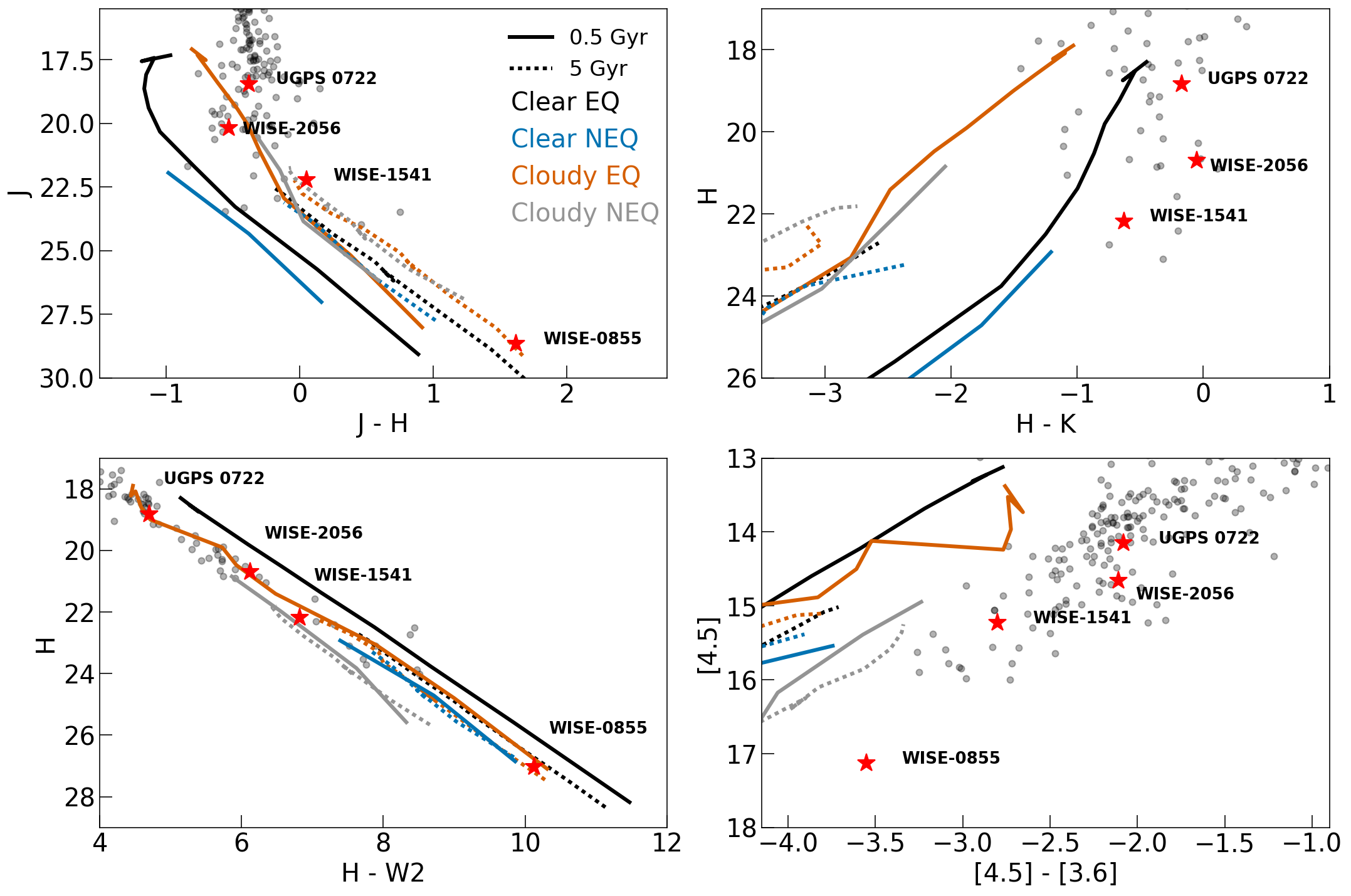}
    \caption{Comparison of cool Brown Dwarf photometry (gray points) against 0.5 and 5 Gyr isochrones generated using the evolutionary models of Burrows 1997 and our various classes of atmospheric+spectral models. Red stars mark the photometric points corresponding to objects for which we also compare spectra. Black lines denote clear models with equilibrium chemistry, orange lines denote clear models with disequilibrium chemistry, blue lines denote models with E10 water clouds, and gray lines denote models with E10 water clouds and disequilibrium chemistry. Dotted lines denote 5 Gyr isochrones and solid lines denote 0.5 Gyr isochrones. Photometric data are taken from the sample presented in \citealt{Leggett2021}.}
    \label{fig:photometry_cmd}
\end{figure}

\begin{figure}
    \centering
    \includegraphics[width=0.9\textwidth]{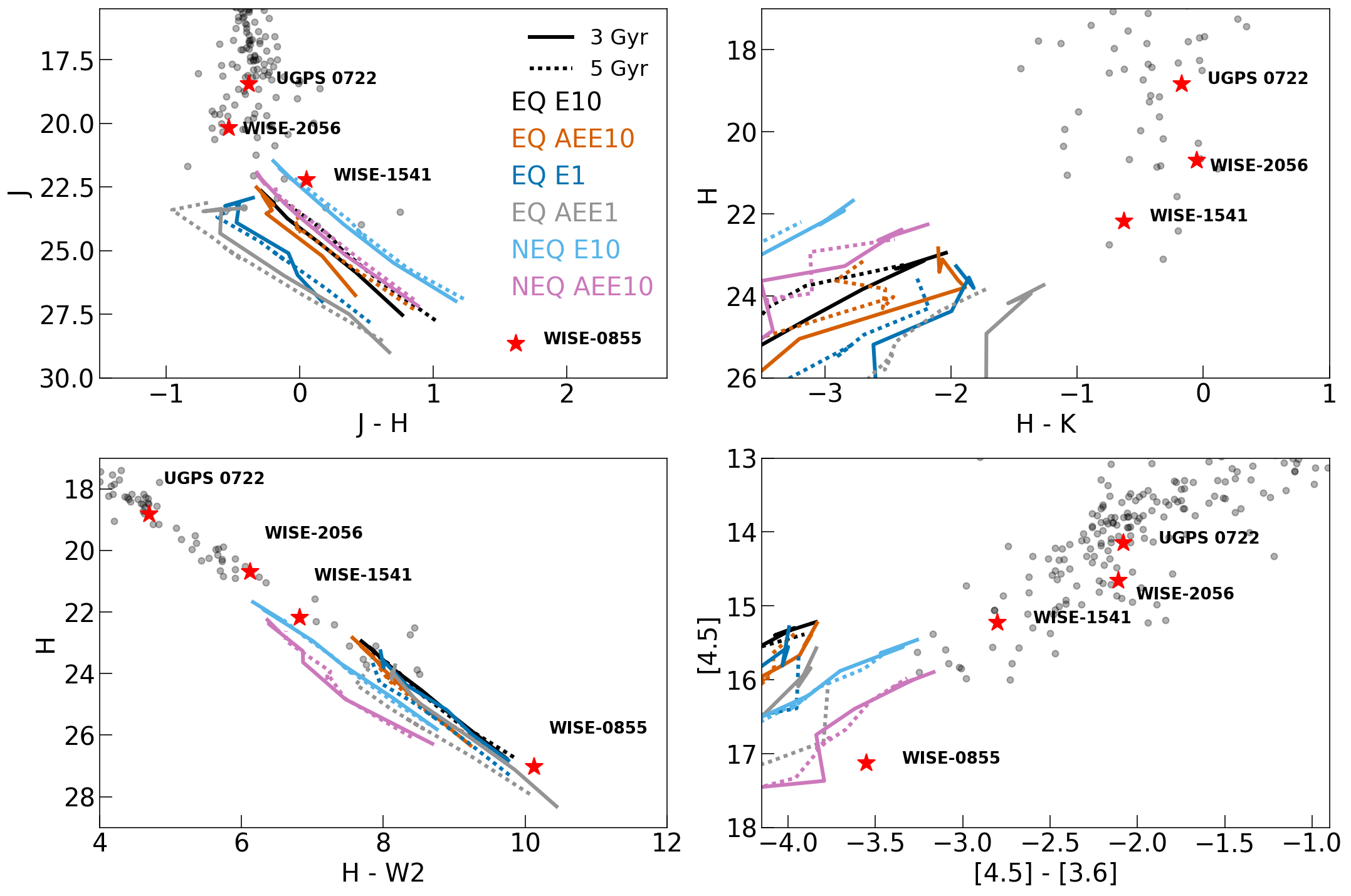}
    \caption{Similar format to Figure \ref{fig:photometry_cmd}, but now with isochrones for 3 and 5 Gyr constructed using models with a wider variety of cloud properties.}
    \label{fig:photometry_cmd_clouds}
\end{figure}

Figure \ref{fig:photometry_cmd} compares isochrones constructed from our model grids and the evolutionary models of \citealt{Burrows1997} against color-magnitude data for cool brown dwarfs. All isochrones should be taken as an approximations, since we did not run a new set of evolutionary models consistently with this grid of atmospheres. Doing so would require extending up to warmer temperatures than those explored here, so it is left to future work. In all cases dotted lines are isochrones for 5 Gyr and solid lines are isochrones for 0.5 Gyr, while colors indicate different cloud and chemistry combinations. Figure \ref{fig:photometry_cmd_clouds} is similar, but with a wider variety of cloud particle size and cloud vertical extent. Since cloudy models reach up to temperatures of only 400 K, we show isochrones for 3 Gyr rather than 0.5 Gyr with the dotted lines. 

Beginning with Figure \ref{fig:photometry_cmd}, one can see that the clear equilibrium models are a little too dim in the H band, and then much too dim in the K band and [3.6] band (black lines in Figure \ref{fig:photometry_cmd}). It's not shown in the color-magnitude diagrams, but, looking ahead to Figure \ref{fig:spectra1}, one can see that the clear equilibrium models tend to be too dim in W3 as well.

Incorporating non-equilibrium chemistry, while keeping a clear atmosphere, tends to brighten the H band enough to improve agreement with data in the J, J-H plane. H-W2 colors are also improved (orange/rust colored lines in \ref{fig:photometry_cmd}). However, the K band is roughly the same brightness, if not a bit dimmer, leading to an even worse H-K fit than the clear equilibrium models. Disequilibrium chemistry brightens the [3.6] band for $T_{\mathrm{eff}}$ $\geq$450 K, but not to the extent needed to agree with observed [4.5]-[3.6] colors. Non-equilibrium chemistry brightens W3 photometry for effective temperatures above 425 K, but for cooler objects it has only a small impact.

Incorporating clouds while maintaining equilibrium chemistry brightens the K band and [3.6] band, but makes NIR peaks in Y, J, and H too bright relative to flux at longer wavelengths (blue lines in \ref{fig:photometry_cmd}). The J - H colors are even bluer than clear equilibrium models making for a worse fit while H - K colors end up similar to clear equilibrium models. H - W2 and [4.5] - [3.6] colors see a slight improvement over clear equilibrium models.

Models with both water clouds and disequilibrium chemistry attain the closest fit to the [4.5] - [3.6] colors, but are still smaller than observed photometry (gray lines in Figure \ref{fig:photometry_cmd}). They brighten the K, [3.6], and W3 bands, but not to the extent needed to agree with photometry. J - H colors agree well, but H - W2 and H - K are both too blue.

In the previous Figure \ref{fig:photometry_cmd}, we presented only the case of E10 type clouds. Microphysical models and observations in the Solar System/on earth indicate that particles forming higher up in the atmosphere ought to have smaller particle sizes and that the vertical extent of a cloud will depend on the details of nucleation and vertical mixing profiles. For example, if water is forming on seed particles of KCl or NaCl it will have a more compact vertical extent than if it is condensing homogeneously via supersaturation \citep{Mang2022}. We chose E10 type clouds as our fiducial clouds because they agree well with CARMA cloud profiles presented in \citealt{Mang2022} for 250 K - 325 K objects. However, given the large degree of uncertainty as to the details of water cloud properties in Y dwarfs, we also computed a set of cloudy models with a larger vertical extent in the atmosphere and a set of models with smaller modal particle sizes, as described in \S \ref{sec:cloudy_eq_models}. 

In Figure \ref{fig:photometry_cmd_clouds}, we compare isochrones constructed from these four permutations of cloud particle size and vertical extent against photometric data. For a given particle size and chemistry, clouds with a greater vertical extent tend to cause a smaller J-H and H-W2 value, but a higher H - K and [4.5] - [3.6]. In J - H and [4.5] - [3.6] the nonequilibrium models with 10-$\mu$m modal particle size perform best, but in H - K and H - W2 they are too small (light blue and pink lines). Models with 1-$\mu$m modal particle sizes and equilibrium chemistry (gray and blue lines) attain better H - K, H - W2, and [4.5] - [3.6] colors than the 10-$\mu$m modal particle size equilibrium chemistry models (black and orange lines), but are far too small in J - H. 

\subsection{Comparison to Spectra}\label{sec:spectra_compare}

From the photometric comparisons above, we can already tell that our models incorporating nonequilibrium chemistry and clouds together have failed to bring models into agreement with observations. We now examine a selection of four objects (UGPS-0722, WISE-2056, WISE-1541, and WISE-0855) with M band spectra in order to better identify the source of these discrepancies: namely K, [3.6]+W1, and W3 being too dim. Note that in UGPS-0722 and WISE-2056, where we have spectra blueward of 1 $\mu$m, it also appears that there is too much potassium absorption in our models. These four objects span a range of temperatures from a late T dwarf at  $\sim$550 K down to the coolest known object at 250 K. 

\begin{figure}
    \centering
    \includegraphics[width=0.95\textwidth]{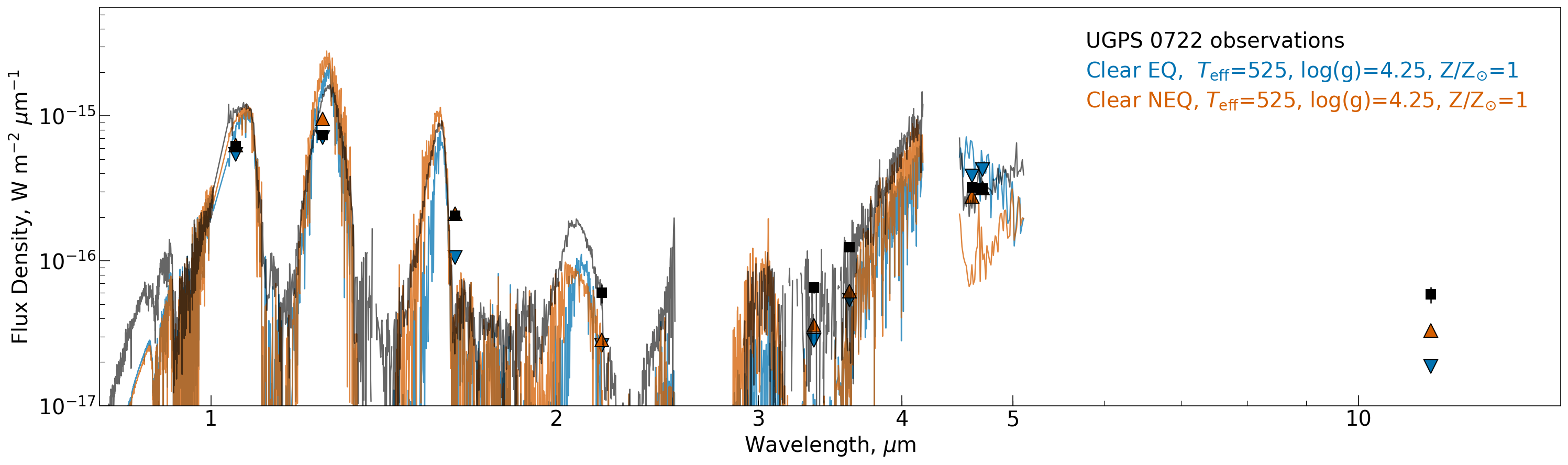}
    \includegraphics[width=0.95\textwidth]{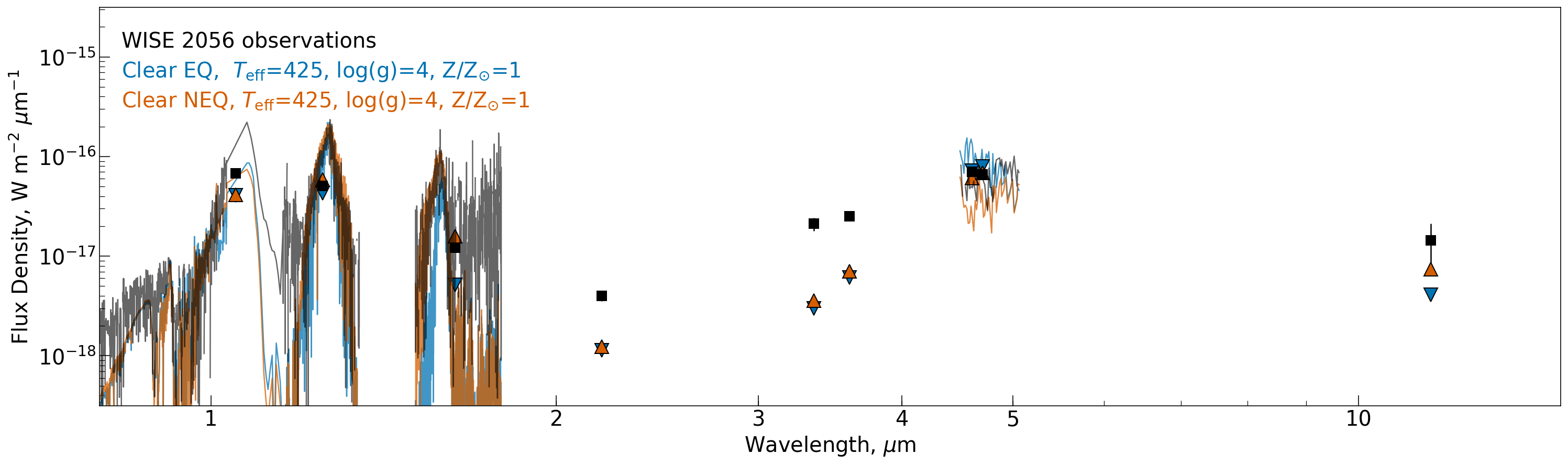}
    \includegraphics[width=0.95\textwidth]{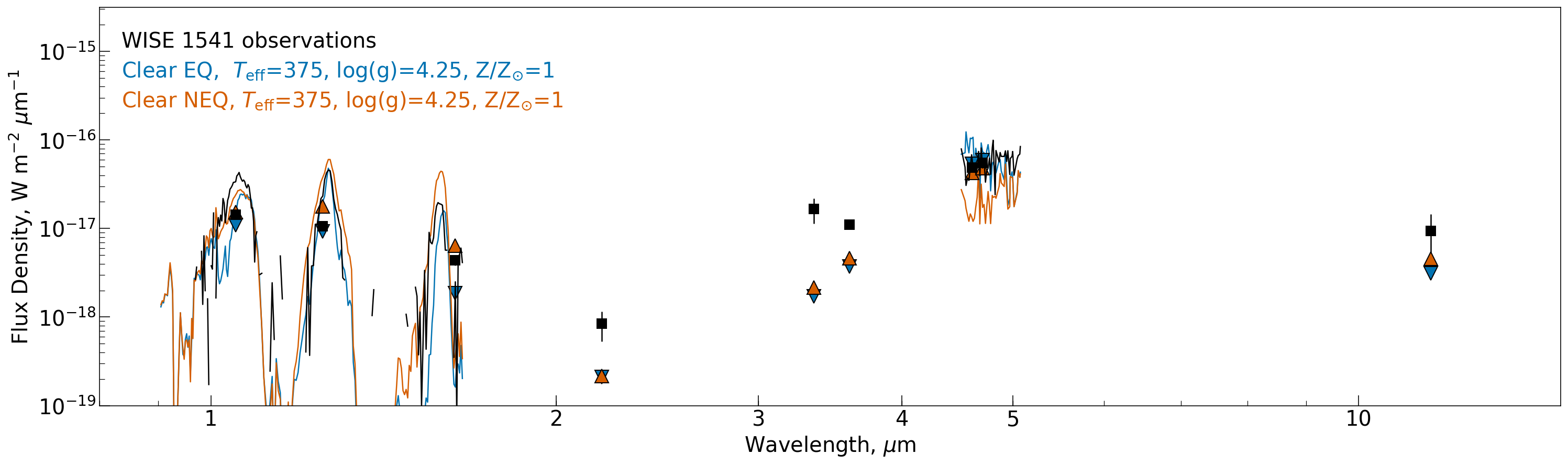}
    \includegraphics[width=0.95\textwidth]{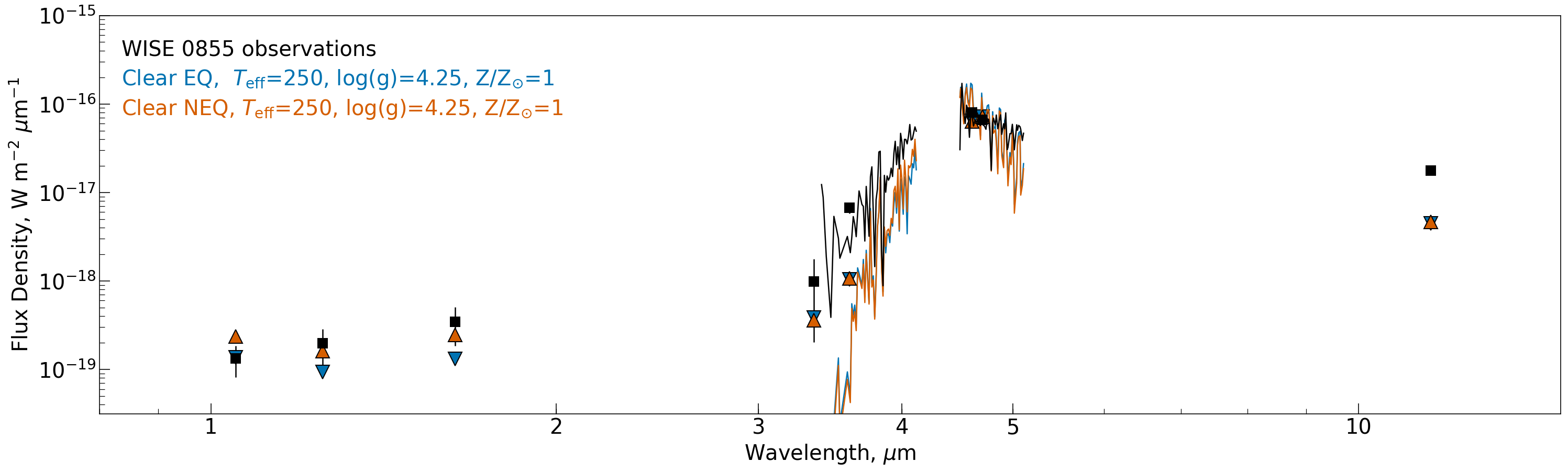}
    \caption{For each object we compare models with clear atmospheres to observations. Black lines and squares correspond to observations, blue lines and downward pointing triangles correspond to equilibrium chemistry and orange lines and upward pointing triangles correspond to nonequilibrium chemistry. Each model is scaled to minimize the $\chi^2$ value rather than using a radius from an evolutionary model and the distance as was done in the comparisons to photometry.}
    \label{fig:spectra1}
\end{figure}

\begin{figure}
    \centering
    \includegraphics[width=0.95\textwidth]{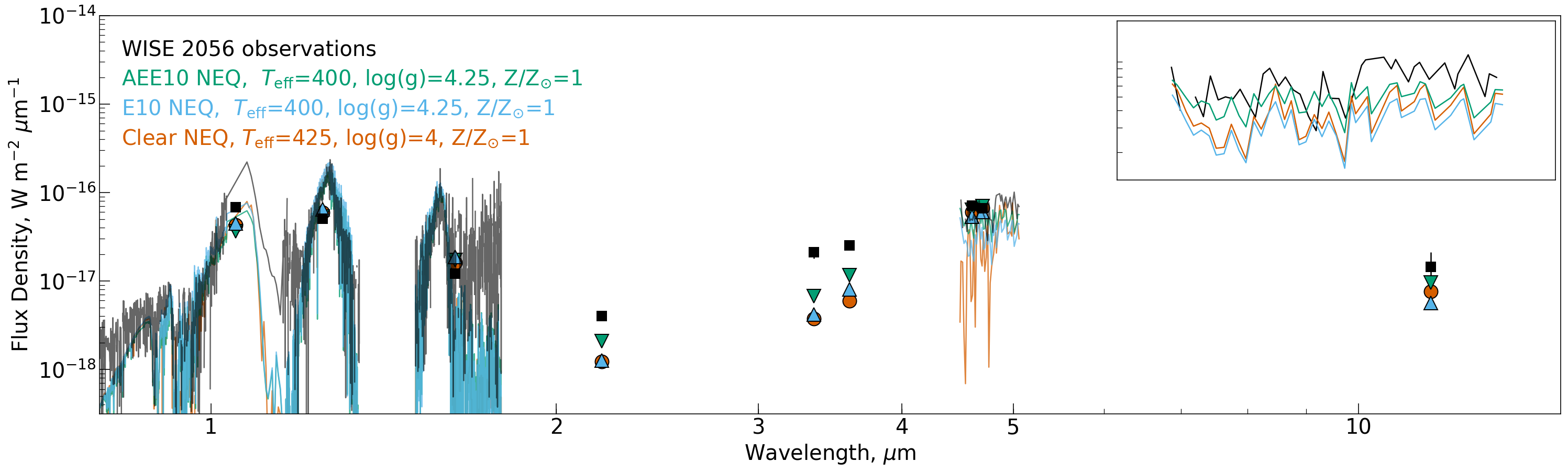}
    \includegraphics[width=0.95\textwidth]{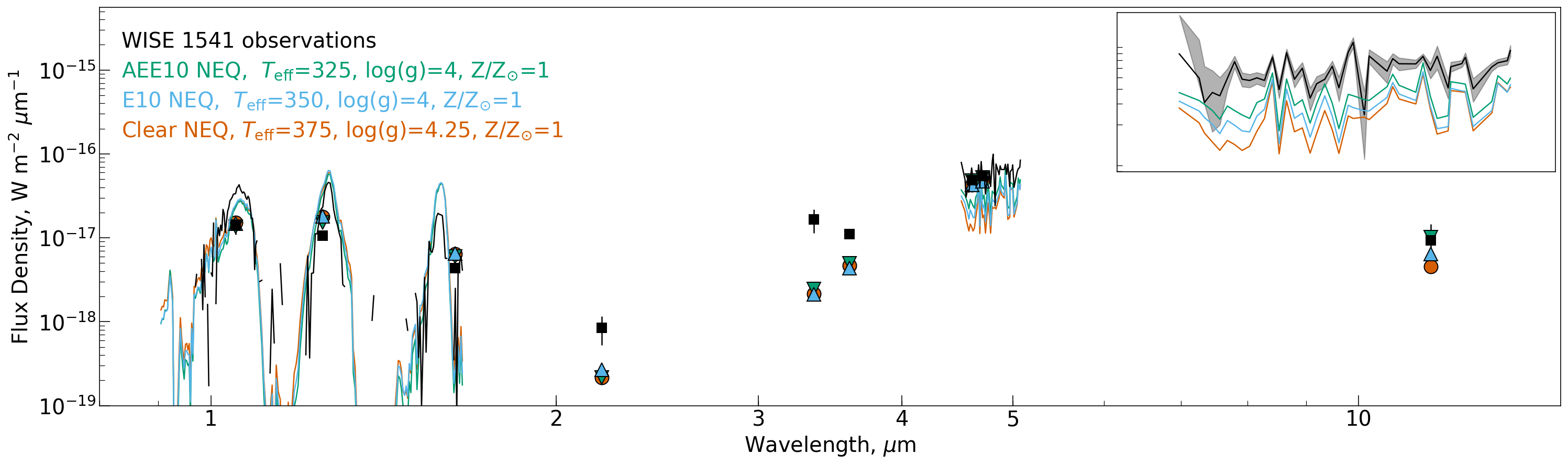}
    \includegraphics[width=0.95\textwidth]{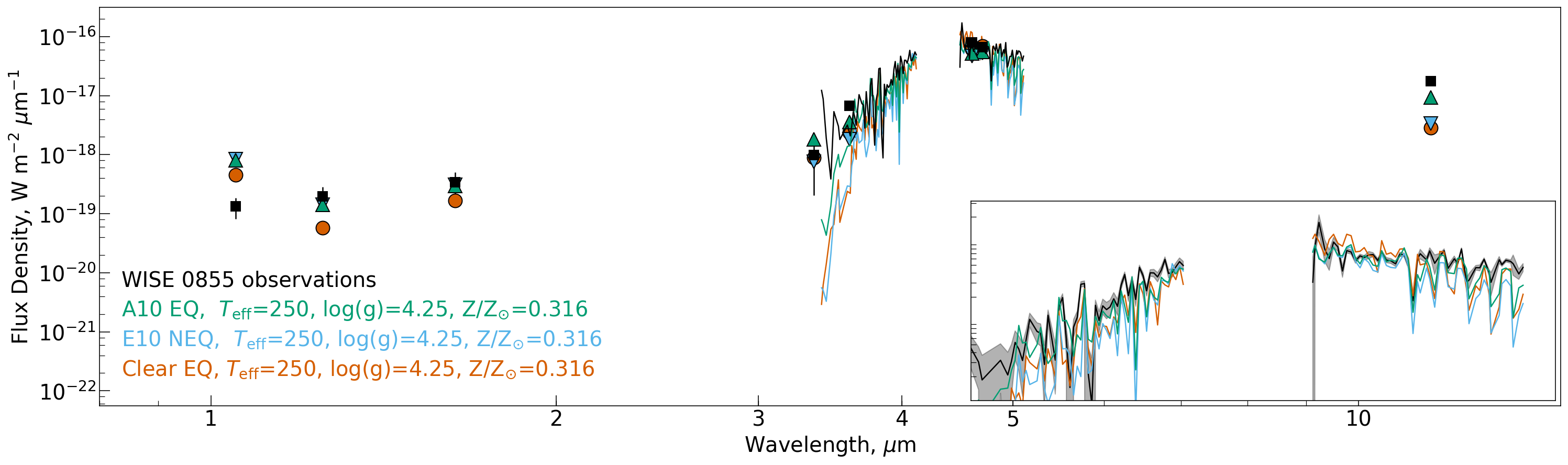}
    \caption{Similar to Figure \ref{fig:spectra1}, but now including cloudy models. The cloud type and chemistry type for each model are indicated by the color code in the upper left of each panel. For WISE 2056 and WISE 1541, green lines and downward triangles indicate a thicker cloud with disequilibrium chemistry, blue lines and upward triangles indicate a thinner cloud and disequilibrium chemistry, and orange lines with circles indicate cloud-free models with disequilibrium chemistry. For WISE 0855, with its cooler effective temperature, we actually find that the data prefer equilibrium chemistry models and a subsolar metallicity. In this panel, green lines and upward triangles correspond to an extremely thick water cloud with equilibrium chemistry, blue lines and downward triangles correspond to a thinner water cloud with nonequilibrium chemistry, and orange lines with circles correspond to clear equilibrium models. The insets in all panels zoom in on the longer wavelength spectral coverage. They have logarithmic y axis and linear x axis.}
    \label{fig:spectra2}
\end{figure}

Table \ref{tab:lit_inferred_properties} summarizes assessments of spectral type for each object, a distance estimate based on parallax from \cite{Kirkpatrick2019}, and the range of temperatures and surface gravities that have been reported in the literature. A selection of available photometric and spectral observations are summarized in Table \ref{tab:photometry}. Figure \ref{fig:spectra1} shows observations, compared to ``better-fitting" cloud-free models with equilibrium and nonequilibirum chemistry. Figure \ref{fig:spectra2} compares cloudy non-equlibrium models to observations of the cooler of the four objects: WISE-2056, WISE-1541, and WISE-0855. 

\begin{table}[]
    \centering
    \begin{tabular}{cccccc}
        Identifier & Spectral  & Dist & $T_{\mathrm{eff}}$ & log$_{10}$($g$)&Discovery \\
         & Type & (pc) & (K) & log$_{10}$(cm/s$^2$)& Ref. \\
         \hline
        UGPS 0722 & T9-Y0 & 4.12(0.04) & 500–550 & 4.0-5.0 & \citealt{2010MNRAS.408L..56L} \\
        WISE 2056 & Y0 &7.23(0.12)& 400-525&4.0-5.0&\citealt{Cushing2011}\\
        WISE 1541 & Y1 &5.98(0.08)&300-425&4.0-5.0& \citealt{Cushing2011}\\
        WISE 0855 &$>$Y2 & 2.27(0.02) &240-260&3.5-4.5& \citealt{Luhman2014}\\
    \end{tabular}
    \caption{Inferred properties reported in the literature for selected Y dwarfs. Distances and spectral types were found using the Y Dwarf Compendium$^{1}$. Effective temperature and surface gravity ranges are based on \cite{Zalesky2019}, \cite{Miles2020}, \cite{Leggett2021}, and \cite{Morley2018}. \footnotesize{$^{1}$https://sites.google.com/view/ydwarfcompendium/}}
    \label{tab:lit_inferred_properties}
\end{table}

\begin{table}[]
    \centering
     \begin{tabular}{c|cccc}
     &UGPS 0722$^{M,*}$& WISE 2056$^{M,*}$& WISE 1541$^{M,*}$ & WISE 0855$^{M}$  \\
    \hline
Y & 17.37(0.02) & 19.77(0.05) & 21.46(0.13)  & 26.54(0.12) \\ 
J & 16.52(0.02) & 19.43(0.04) & 21.12(0.06)  & 25.45(0.12) \\
H & 16.9(0.02) & 19.96(0.04) & 21.07(0.07)  & 23.83(0.2) \\ 
K & 17.07(0.08) & 20.01(0.06) & 21.7(0.2)  & --- \\ 
$[$3.6$]$ & 14.3(0.02) & 16.03(0.03) & 16.92(0.04)  & 17.47(0.1) \\ 
$[$4.5$]$ & 12.22(0.02) & 13.92(0.02) & 14.12(0.02)  & 13.92(0.02) \\ 
W1 & 15.25(0.05) & 16.48(0.08) & 16.74(0.17)  & 19.81(0.2) \\ 
W2 & 12.2(0.02) & 13.84(0.04) & 14.25(0.06)  & 13.7(0.04) \\ 
W3 & 10.21(0.07) & 11.73(0.25) & 12.2(0.3)  & 11.51(0.28) \\ 
    \end{tabular}
    \caption{Photometry for the main objects against which we compare models, taken from the compilation in \citealt{Leggett2021}. A $^{*}$ indicates that this object also has a NIR HST spectrum from Schneider et al 2015. A $^{M}$ indicates that an object also has an M band spectrum from Miles et al 2020. Other references for photometry include: \citealt{Leggett2013}; \citealt{Leggett2015}; \citealt{Leggett2017}; \citealt{Kirkpatrick2019}; \citealt{Cutri2013}; Schneider et al 2013; Luhman \& Esplin 2016.}
    \label{tab:photometry}
\end{table}

The [3.6]+W1 bands are largely sensitive to methane, but also include portions of the spectrum dominated by gaseous water and ammonia absorption and even a water cloud feature if particle sizes get small enough ($\sim$1 $\mu$m). The tendency of this portion of the spectrum to be too dim in models thus hints that water clouds, if present, cannot have uniformly small particle sizes, and that there might be less methane and ammonia in the upper atmosphere than our non-equilibrium quenching approximation is allowing. Models with higher surface gravity and lower metallicity tend to be a bit brighter in the [3.6]+W1 bands.

The W3 band spans methane, ammonia and water absorption. Due to this, non-equilibrium models with less ammonia and methane in the upper atmosphere can brighten the W3 band relative to clear models. Since our non-equilibrium models remain too dim in the W3 band, this again hints again that a larger degree of depletion of methane and ammonia is needed to match observations. Metallicity and surface gravity have only a very small effect in this band.

The K band is a bit more complicated since it is sensitive to a variety of opacity sources including methane, ammonia, collision-induced absorption, and, depending on disequilibrium levels, CO. Models with a lower surface gravity and higher metallicity tend to have a brighter K band, since they have photospheres at lower pressures and collision-induced opacity scales with density squared. Non-equilibrium chemistry with our quenching approximation changes the shape and central wavelength of the K band at 450 K, keeping the total flux roughly the same. At cooler temperatures of 350 K, non-equilibrium chemistry leads to a dimmer K band. 

Because K band and the [3.6]+W1 bands have opposite scalings with metallicity, surface gravity, and the presence of nonequilibrium chemistry, it is not straightforward to brighten all three regions of Y dwarf spectra simultaneously by adjusting the fundamental parameters of our models (effective temperature, surface gravity, metallicity) or adopting disequilibrium chemistry. For example, for UGPS 0722, nonequilibrium models with low surface gravity and supersolar metallicity can bring the K band photometry and W3 photometry into agreement, but then disagree with the [3.6]+W1 band photometry and the L band region of the spectrum, and tend to have too much CO absorption around 4.5-5 $\mu$m.

Looking at the model spectra compared to observations for the warmest three objects, one can see that the nonequilibrium models  capture the presence of CO, but there is too much of it. They also do a better job on the blue edge of the H band, where ammonia dominates. For UGPS 0722, the reduced CH$_4$ from 3-4 $\mu$m improves the fit, but does not reduce CH$_4$ enough, and for WISE 2056 and WISE 1541 there is little change between the nonequilibrium and equilibrium models for that region of the spectrum. Reduced NH$_3$ brings W3 photometry closer to observations in all three cases, but is still far from bright enough.

All of this raises the question-- can an adjustment to the standard quenching approximation we've applied improve agreement? The effective reaction rates used for chemical timescales in our quenching approximation (\citealt{Hubeny2007}) were shown to replicate results of a chemical kinetics code down to temperatures of 500 K (\citealt{Zahnle2014}). Notably, models for UGPS 0722, come much closer to being bright enough in the K, [3.6], and W3 bands than WISE 2056 and WISE 1541. This could indicate that any necessary adjustments to chemistry take effect at $T_{\mathrm{eff}}$ below 500 K, that we are not capturing something associated with water clouds correctly, some combination of the two, or it could be coincidental. Given this, and looking at dominant opacity sources in the K, [3.6], and W3 bands, initially it seems plausible that some adjustments to chemical or mixing timescales could improve fits.

A preliminary investigation indicated that simply nudging the quenching point to occur deeper in the atmosphere in order to reduce methane abundances (aiming to brighten the [3.6] band) is not an adequate solution. Physically this deeper quenching point would correspond to either slower chemical timescales or faster mixing timescales. We found that the CO absorption in the 4.5-5 $\mu$m regions and K band becomes too strong before the reduction in CH$_4$ attains the low abundance necessary to brighten [3.6] band flux to the levels seen in WISE 2056 and WISE 1541. Furthermore, the level of CH$_4$ reduction needed to brighten the [3.6] band sufficiently ends up changing the shapes of the Y, J, and H peaks in a manner inconsistent with observations. Reducing NH$_3$ and CH$_4$ can only brighten the K band to a point before CIA starts to dominate opacity in that portion of the spectrum. Our brief exploration thus indicated that a change to chemistry beyond just slower chemical timescales or faster mixing timescales would be needed to replicate observations.

Incorporating water clouds with large particle sizes tends to brighten all of the problem spectral regions, at least a bit. However, for most of the models we ran, while the cloudy K, [3.6]+W1, and W3 bands are brighter than cloud-free, they are not as bright as needed to match observations. In addition, in some cases the warming effect of clouds makes Y J and H photometry too bright (see e.g. the green cloudy model for WISE 1541 in Figure \ref{fig:spectra2}). 

\begin{figure}
    \centering
    \includegraphics[width=0.5\textwidth]{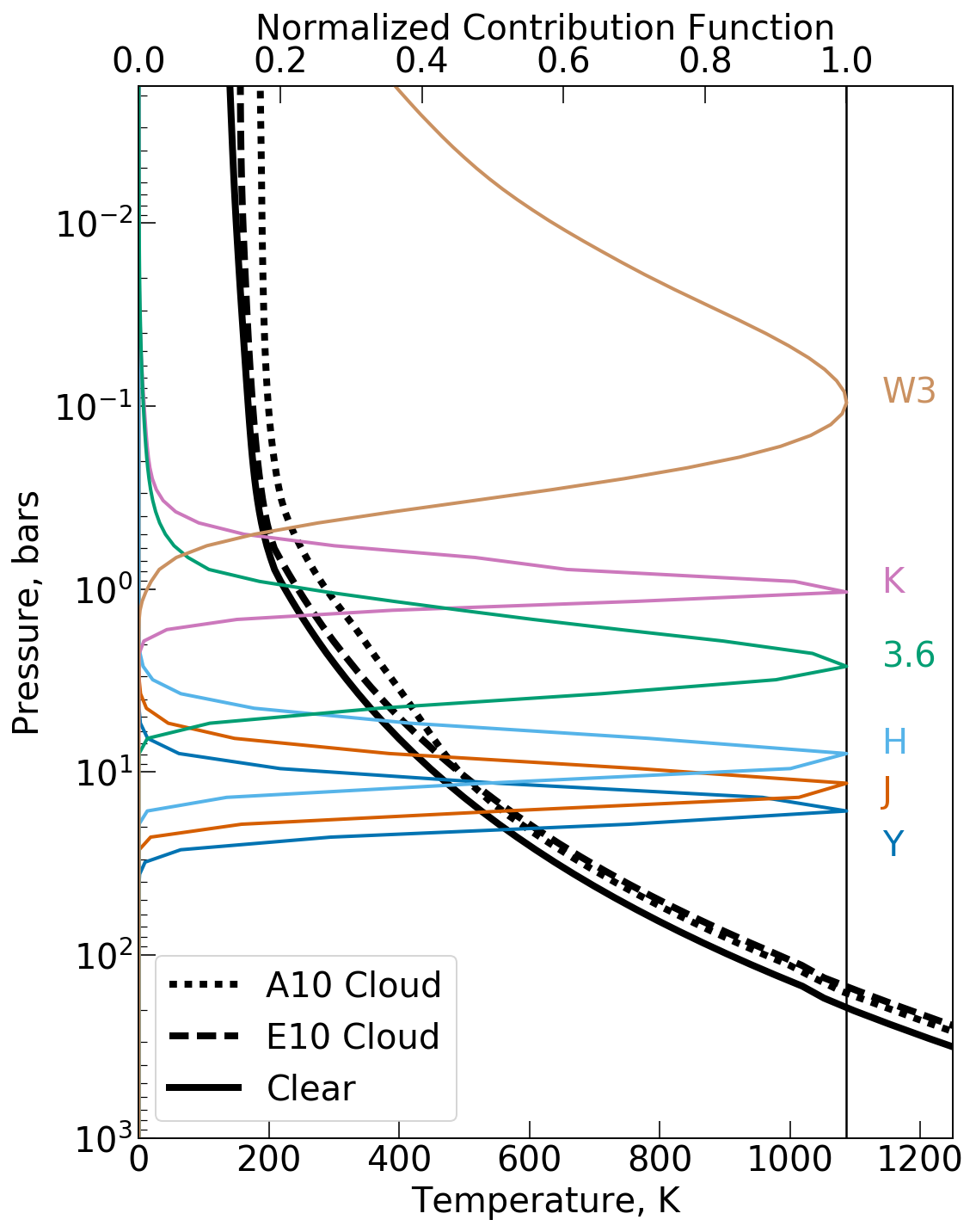}
    \caption{An illustration of which layers of atmosphere produce the majority of the flux in various photometric bandpasses, and why clouds alter their relative brightnesses. The solid black line denotes the P-T profile for a clear atmosphere, the dashed line one with a vertically thin cloud, and the dotted line one with an extremely vertically extended cloud. All models have equilibrium chemistry, $T_{\mathrm{eff}}$ = 250 K, log$_{10}$(g) = 4.25, and Z/Z$_{\odot}$=0.316. The top x-axis corresponds to the normalized contribution function to the clear atmosphere for a selection of photometric bandpasses labeled on the right. Where the contribution function is higher, those atmospheric layers contribute more photons to the emergent spectrum in that bandpass. These shift a bit for the different P-T profiles, but not drastically.}
    \label{fig:A10_TP_profile}
\end{figure}

We ran a wider variety of cloudy models for WISE 0855 at an effective temperature of 250 K, log$_{10}$ surface gravity of 4.25, and sub-solar metallicity to investigate whether a better fit could be found for this coolest object. These included modal particle sizes of 1 $\mu$m, 3 $\mu$m, and 10 $\mu$m, extents of $TCUP$ = 6, 2, 1, and 0, and $S_{cloud}$ of 0.01 and 0.005. We find that models with vertically extended water clouds best match the W3 band photometry and the M band slope (see green lines and upward triangles in the bottom panel of Figure \ref{fig:spectra2}). The significant additional opacity in the upper atmosphere induces a detached convective zone and warms layers where W3 band and K band flux emerges relative to the deeper atmospheric layers where Y, J, and H band flux emerges. This change in P-T shape relative to the atmosphere levels where different photometric bands emerge is illustrated in Figure \ref{fig:A10_TP_profile}. Note that this favorable model emerged only for a particular combination of cloud properties. Smaller modal particle sizes of 3 and 1 $\mu$m with such a large vertical extent would lead to water cloud absorption features that are not seen in the data. A larger supersaturation parameter with the same 10-$\mu$m modal particle size leads to too much dimming in the 4-5 $\mu$m range.

This is an intriguing result, but not entirely conclusive. Such large particle sizes should not persist so high in the atmosphere (Mang, private correspondence). Further work is warranted, to see whether a self-consistent atmospheric structure and physically realistic cloud can be made to produce a similar result.

This finding for WISE 0855 further raises the question of whether it is possible for water clouds in temperature ranges closer to WISE 2056 and WISE 1541 (350-450 K) to warm the upper atmosphere enough to brighten K, [3.6], and W3 bands, without over-brightening Y, J, and H bands and over-dimming the [4.5] band. Unfortunately, models in this range prove quite challenging to converge as the atmospheric profile and the Clausius-Clapeyron line start to run parallel to each other, and non-equilibrium effects must be accounted for along with the clouds. As such, we did not complete a neat grid of cloud's particle size, supersaturation parameter, and vertical extent of the thickest cloud type "A". However, a preliminary investigation along these lines indicated that, even when clouds were opaque enough to induce a detached convective region, it did not suffice to brighten K band, [3.6], and W3 band photometry to the degree necessary to replicate observations.

\section{Summary and Conclusions}\label{sec:conclusions}

We have presented a new suite of 1d radiative convective equilibrium models for Y dwarf atmospheres spanning temperatures of $\sim$200-600 K, surface gravities of log$_{10}$(g/g$_0$) = 3.5-5.0 where g$_0$ = 1 cm/s$^2$, equilibrium and non-equilibrium chemistry, and metallicities of 0.316, 1.0 and 3.16$\times$Z$_{\odot}$, or -[0.5], [0.0], and [0.5] dex. For effective temperatures below 400 K, we also compute models which simultaneously account for both water clouds and nonequilibrium chemistry. 

We began by illustrating the model dependence on fundamental parameters of 1d radiative convective equilibrium models: effective temperature, surface gravity, and metallicity. Then, we demonstrated the effects of incorporating water clouds, nonequilibrium chemistry arising from vertical mixing, and both together. After that, we compared our models to observations of a  sample of Y dwarfs. Comparisons to yielded consistent results with previous studies (\citealt{Phillips2020}): disequilibrium CO-CH$_4$ chemistry and N$_2$-NH$_3$ chemistry seem to replicate qualitative trends of features that are seen in the observations better than equilibrium models, but do not quantitatively agree. For the coolest objects, water clouds are favored. It was hypothesized that including the effects of both water clouds and disequilibrium chemistry in self-consistent models might produce a better match to observations than previous works. However this was not borne out. 

In general all models which fit Y, J, H, and [4.5]+W2 photometry are much too dim in K band, [3.6]+W1 bands, and the W3 band. Most of the parameters we can vary in our models do not simultaneously brighten K band, [3.6]+W1 bands, and W3 bands. Lower surface gravities brighten K but dim [3.6]+W1 and leave W3 roughly the same. Lower metallicities brighten [3.6]+W1, but dramatically dim K band and again leave W3 roughly the same. Nonequilibrium chemistry brightens [3.6]+W1 and W3 a bit relative to equilibrium chemistry models, but has little effect on K band magnitude, or even dims is depending on the temperature-gravity combination. 
 
 Adding clouds does actually brighten all three problem regions of the spectrum (K band, [3.6]+W1, and W3), but, in general, the clouds' warming effect also severely over-brightens Y, J, and H flux. There is one exception: for WISE 0855 we ran a mini-grid of additional cloudy variations, and we found that incorporating water clouds which extended all the way to the top of the atmosphere warmed the upper atmosphere sufficiently to brighten K, [3.6]+W1, and W3, and induced a detached convective zone which allows the lower atmosphere to remain relatively cool, keeping Y, J, and H flux closer to their cloud-free values. This structural change, with warmer upper atmosphere and cooler deep atmosphere is the key change that \citealt{Leggett2021} induced to achieve agreement with a number of objects. 
 
 It could be the case that a broader exploration of cloud properties, chemical mixing timescales and vertical mixing time scales might be able to identify a combination which reproduces this P-T structure for objects in the 250 - 400 K range. A key pattern that emerges in the novel cloudy disequilibrium grid is that the presence of water clouds enhances the spectral signatures of vertical mixing by shifting atmosphere profiles to warmer temperatures for a given effective temperature and surface gravity combination. Quench points thus tend to correspond to higher CO/N2 abundances and lower CH$_4$/NH$_3$ abundances than clear disequilibrium counterparts. This drives home the importance of accounting for both effects together. 
 
 Our choices for water cloud treatment and disequilibrium chemistry align with previous published work and are reasonable, but they are by no means authoritative or exhaustive in terms of what is physically plausible. When incorporating water clouds, we assume a single particle size distribution for the whole atmosphere, but it is expected that larger particles should form near the base of the clouds,  while smaller particles will extend further up to lower pressures. Our disequilibrium chemistry formulation could be updated to the exact reaction rates recommended by \citealt{Zahnle2014}, or perhaps a similar study extending to cooler temperatures is warranted for this regime. We consider only models with a mixing length equal to one pressure scale height, as is typical. However, varying this has been shown to impact the emergence of detached radiative zones and the relative abundances of disequilibrium species with the quenching approximation (\citealt{Mukherjee2022}). We also have not considered the effects of variation in C:O ratio. There is no reason to think that Y dwarfs will be systematically skewed toward high or low C:O ratio relative to solar, but some natural spread of values should occur.  
 
 Of course, it is entirely possible that tweaks to quenching chemistry and cloud properties are not the whole story. As has been pointed out in previous studies, Y dwarf atmospheres and their emergent spectra could be affected by a number of processes not captured in our models: compositional gradients, breaking gravity waves, deviations from adiabaticity due to 2d/3d flows, or inhomogeneous surface features (\citealt{Tremblin2016}; \citealt{Leggett2021}; \citealt{Morley2014}; \citealt{Morley2018}). One clue which points towards such additional processes is that UGPS 0722 is almost certainly too warm for water clouds to have a large impact on its atmospheric structure and emergent spectrum. It is beyond the scope of this paper to explore all such possible variations.

JWST has already collected the first of many new Y dwarf observations that are planned. These higher resolution, higher-precision panchromatic data should help us to refine models and form a better understanding of the physical conditions in the coolest readily observable substellar extrasolar objects.

\acknowledgments
The authors would like to acknowledge support for this research under NASA 
WFIRST-SIT award \# NNG16PJ24C and NASA Grant NNX15AE19G, and from the Heising-Simons Foundation via the 51 Pegasi b Fellowship. Work presented here made use of the SVO Filter Profile Service (http://svo2.cab.inta-csic.es/theory/fps/) supported from the Spanish MINECO through grant AYA2017-84089. The authors would also like to thank Dr. Nicole Allard for providing data and an interpolation code used to incorporate Na I and K I absorption cross sections into our models. This research was expedited and improved thanks to fruitful conversations with Dr. Caroline Morley and her research group at UT Austin, in particular Ph. D. student James Mang who consulted on water-cloud properties in microphysical models. 

\vspace{5mm}

\appendix
\section{Updated Opacities}\label{sec:opacity_appendix}

In this appendix, we include a more detailed accounting of the opacities used to compute all of our models. In general we follow the opacity calculation procedures and the chemistry calculations described in \cite{Sharp2007}, but introduce data updated in the intervening 15 years where available. 

\subsection{Molecular Line Absorption}

The approach used for computing molecular absorption cross sections draws heavily from the recommendations of two recent publications by the field's leading experts: ExomolOP \citep{Chubb2021} and EXOPLINES \citep{exoplines}. ExomolOP was created by members of the ExoMol \citep{ExoMol2020} team and includes absorption cross sections for 80 of the database's molecules. Their main purpose was to provide cross sections for a broad array of species formatted in a manner that can be input to four publicly available retrieval codes. EXOPLINES focused specifically on the species relevant for hot Jupiter atmospheres: metal hydrides, SiO, TiO, VO, and water. Their main innovation was working out a more detailed pressure-broadening scheme for metal hydrides, SiO, TiO and VO, although it is still an approximation.

\begin{table}[]
\small
    \centering
    \begin{tabular}{ll}
        \texttt{exocross} key word & setting used in these calculations\\
        \hline
        Wavenumber grid &  200 - 33333 cm$^{-1}$ with R 20000\\
        Profile &  Voigt\\
        Broadening & H$_2$ and He \\
        Integration & default scheme for Voigt: \\
        Line wing cut-off & for pressures $<$ 200 bars: 30 cm$^{-1}$\\
        ~~~(key word ``Offset") & for pressures $>$ 200 bars:  150 cm$^{-1}$\\
        Line strength threshhold &  10$^{-50}$\\
        ~~~(key word ``Threshold") & \\
        Mass & took value from .def file provided by ExoMol \\
        \hline
    \end{tabular}
    \caption{\texttt{exocross} Parameter Settings}
    \label{chpt4_tab:exocross_parameters}
\end{table}

The ExoMol team has provided a publicly available Fortran code called \texttt{exocross}\footnote{https://exocross.readthedocs.io/en/latest/} which can compute a variety of quantities from molecular line lists, including generating spectra (absorption and emission), absorption cross sections, and thermodynamic properties \citep{exocross}. We used \texttt{exocross} to compute absorption cross sections over a grid of temperatures, pressures, and wavenumbers suitable for modeling substellar atmospheres. This grid covers a temperature range from 50-5000 K with 80 points spaced evenly in log, a pressure range from 3000 bars to 10$^{-6}$ bars with 30 points spaced evenly in log, and a wavelength range from 0.3-50 $\mu$m, which corresponds to wavenumbers from 200 - 33,333 cm$^{-1}$. The wavelength grid corresponds to a spectral resolution of R=$\frac{\lambda}{\Delta\lambda}$=20,000. We chose this resolution since it is a bit higher than the resolving power of R=$\frac{\lambda}{\Delta\lambda}$=15,000 deemed necessary for retrieval forward models used to analyze JWST spectra, \citep{Rocchetto2016}, although still insufficient for high-resolution Doppler shift studies which require R$\sim$100,000 \citep{Brogi2012}. 

User-defined aspects of \texttt{exocross} absorption cross-section calculations include: line list, line profile with pressure-dependent broadening if desired, line-wing cutoff, and an optional line strength threshold (to exclude weak lines and speed up the calculation). The code then handles summing up the contributions from every line into the specified wavelength bins. For molecules like PH$_3$, H$_2$O and CH$_4$ these calculations are very computationally intensive due to the large number of lines (e.g. taking 10's of hours for a single P-T point parallelized on 28 threads). For simpler diatomics, like the metal hydrides and metal oxides, these calculations are much quicker (e.g. taking minutes for a single P-T point parallelized on 28 threads). We will now walk through the details of the choices made for line lists, line shape and pressure broadening, line wing cut-off, and line strength threshold for each molecule, which are also summarized in Table \ref{chpt4_tab:exocross_parameters}.

\begin{table}[]
\small
    \centering
    \begin{tabular}{cccccccl}
        Molecule & \# Iso.&Wavelengths & \# lines & T$_{max}$&Broadening&Line List & Line List Ref.\\
         & & & & & & Name & \\
        \hline
         SiO & 5&1.65-100 & 250,000 &9000 &(3)&EBJT&\citealt{Barton2013}\\
         TiO & 5&0.33-100 &30 million & 5000&(3)&TOTO&\citealt{McKemmish2019}\\
         VO &1 &0.29-100 & 27 million & 5000 &(3)&VOMYT&\citealt{McKemmish2016}\\
         CO &6 &0.43-100 &145,000 &5000 &(1)&Li2015&\citealt{Li2015}\\
         *CO2& 7&0.5-100 & 8 billion& 5000 &(2)&UCL-4000&\citealt{Yurchenko2020}\\
         H2O& 4&0.24-100 &6 billion & 5000&(1)&POKAZATEL&\citealt{Polyanksy2018}\\
          & & & & & &  HotWat78&\citealt{Polyansky2017}\\
         *H$_2$S&1 &0.91-100 & 115 million& 2000&(2)&AYT2&\citealt{Azzam2016}; \citealt{Chubb2018}\\ 
         *HCN&2 &0.56-100 &34.4 million & 4000&(1)&Harris &\citealt{Harris2006}; \citealt{Barber2014}\\
         &&&&&&Larner&\citealt{Harris2008}\\
         CaH& 1&0.45-100 &6000 &5000 &(3)&MoLList&\citealt{Bernath2020}\\
         CrH& 1&0.69-100 &13,800 &5000 &(3)&MoLList&\citealt{Bernath2020}\\
         FeH& 1&0.67-100 &93,000 &5000 &(3)&MoLList&\citealt{Bernath2020}\\
         *MgH&3 &0.34-100 &14,200 &2000 &(3)&MoLList&\citealt{Bernath2020}\\
         TiH& 1&0.42-100 &200,000 &5000 &(3)&MoLList&\citealt{Bernath2020}\\
        *CH4&1 &0.56-100 &34 billion &2000 &(1)&YT34-to-10&\citealt{Yurchenko2017}\\
        *NH3& 2&0.5-100 &16.9 billion &1500 &(1)&CoYuTe&\citealt{Coles2019}; \citealt{AlDerzi2015}\\
        &&&&&&BYTe&\citealt{Yurchenko2015}\\
        *PH3& 1&1.00-100 & 16.8 billion &1500 &(1)&SAITY&\citealt{Sousa-Silva2014}\\
        H$_2$&2 &0.28 100 & 4700 &5000 &(2)&RACPPK&\citealt{Roueff2019}\\
        \hline
    \end{tabular}
    \caption{Summary of molecule line list properties. *draws attention to species for which T$_{max}$ $<$ 5000 K. For CO$_2$, the main isotopologue is valid up to 5000 K, but the less abundant isotopologues are valid only at room temperature. For the broadening column, (1) indicates that H and He broadening coefficients are available in the ExoMol database, (2) indicates that we chose to follow the same approach as \cite{Chubb2021}, and (3) indicates we followed \cite{exoplines}.}
    \label{chpt4_tab:molecular_linelists}
\end{table}

\begin{table}
\small
\label{chpt4_tab:isotopologues}
\begin{center}
\begin{tabular}{lc}
    Molecule & Natural Abundance (\%)\\
    \hline
     $^{1}$H$_2-^{16}$O &  99.74 \\
    $^{1}$H$_2-^{18}$O &  0.2 \\
     $^{1}$H$_2-^{17}$O  &  0.03 \\
    $^{1}$H$-^{2}$H-$^{16}$O &  0.01 \\
    \hline
     $^{1}$H-$^{12}$C-$^{14}$N & 98.56 \\   
      $^{1}$H-$^{13}$C-$^{14}$N  & \\  
    \hline
     $^{1}$H$_2$  & 99.99 \\      
      $^{1}$H-$^{2}$H &0.01 \\    
    \hline
     $^{12}$C-$^{16}$O$_2$   & 98.45 \\
    $^{13}$C-$^{16}$O$_2$     & 1.1\\
    \hline
     $^{12}$C-$^{16}$O & 98.7 \\
     $^{13}$C-$^{16}$O &  1.1 \\
     $^{12}$C-$^{18}$O &   2.0 $\times$ 10$^{-3}$\\
     $^{12}$C-$^{17}$O & 3.7 $\times$ 10$^{-4}$\\
     $^{13}$C-$^{18}$O &  2.2 $\times$ 10$^{-5}$\\
     $^{13}$C-$^{17}$O &  4.1 $\times$ 10$^{-6}$\\
    \hline
     $^{14}$N-$^{1}$H$_3$ &   99.6\\
    $^{15}$N-$^{1}$H$_3$ &  0.4\\
    \hline
     $^{28}$Si-$^{16}$O  &  92.00\\
     $^{29}$Si-$^{16}$O & 4.67\\
     $^{30}$Si-$^{16}$O &  3.08\\
     $^{28}$Si-$^{18}$O &  0.19\\
     $^{28}$Si-$^{17}$O &  0.03\\
    \hline
     $^{48}$Ti-$^{16}$O &  73.7 \\
     $^{46}$Ti-$^{16}$O &  8.3 \\
     $^{47}$Ti-$^{16}$O &  7.4 \\
     $^{49}$Ti-$^{16}$O &  5.4 \\
     $^{50}$Ti-$^{16}$O &  5.2 \\
    \hline     
     $^{24}$Mg-$^{1}$H &  79.0\\
    $^{25}$Mg-$^{1}$H &  10.0\\
     $^{26}$Mg-$^{1}$H & 11.0 \\ 
    \hline
\end{tabular}
\caption{Natural abundances for species where multiple isotopologues were available in ExoMol. Data are taken from NIST.}
\end{center}
\end{table}

 For each molecule, we use the line list recommended by the ExoMol team as indicated on their online database\footnote{http://www.ExoMol.com/} at the time of the update ($\sim$winter 2020-2021). These are summarized in Table \ref{chpt4_tab:molecular_linelists}. This includes some line lists created by the ExoMol team and some line lists converted to the ExoMol format from other sources by the ExoMol team. For more details on the ExoMol line list format, see \citealt{Tennyson2013}, \citeyear{Tennyson2016}, \citeyear{Tennyson2020}. A few asides are warranted on these line lists. It should be noted that not all of the molecules of interest have data spanning this full wavelength range, and a few molecules have line lists which are only intended for use up to temperatures around 2000 K (again see \ref{chpt4_tab:molecular_linelists}). 
 
 The maximum temperature of completeness for all the MoLLIST (Bernath 2020) molecules (CaH, CrH, FeH, TiH) is listed as 5000 K since in some portions of the spectrum this is true. However, in other portions of the spectrum they are not complete to 5000 K or even lower temperatures. Nevertheless, they were the best data available for these molecules at the time we began this study. New line lists for CaH and MgH have recently become available and will be replaced in our database at a future time. The previous line list used for CH$_4$ was only complete up to 1500 K  (\citealt{Yurchenko2014}, the new one we adopt is complete up to around 2000 K. The YT34-to-10 CH$_4$ wavelength coverage does not reach blueward of 0.56 $\mu$m. We supplement this with empirical CH$_4$ absorption cross sections inferred from Jupiter's spectrum by Karkoschka 1994. SiO, CrH, and FeH, are all expected to have non-negligible opacity shortward of 0.67 $\mu$m. This means that for hotter objects, we must be cautious in interpreting this wavelength region. The ExoMol project reports that it is planning further work on this problem which will need to consider bound-free (photodissociation) in addition to bound-bound processes, and that it has already begun on SiO. 
 
Recently, \cite{Tannock2022} conducted a careful comparison of a high resolution brown dwarf spectra to models computed with various line lists for H$_2$O, CO, CH$_4$, NH$_3$, and H$_2$S, providing some insight into how well these line lists are performing. Their results support our choice of line lists for H$_2$O, CO, NH$_3$, and H$_2$S, but did not include comparisons to models which use our choice for CH$_4$. They recommend HITEMP's Hargreaves2020 methane line list \citealt{Hargreaves2020}, which we anticipate replacing our methane with in the future.

For most pressures in our grid, we set \texttt{exocross} to use Voigt profiles with line-widths calculated according to a sum of the Lorentzian HWHM for H$_2$ and He collisions following:
\begin{equation}
    \Delta\bar{\nu} = 2 \times \sum_b \Big( \frac{T}{296 K} \Big) ^{-n_{T,b}} \gamma_{L,b} p_b \quad ,
\end{equation}
where $\gamma_{L,b}$ is the Lorentz coefficient of an absorber perturbed by collisions with species $b$ (HWHM, unit: cm$^{-1}$/bar), $p_b$ is the broadener's partial pressure (unit: bar), and $n_{T,b}$ is dimensionless. We assume a background atmosphere composed by number of 85\% H2 and 15\% He, as would be typical for Jupiter. 

$\gamma_{L,b}$ and $n_{T,b}$ are controlled by broadener-absorber interactions. Ideally, these are determined through experimental data or detailed ab initio calculations. ExoMol includes H$_2$ and He pressure broadening coefficients for only a small subset of the molecules: H$_2$O,  CH$_4$, CO, HCN, NH$_3$, PH$_3$, and only 13 species in total. These are formatted as a list of $\gamma_{L,b}$ and $n_{T,b}$ in terms of the J values of the upper and lower states involved in a transition (\citealt{Barton2017}). References are: H$_2$O (\citealt{Voronin2010}; \citealt{Voronin2012}; \citealt{Barton2017}; \citealt{Solodov2008}; \citealt{Solodov2009}; \citealt{Petrova2013}; \citealt{Petrova2012}; \citealt{Petrova2016}), HCN (\citealt{Mehrotra1985}; \citealt{Cohen1973}; \citealt{Charron1980}), CH$_4$ (\citealt{Varanasi1990}; \citealt{Pine1992}; \citealt{Fox1998}; \citealt{Varansi1972}; \citealt{Varanasi1989}; \citealt{Grigoriev2001}; \citealt{Gabard2004}; \citealt{Manne2017}; \citealt{Vispoel2019}; \citealt{Lyulin2014}; \citealt{Laurent2014}), CO (\citealt{Faure2013}; \citealt{Gordon2017}), NH$_3$ (\citealt{Wilzewski2016}), and PH$_3$ (\citealt{Kleiner2003}; \citealt{Levy1993}; \citealt{Sergent-Rozey1988}; \citealt{Salem2004}; \citealt{Salem2005}). 

\cite{exoplines} estimate J-dependent broadening for additional species important to hot Jupiter atmospheres: TiO, VO, SiO, FeH, CaH, CrH, TiH, and MgH. Rather than ab initio calculations or laboratory studies, \cite{exoplines} approximate the necessary quantities through a combination of classical collision theory and extrapolation from HCl and CO behavior, since these are the most similar molecules with available data. For the remaining three molecules (H$_2$, H$_2$S, CO$_2$), we adopt the same choices as \cite{Chubb2021}. They use the average $\gamma_{L,b}$ and $n_{T,b}$ of an available molecule deemed most similar in structure to the molecule of interest, rationalizing along the same lines as \cite{exoplines}. This approach incorporates a temperature and pressure dependence, but neglects any J-dependence in light of the large uncertainties. When deciding which species is most similar from a broadening point of view they consider the dipole moment and quadrupole moment, they look at the general structure (e.g. linear, non-linear, diatomic), and they consider other molecular properties such as the centre of symmetry and interatomic distances. This procedure does not always result in an unambiguous choice. For example, the dipole moment of H$_2$S is closest to HCl. However, the intermolecular distances and potential energy surface of H$_2$S are more similar to OCS than HCl (Johnson-III 2019), so it is not clear which broadening parameters should be used. 
 
 For pressures less than 10$^{-6}$ bars, where collisions are rare, pressure broadening becomes negligible. For the lowest pressure grid point, we thus employ a Doppler profile with FWHM: 
 \begin{equation}
     \Delta\bar{\nu} = 3.581 \times 10^{-7}\sqrt{\frac{T}{M}}\bar{\nu} \quad ,
 \end{equation}
where $M$ is an absorber's molecular mass in amu (atomic mass unit), $T$ is the temperature (in K), and $\bar{\nu}$ is the line position in wavenumber (cm$^{-1}$). 
Doppler profiles are formed by the cumulative Doppler shifts of many particles with a range of velocities following a Maxwellian distribution, which depends only on temperature not on pressure. Later, when these molecular absorption cross sections are incorporated into a model atmosphere extending to pressures lower then 10$^{-6}$ bars (opacity grids used in \textit{coolTLUSTY} extend to 10$^{-9}$ bars in places), we simply extrapolate using this final grid point.

The line-wing cutoff, $d$, is an ad-hoc fix to the reality that we don't have a good understanding of line shapes far from the line center. It is common practice to fix it at 100 cm$^{-1}$ away from line center, or to apply a multiple of the Voigt width, as we do for atomic species. In these calculations line wings were cut off at 150 cm$^{-1}$ for pressures over 200 bars and at 30 cm$^{-1}$ for pressures lower than 200 bars. \cite{exoplines} made this choice in light of several laboratory and theoretical studies which show that these result in a more accurate opacity continuum for water and methane broadened by H$_2$ collisions (\citealt{Ngo2012}; \citealt{Hartmann2002}). We adopt the \cite{exoplines} recommendations again for the line-strength threshold, leaving out lines weaker than 10$^{-50}$ for all molecules. The inclusion of weaker lines does not impact results at our sampling, and their exclusion speeds up calculations.  

For some species, it is important to account for different isotopologues of the same species. Where ExoMol included data for multiple isotopologues, we computed separate absorption cross sections for each and then combined them weighted by their natural abundances (taken from NIST, listed in Table \ref{chpt4_tab:isotopologues}). Where data are not available for less abundant isotopologues, but it is expected that there will be a significant effect, line lists for the main isotopologue can be extended to the other isotopologues using the method outlined in \cite{Polyansky2017} or \cite{Sharp2007}. 

\subsubsection{A Note on HCN and CO$_2$}
HCN and CO$_2$ are not saved in the heritage chemical abundance tables described in \cite{Sharp2007}. Like PH$_3$, they are not usually dominant opacity sources at thermochemical equilibrium abundances, but they are expected to help characterize disequilibrium chemistry, based on what we have seen in the Solar System. We do not include them at all in this study, but they are present in the tables in this appendix so that they can be referenced in future work.

\subsection{Atomic Line Absorption}

We include absorption by Na, K, Li, Rb, Cs, and Fe. The Na I resonant doublet around 0.59 $\mu$m and K I resonant doublet around 0.77 $\mu$m are the dominant opacity sources across the visual and into the NIR for temperatures below $\sim$1200 K (\citealt{Burrows2000}), where metal hydrides, TiO and VO have rained out. In the presence of H$_2$ and He collisions, these strong lines are intensely broadened and exhibit non-Lorentzian behavior in their wings, warranting more careful attention in defining their line shapes than a simple Lorentzian. Line profiles based on detailed quantum chemistry calculations of the H$_2$-K and H$_2$-Na collisional systems are available covering temperatures of around 600 - 3000 K and for H$_2$ number densities up to 10$^{21}$ cm$^{-3}$ (\citealt{Allard2016}\footnote{https://cdsarc.unistra.fr/viz-bin/cat/J/A+A/589/A21}; \citealt{Allard2019}\footnote{https://cdsarc.unistra.fr/viz-bin/cat/J/A+A/628/A120}, see also \citealt{Burrows2003}). We incorporate opacity for these lines by interpolation within that range. For higher pressures and cooler temperatures than the available range, these lines are less important because Na and K are less abundant. More recently \cite{Peach2020} has presented work with He broadening, but this has not yet been incorporated into our opacities.

Aside from the Na and K resonant lines and Fe in the UV, molecular opacities dominate over these atomic opacities across most of the spectral range, so their inclusion does not have a large effect on the energy budget. They can, however, be useful diagnostics if one can obtain sufficiently high resolution data.

For the rest of the atomic lines,  we used line lists from the National Institute of Standards and Technology (NIST)\footnote{https://www.nist.gov/pml/atomic-spectra-database} and used equations \ref{chpt4_eq:LineStrength} and \ref{chpt4_eq:PartitionFunction} to compute integrated line strengths at all the temperatures of interest. 

In local thermodynamic equilibrium the integrated strength, $S$, of a single spectral line in units cm$^2$ s$^{-1}$ species$^{-1}$ at a given temperature is:
\begin{equation}\label{chpt4_eq:LineStrength}
    S = \frac{\pi e^2 g_i f_{ij}}{m_e c} \frac{e^{-hcF_i/kT}}{Q(T)} (1-e^{-hc(F_j - F_i)/kT})
\end{equation}
where $g_i$ is the statistical weight of the ith energy level (2J +1), $f_{ij}$ is the oscillator strength for a transition from level i to a higher level j, $F_i$ and $F_j$ are the excitation energies of levels i and j in cm$^{-1}$, $Q(T)$ is the value of the partition function at temperature $T$, and the other symbols are the usual physical constants. Note that the first term is the line strength in cm$^2$ s$^{-1}$ absorber$^{-1}$, the next term is the Boltzmann factor which scales from per absorber to per species, and the final term is a correction for stimulated emission \citep{Sharp2007}. $Q$ can be calculated from:
\begin{equation}\label{chpt4_eq:PartitionFunction}
    Q(T) = \sum_{i=1}^{n}g_ie^{-hcF_i/kT} \quad .
\end{equation}
In the temperature and pressure ranges of interest for substellar atmospheres, it is generally sufficient to truncate the summation at term values of $F_i$ $\sim$ 20,000 cm$^{-1}$ \citep{Sharp2007}.

The final ingredient to translate from line strength to an absorption cross section is the line shape. Rather than a delta function of absorption precisely at the central wavelength of a transition, each spectral line has a natural width due to the finite lifetimes of the upper and lower energy levels of the absorber, and the Heisenberg uncertainty principal. The natural width is negligible compared to Doppler broadening (that is broadening from thermal motions of a group of atoms/molecules) and collisional or pressure broadening (from perturbations to the radiating species' potential by passing H$_2$ and He). The convolution of Doppler and collisional broadening gives a Voigt profile if one makes the assumption of ideal gas behavior \citep{Humlicek1979}. In the calculations done for molecular species we use a Voigt profile. In calculations done for atoms, we mainly use a Lorentz profile since collisional broadening tends to dominate over thermal broadening at the temperatures and pressures with which we are most concerned, and the uncertainties in collisional broadening properties outweigh the improved precision from a Voigt profile. The monochromatic absorption cross section in cm$^2$ species$^{-1}$ for a Lorentz profile can be expressed as: 
\begin{equation}\label{chpt4_eq:AtomProfile}
\sigma(\bar{\nu}) = \Big(\frac{Sb}{c}\Big) \frac{\Delta\bar{\nu}/2\pi}{(\Delta\bar{\nu}/2)^2 + (\bar{\nu} - \bar{\nu_0})^2} \quad ,
\end{equation}
where $b$ is a normalization factor to account for the fact that you must truncate the line wings, $c$ is the speed of light, $\bar{\nu}{_0}$ is the line center in cm$^{-1}$, $S$ is the integrated line strength given by expression \ref{chpt4_eq:LineStrength}, and $\Delta\bar{\nu}$ is the Lorentzian's full width half max in cm$^{-1}$. Note that in moving to monochromatic absorption cross sections in terms of wavenumber, $\bar{\nu}$ rather than frequency $\nu$, we have divided $S$ by $c$. This follows from the relation $\nu=c\bar{\nu}$; therefore, $d\nu=cd\bar{\nu}$. This requires that $\sigma(\bar{\nu})$ satisfy $S=\int\sigma(\nu)d\nu=\int\sigma(\bar{\nu})cd\bar{\nu}$.

In calculating the total monochromatic absorption cross section of a molecular or atomic species, one must sum up the contributions from all the different lines at each wavenumber point. 

To compute $b$, one can use the following if the line is well-sampled by the wavenumber grid:
\begin{equation}
    b = \Big[\Big(\frac{2}{\pi}\Big)arctan\Big(\frac{2d}{\Delta\bar{\nu}}\Big)\Big]^{-1} \quad ,
\end{equation}
where $d$ is the distance away from line center $\bar{\nu_0}$ that you have truncated. This gives a quantity $b$ $>$ 1, provided $2d > \Delta\bar{\nu}$. On the other hand, if the wavenumber grid is coarse compared to the line width, \cite{Sharp2007} recommend shifting the line center to the nearest grid point and using:
\begin{equation}
    b = \big(\frac{\pi}{4}\big) \big(\frac{\Delta\bar{\nu}}{w}\big) \big[\frac{4+(\Delta\bar{\nu}/w)^2}{2+(\Delta\bar{\nu}/w)^2}\big] \quad ,
\end{equation}
where $w$ is the grid interval in cm$^{-1}$, and the resulting value of $b$ will be less than 1, provided $\Delta\bar{\nu}<w^2/2$. 

In addition to broadening, collisions also shift the central wavelength of a line \citep{Bernath1996}. For our purposes of modeling atmospheric structure and moderate resolution spectra, this effect can be safely neglected, but cross-correlation studies using higher resolutions might need to consider it (\citealt{Gandhi2020}).

The Lorentzian FWHM $\Delta\bar{\nu}$ of each line depends on the temperature and pressure of the surrounding gas, the abundances and properties of perturbing species, and the quantum properties of the absorber. An experimentally measured or carefully calculated line width is not readily available for every line perturbed by H$_2$ and He. We adopt two different approaches to estimate the atomic line widths. For Na, K and Li lines, excluding the Na I doublet and K I doublet discussed above, we used the line widths inferred from \cite{Allard2007b}. For Fe, Rb, and Cs, we estimate the line width based on Van der Waals interactions using the methods outlined in \cite{Schweitzer1996}. The Van der Waals damping constant can be calculated from:
\begin{equation}\label{chpt4_eq:vanderwaals}
    \gamma_{vw}^{p} = 17 C_6^{2/5} \bar{v}^{3/5} N_p \quad ,
\end{equation}
where $C_6$ is the Van der Waals interaction coefficient, N$_p$ is the number density of perturbers, and $\bar{v}$ is the average relative velocity of perturber and absorber, which, assuming an ideal gas, is given by:
\begin{equation}
    \bar{v} = \Big(\frac{8k_B T}{\pi \mu}\Big)^{1/2} \quad ,
\end{equation}
where $\mu$ = $m_pm/(m_p+m)$ is the reduced mass of the perturber-absorber system. We approximate values for $C_6$ by scaling the $C_6$ value for interactions with hydrogen by the polarizability of each perturber:
\begin{equation}\label{chpt4_eq:c6}
    C_6 = \frac{\alpha_p}{\alpha_H} 1.01 \times 10^{-32} (Z+1)^2 \Big[ \frac{E_H^2}{(E-E_i)^2} - \frac{E_H^2}{(E-E_j)^2} \Big] ~ cm^6 s^{-1} \quad ,
\end{equation}
where $\alpha_p$ is the polarizability of the perturber, $Z$ is the charge of the absorber (here, we are dealing with neutral atoms, so Z is always equal to 1), $E$ is the ionization potential of the absorber in eV, $E_i$ is the lower and $E_j$ is the upper excitation energy of the absorber in eV, $\alpha_H$ is the polarizability of hydrogen, and $E_H$ = 13.6 eV. The total Van der Waals damping constant is the sum of the effects from all the different perturbers:
\begin{equation}
    \gamma_{vw}^{tot} = \sum_p \gamma_{vw}^{p} \quad .
\end{equation}
Finally we convert the HWHM $\gamma_{vw}^{tot}$, which is in terms of a frequency width, to the desired FWHM in terms of wavenumber to use back in Equation \ref{chpt4_eq:AtomProfile}:
\begin{equation}\label{chpt4_eq:vanderwaal}
    \Delta\bar{\nu} = \frac{2\gamma_{vw}^{tot}}{c} \quad .
\end{equation} Finally, we apply a line-wing cutoff at $d$ = min(25$P$, 100 cm$^{-1}$), where $P$ is the total gas pressure in atmospheres \citep{Sharp2007}.

\subsection{Collision-Induced Absorption}
Collision induced absorption occurs when a close encounter with another molecule or atom induces a transient dipole moment which allows a rotational or ro-vibrational transition to occur which would not otherwise be possible. This manifests as a continuum opacity source rather than distinct spectral lines, although there are still broad spectral features corresponding to resonances. 

\begin{table}[]
\small
    \centering
    \begin{tabular}{cccc}
        Species & Temperature Range & Wavenumbers & Reference \\
         &  & (cm$^{-1}$) &  \\        
        \hline
        H$_2$-H$_2$ & 200 - 9000 K & 0-20,000 & HITRAN: \citealt{Abel2012}\\
        H$_2$-H$_2$ & 40-400 K & & HITRAN: \citealt{Fletcher2018}  \\
        H$_2$-He & 200 - 9000 K & 0-20,000 & HITRAN: \citealt{Abel2011}\\
        \hline
    \end{tabular}
    \caption{Sources of collision-induced-absorption included, and the corresponding references.}
    \label{chpt4_tab:CIA}
\end{table}

For giant exoplanets and brown dwarfs the primary sources of CIA are H$_2$ pairs and H$_2$-He pairs. We incorporate H$_2$-H$_2$ and H$_2$-He CIA into our opacity database using data from HITRAN\footnote{https://hitran.org/cia/}. The HITRAN data are in the form of CIA absorption coefficients with units cm$^{5}$ molecules$^{-2}$. These must be multiplied by the number density of each colliding species to get an inverse path length, and then divided by the overall mass density to get the monochromatic mass opacity utilized by radiative transfer calculations. A correction is also made for stimulated emission before combining with the other opacity sources. Comparison between new and old total opacity tables indicates that the HITRAN data are only slightly different than that used in \cite{Sharp2007}. 

The temperature and wavelength coverage of this data is summarized in Table \ref{chpt4_tab:CIA}. Data are also available for other types of CIA, but they are not important for most the environments of interest, aside from perhaps H$_2$-CH$_4$. Data for this is also available from HITRAN and may be included in future work.

\subsection{H- bound-free and free-free Absorption}

The negative hydrogen ion, $H^{-}$, is a major opacity source in cool stars and hot gaseous exoplanets which have temperatures between 2500 and 8000 K \citep{Chandrasekhar1945, Wildt1939}. $H^{-}$ absorbs through two continuum processes: (1) Bound-free interactions where a sufficiently energetic photon detaches the extra electron from the ion, h$\nu$ + H$^{-}$ $\rightarrow$ H + e$^{-}$, and (2) free-free interactions where a free electron passing near a hydrogen atom alters its potential, leading to continuum absorption of photons rather than distinct levels, h$\nu$ + H + e$^{-}$ $\rightarrow$ H + e$^{-}$. 

We include $H^{-}$ continuum absorption using the approach described in \cite{John1988}. This work is actually drawing from the calculations made in \cite{Bell1987} for free-free absorption and \cite{Wishart1979} for bound-free absorption. These data are identical to that used in \cite{Sharp2007}. It has long been available to high accuracy. 

\section{Comparison with other published grid}
\label{sec:literature}
In this appendix we compare our models against the Sonora-Bobcat grid (\citealt{Marley2021}). We consider only equilibrium chemistry and cloud-free models.

\begin{figure}
    \centering
    \includegraphics[width=0.75\textwidth]{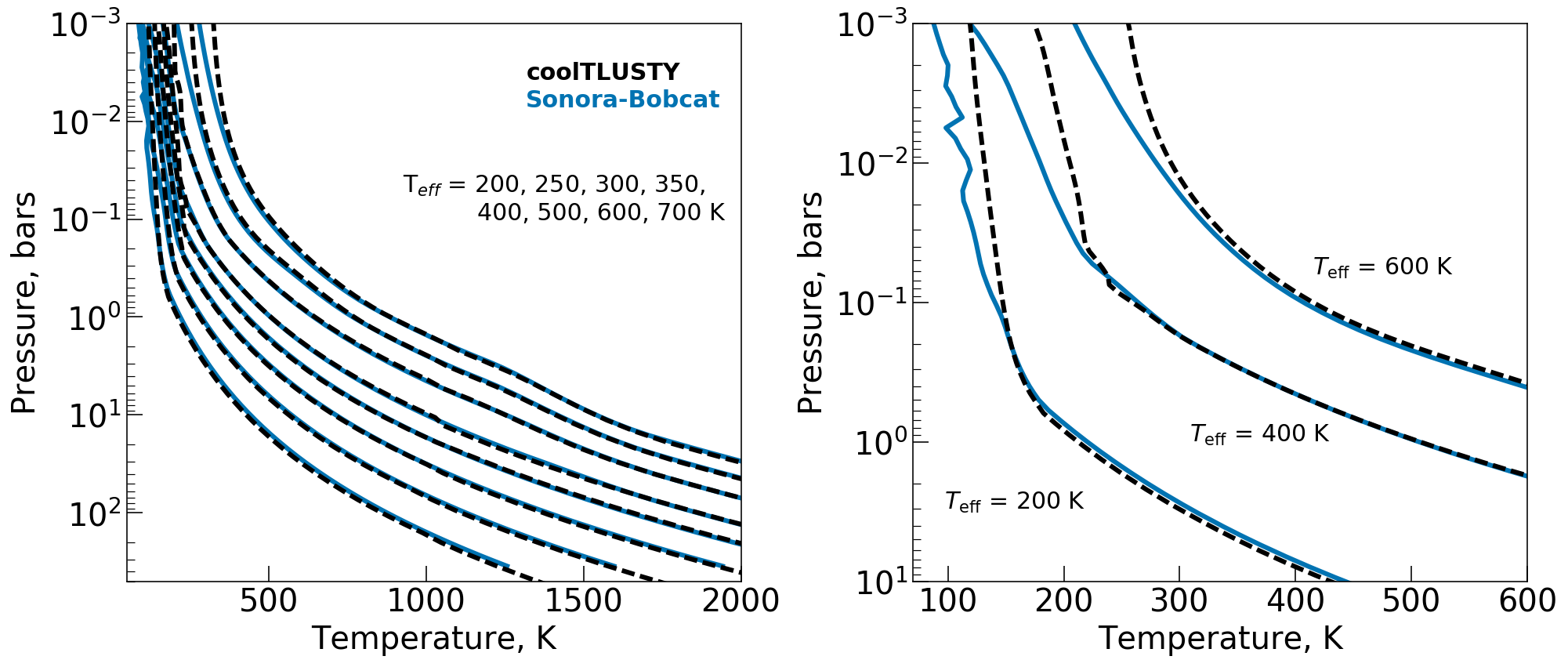}
    \caption{Left panel: P-T profiles for 1D radiative-convective equilibrium model atmospheres with $T_{\mathrm{eff}}$=200, 250, 300, 350, 400, 500, 600 and 700 K from left to right, all assuming a surface gravity of log$_{10}$($g$)=4. We compare the Sonora-Bobcat grid described in \cite{Marley2021} (blue lines), against our models (black dashed lines). One can see that the agreement deep in the atmosphere up through the typical photospheric layers is good, but there are deviations above 0.1 bars. The right panel zooms in on lower pressures for $T_{\mathrm{eff}}$=200, 400, and 600 K. \textit{coolTLUSTY} models converge to a more isothermal upper atmosphere than Sonoroa-Bobcat models at all effective temperatures. The squiggles in our 400 K model occur around the Clausius-Clapeyron line for water rainout. Because this deviation occurs at lower pressures above most of the photosphere, resulting spectra still agree (see Figure \ref{fig:other_grid_spec}).}
    \label{fig:other_grid_tp}
\end{figure}

First, Figure \ref{fig:other_grid_tp}, shows P-T profiles for a range of effective temperatures, all with a surface gravity of log$_{10}$($g$)=4 and solar metallicity. cool\textit{TLUSTY} is shown in black dashed lines, Sonora-Bobcat in blue solid lines. The right panel zooms in on the upper atmosphere for just $T_{\mathrm{eff}}$=200, 400 and 600 K to show the differences more clearly. Around $\sim$0.1 bars the models differ. Our profiles are warmer and closer to isothermal. This is suspected to be due to the differing treatment of opacities (k-tables vs. line-by-line), and to a difference in radiation convergence criterion in the upper atmosphere for cool\textit{TLUSTY}. Since the photosphere in these objects tends towards deeper pressures where the two model grids agree, there is little difference in the associated emergent spectra.

\begin{figure}
    \centering
    \includegraphics[width=0.9\textwidth]{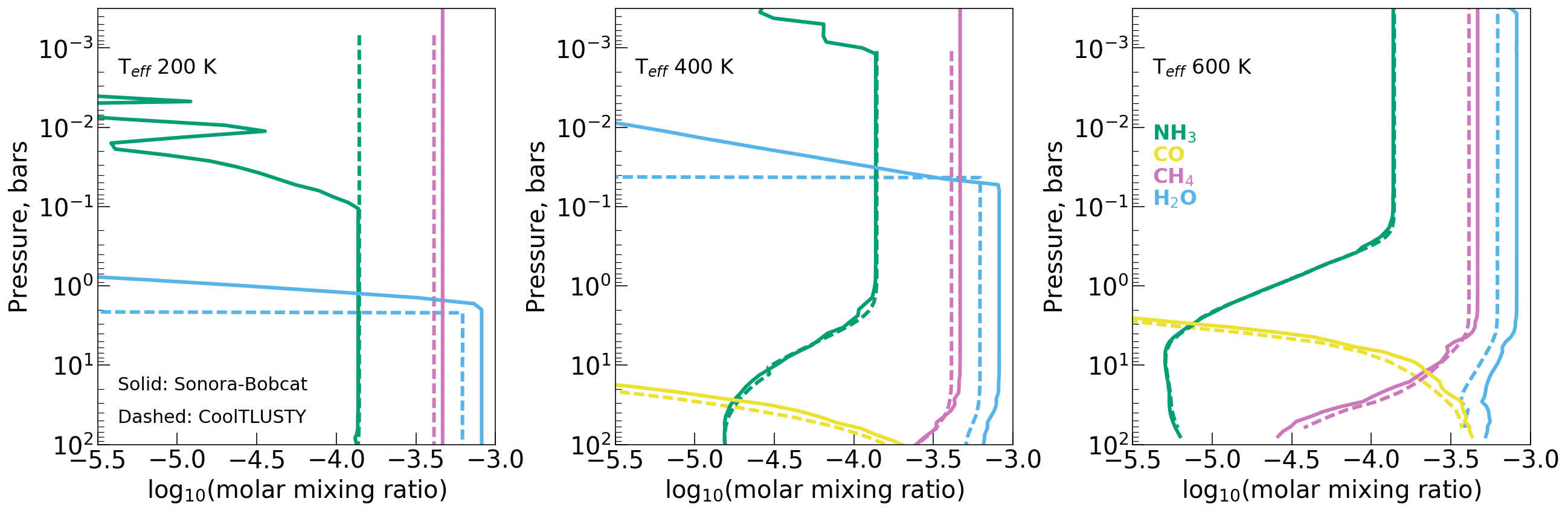}
    \caption{Abundance profiles of H$_2$O, CH$_4$, CO, and NH$_3$ for the same models as the right panel of Figure \ref{fig:other_grid_tp}. Solid lines show abundances for Sonora Bobcat and dashed lines show abundances for this work.}
    \label{fig:other_grid_chem}
\end{figure}

These deviations at low pressure could also be caused by different treatments of condensation and rainout of important opacity sources, namely water. Figure \ref{fig:other_grid_chem} shows abundance profiles of H$_2$O (blue lines), CH$_4$ (red lines), CO (green lines), and NH$_3$ (gray lines) for the same models as Figure \ref{fig:other_grid_tp}. The rainout chemistry calculations used with \textit{coolTLUSTY} used a very steep rate of water condensation compared to more gradual rates of Sonora-Bobcat. 

\begin{figure}
    \centering
    \includegraphics[width=0.9\textwidth]{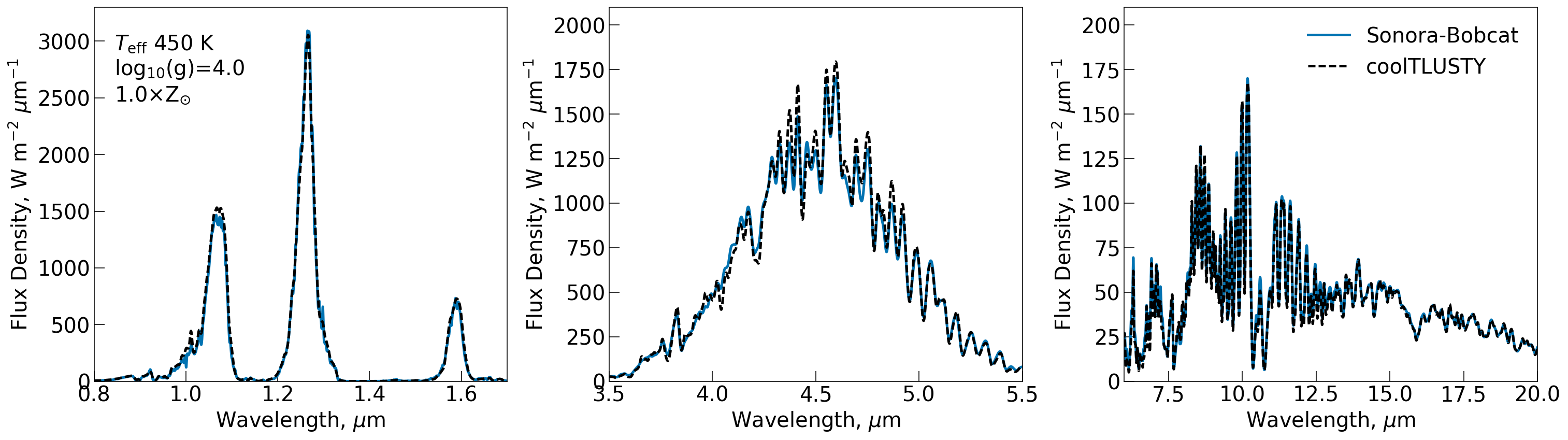}
    \caption{Comparison of spectra computed by \textit{coolTLUSTY} (black dashed lines) and Sonora-Bobcat (blue lines). Models are shown for atmospheres with an effective temperature 450 K, a surface gravity of log$_{10}$(g)=4, solar metallicity, equilibrium chemistry, and no clouds. Note the different Y axes between columns. Fluxes are the emergent flux from the surface. Spectra have been convolved to have roughly equivalent spectral resolution.}
    \label{fig:other_grid_spec}
\end{figure}

Figure \ref{fig:other_grid_spec} shows the associated spectra for these models, convolved to similar resolutions.

\end{document}